\documentclass[twocolumn, trackchanges]{aastex701}

\definecolor{green}{rgb}{0.0, 0.8, 0.1}

\begin{document}

% \title{Template \aastex Article with Examples: 
% v6.3.1\footnote{Released on March, 1st, 2021}}
\title{Rotation Periods of Candidate Single Late-M Dwarfs in TESS}

\author[0000-0003-3219-4818]{Samantha Lambier}
\affiliation{Department of Physics and Astronomy, The University of Western Ontario, 1151 Richmond St, \\London, Ontario, N6A 3K7, Canada}
\affiliation{Institute for Earth and Space Exploration, The University of Western Ontario, 1151 Richmond St, \\London, Ontario, N6A 3K7, Canada}
\email[show]{slambier@uwo.ca}

\author[0000-0003-3050-8203]{Stanimir Metchev}
\affiliation{Department of Physics and Astronomy, The University of Western Ontario, 1151 Richmond St, \\London, Ontario, N6A 3K7, Canada}
\affiliation{Institute for Earth and Space Exploration, The University of Western Ontario, 1151 Richmond St, \\London, Ontario, N6A 3K7, Canada}
\email{smetchev@uwo.ca}

\author[0000-0003-2446-8882]{Paulo Miles-P\'{a}ez}
\affiliation{Centro de Astrobiolog\'ia, CSIC-INTA, Camino Bajo del Castillo s/n, 28692 Villanueva de la Ca\~nada, Madrid, Spain}
\email{pamiles@cab.inta-csic.es}

\author[0000-0001-7171-5538]{Leslie Moranta}
\affiliation{Plan\'{e}tarium Rio Tinto Alcan, Espace pour la Vie, 4801 av. Pierre-de Coubertin, Montr\'{e}al, Qu\'{e}bec, Canada}
\affiliation{Institute for Research on Exoplanets, Universit\'{e} de Montr\'{e}al, D\'{e}partement de Physique, C.P. 6128 Succ. Centre-ville, \\Montr\'{e}al, QC H3C 3J7, Canada}
\email{leslie.moranta@umontreal.ca}

\author{Dakota Wolfe}
\affiliation{Department of Physics and Astronomy, The University of Western Ontario, 1151 Richmond St, \\London, Ontario, N6A 3K7, Canada}
\email{dwolfe5@uwo.ca}

\author{Joelene Hales}
\affiliation{Department of Physics and Astronomy, The University of Western Ontario, 1151 Richmond St, \\London, Ontario, N6A 3K7, Canada}
\email{jhales5@uwo.ca}

\author{Jeffrey Martinovic}
\affiliation{Department of Physics and Astronomy, The University of Western Ontario, 1151 Richmond St, \\London, Ontario, N6A 3K7, Canada}
\email{jmart487@uwo.ca}



\begin{abstract}

% This example manuscript is intended to serve as a tutorial and template for
% authors to use when writing their own AAS Journal articles. The manuscript
% includes a history of \aastex\ and includes figure and table examples to illustrate these features. Information on features not explicitly mentioned in the article can be viewed in the manuscript comments or more extensive online
% documentation. Authors are welcome replace the text, tables, figures, and
% bibliography with their own and submit the resulting manuscript to the AAS
% Journals peer review system.  The first lesson in the tutorial is to remind
% authors that the AAS Journals, the Astrophysical Journal (ApJ), the
% Astrophysical Journal Letters (ApJL), the Astronomical Journal (AJ), and
% the Planetary Science Journal (PSJ) all have a 250 word limit for the 
% abstract\footnote{Abstracts for Research Notes of the American Astronomical 
% Society (RNAAS) are limited to 150 words}.  If you exceed this length the
% Editorial office will ask you to shorten it. This abstract has 161 words.

Recent studies suggest that the angular momentum evolution of late-M and brown dwarfs differs from the well-known spin-down evolution of hotter stars. Characterizing the distribution of rotation periods of these objects in the solar neighborhood can help elucidate this evolutionary pathway just above, at, and below the hydrogen burning limit. In this paper, we examine 399 candidate single late-M dwarfs with $G - G_{RP} \geq 1.4$ mag ($\gtrsim$M6) using TESS light curves. To determine rotation periods, we employed Lomb-Scargle Periodograms to provide a first estimate of the period, then refined them with a Gaussian Process approach, requiring multi-sector confirmation when available. 
We found 133 rotation periods, ranging from 2 hours to 6 days, and amplitudes between 0.08\% and 2.71\%. We find that the observed variability fraction in late-M dwarfs rises with the number of available TESS sectors, approaching an apparent ceiling of $\sim 50\%$. This likely reflects a detection limit determined by viewing geometry and supports the idea that spot-induced variability is common across the late-M and brown dwarf population. In our comparison with previously published late-M dwarf rotation periods reported, we found consistent results, confirming or updating 31 periods. Our findings expand the number of previously known late-M dwarf periods under 1 day by 76\%. Combined with published rotation periods for a broader range of spectral types, we find a lower envelope on the rotation period decreasing from 5 hours at early-M dwarfs to 1 hour at L, T, and Y dwarfs. 

\end{abstract}

\keywords{\uat{M dwarf stars}{982} --- \uat{Observational astronomy}{1145} ---  \uat{Periodic variable stars}{1213} --- \uat{Stellar rotation}{1629}}


\section{Introduction} \label{sec:intro}

% Late-M dwarfs ($\gtrsim$M6, $\leq$0.1M$_\odot$) span the range from the onset of full convection in low-mass stars to the stellar–substellar boundary.  \textcolor{red}{(SM: This is a nice intro sentence. I would add a spectral type and mass range. However, this sentence sort of repeats the 3rd sentence. I suggest consolidating them.}
%Ultracool dwarfs \citep[UCDs; $\gtrsim$M7,][]{kirkpatrick1995}  span the boundary between stars and substellar objects. 
Late-M dwarfs ($\gtrsim$M6, $\leq$0.1M$_\odot$) can offer valuable insight into the transition between the mechanisms that govern the formation and evolution of stars and brown dwarfs. Late-M dwarfs are fully convective, contrasting with higher-mass stars (M3, $>$0.035M$_\odot$) that have a radiative core and convective envelope \citep{dorman1989, baraffe1997, chabrier1997, jao2018}. 

Below the convective boundary, stars must undergo a different angular momentum evolution than the ``Skumanich-style" spin-down of hotter stars \citep{skumanich1972}. This is because the Skumanich spin-down is driven by magnetized stellar winds with magnetic fields organized in the tachocline, the transition region between the radiative core and convective envelope \citep[e.g.][]{spiegel1992, dikpati1999, miesch2005}. Fully convective stars, lacking a tachocline, might not behave in the same manner.

It has been shown that fully convective stars close to the convective boundary lose angular momentum faster than partially convective stars on the other side of the boundary, with torques $\sim$1.5 higher for a given angular velocity \citep{lu2024}. %\sout{Furthermore,} \textcolor{red}{(SM: The use of introductory words in consecutuve sentences burdens the logic flow.)} 
This switch between angular momentum loss mechanisms may help to explain the bimodal nature of mid-M stellar rotation \citep[e.g.][]{newton2017, newton2018, lu2022, lu2024}. 
However, late-M dwarfs, which have not been studied in as much detail, may behave differently still. \citet{pass2022} found that some mid-M dwarfs might not experience spin-down, while \citet{tannock2021} suggested that some brown dwarfs may even spin up due to reduced stellar winds or disk-driven momentum loss. Moreover, the $\sim$1 hour limit to brown dwarf rotation proposed by \citet{tannock2021} appears to mark a lower envelope of rotation periods which continue out to approximately 2 hours for late-M dwarfs \citep{miles2023}. A larger dataset, like that provided in this work, would better refine the lower envelope of late-M dwarf rotation.

All-sky synoptic surveys are instrumental in determining the rotation periods of large numbers of stars. The Transiting Exoplanet Survey Satellite \citep[TESS;][]{ricker2015}, in particular, is an excellent tool as it provides light curves for hundreds of thousands of stars. However, as a red-optical instrument, TESS is not optimally sensitive to late-M dwarfs whose spectral energy distributions peak in the near-infrared, and late-M dwarf periods have been under-represented in most TESS-based M dwarf rotation studies (see Section~\ref{sec:literature}). With the present study, we aim to extract the maximum possible information, specifically for late-M dwarf rotations from TESS, extending earlier work by \citet{miles2023}. %TESS, however, is a red-optical instrument, and thus, this paper on UCDs pushes the limits of TESS's capability to detect rotation periods of optically faint stars.

%To detect the rotation period of a UCD, we rely on variations in the light curve of a star.
Near the hydrogen-burning limit, variability is likely to be caused by starspots and flares caused by magnetic activity \citep{hooten1990}. Dust condensates, in the form of hazes or clouds, can also create inhomogeneities and produce variability. Dust particles such as corundum (Al$_2$O$_3$) begin to condense in the stellar atmosphere around a spectral type of M7 \citep{tsuji1996}. Regardless of the nature of the inhomogeneities, their rotation in and out of view as the star spins causes quasi-sinusoidal variations that can track the stellar rotation period \citep[e.g.][]{tinney1999, bailer2002}. 
%As such, a model to fit the quasi-periodic nature of these light curves is more robust than a purely periodic one in determining rotation periods \citep{angus2018}. 

\section{Input Sample}
\label{sec:input_sample}
% \subsection{Initial selection from Gaia DR3}
% \textbf{Taken directly from thesis, needs major modification}.
Our goal is to determine the rotation periods of all single M6-type or later dwarf stars in TESS. We started by selecting targets from the Gaia Catalogue of Nearby Stars \citep[GCNS;][]{smart2021}, the 100 pc subset of Gaia EDR3 \citep{gaiacollaboration2020d}, and updated their $G$ magnitudes from Gaia DR3 \citep{vallenari2023}.
%a subset of the entire Gaia DR3 within 100 pc, and matched them against the Gaia DR3 \citep{vallenari2023} catalogue to obtain updated $G$ magnitudes. 
The GCNS catalogue also contains photometric information from 2MASS \citep{skrutskie2006}, which was utilized in the selection criteria.

We used a $G - G_{RP} \geq 1.40$ mag color criterion to select $\geq$M6 dwarfs from Gaia. This is slightly less stringent than what may be required to select strictly $\geq$M6 dwarfs \citep[$G - G_{RP} \geq 1.43$ mag;][]{pecaut2013}. We opted for the relaxed color cut to allow for completeness over precision in our sample. The $G - G_{RP} \geq 1.40$ mag color selection includes %, which corresponds to 
stars approximately M5.5 or cooler, and a few particularly red M5 dwarfs \citep{pecaut2013}. We did not set a red limit on the $G - G_{RP}$ color, as we would like to determine rotation periods for as late a spectral type TESS would allow. We set the absolute magnitude cutoff to M$_{RP} \geq 10.2$ mag to limit our selection to dwarf stars. As we discuss in Section~\ref{sec:spt}, we refer to this sample as $\geq$M6 dwarfs or ``late-M dwarfs'', while acknowledging that it includes dwarfs as early as M5 and eight L dwarfs as late as L9.

This first estimate of the population of late-M dwarfs in Gaia includes 42,165 objects. In the following subsections, we refine our sample selection to minimize unreliable results by addressing %However, it ignores 
several confounding issues, including: %such as 
unreliable Gaia colors for extremely red objects, the sensitivity limits of TESS, unresolved binaries in Gaia, %and---specific to our TESS analysis---the sensitivity limitations of TESS, 
contamination by nearby sources in TESS, and the availability of TESS time-series photometry. %In the following subsections, we describe how we address these factors to minimize ambiguous or unreliable results. 

\subsection{Removing Abnormal Color Excess}
\label{sec:excess}

\begin{figure}
	\centering
	\includegraphics[width=0.9\linewidth]{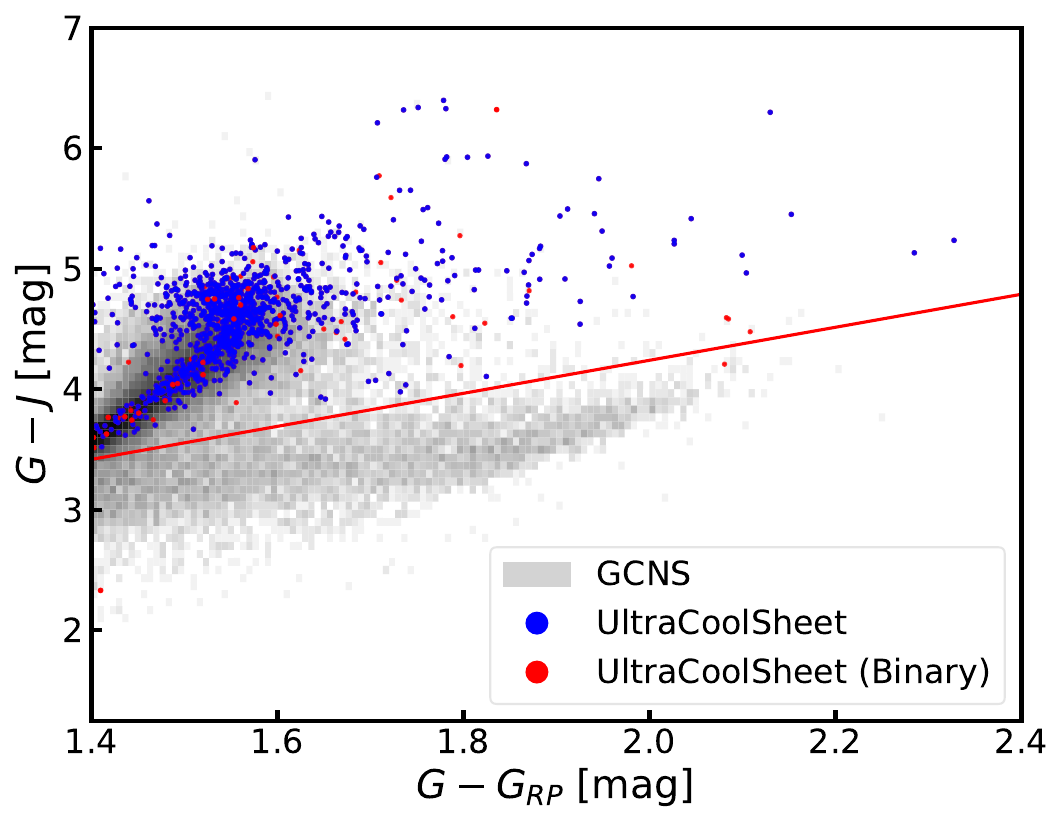}
	\caption{Gaia $G$ - 2MASS $J$ color vs.\ Gaia $G - G_{RP}$ color. The Gaia Catalogue of Nearby Stars is plotted as a density map in gray, with spectroscopically confirmed single field dwarfs from the UltraCoolSheet \citep{best2024} plotted in blue and binary field dwarfs in red. The red line marks the cutoff above which we select stars with reliable Gaia photometry, as computed in Section \ref{sec:excess}.}
	\label{fig:gj_grp}
\end{figure}

Gaia $G_{BP}$ and $G_{RP}$ photometry is sensitive to background overestimation, which can lead to spurious results using only simple color and absolute magnitude cutoffs. This is because the $G_{BP}$ and $G_{RP}$ flux is the total flux in a 3.5$\times$2.1 arcsec$^2$ field, as opposed to $G$-band flux, which is fit to a profile \citep{evans2018}. The standard metric to detect overestimation is the corrected color excess factor $C^*$ as provided by \citet{riello2021}. The corrected color excess factor is positive when the sum of the $G_{BP}$ and $G_{RP}$ flux is greater than the $G$ flux and negative when the opposite is true.

As late-M dwarfs and brown dwarfs are faint, red objects, they are especially sensitive to background overestimation at $G_{BP}$ %in the G$_{BP}$-band 
due to low flux in that band. This can inflate $C^*$ values for the faintest and reddest stars to levels above the recommended threshold for other stars \citep{reyle2018}. As such, we do not use $C^*$ as a cutoff. 

Following work by \citet{reyle2018}, we instead trace color excess by plotting $G - G_{RP}$ against $G - J$ (Figure \ref{fig:gj_grp}) to remove $G_{BP}$ as a factor.  We overlay all of the spectroscopically confirmed objects from the UltraCoolSheet \citep{best2024}, which is composed of objects with spectral types M6 and later.  As seen in Figure \ref{fig:gj_grp}, we find that they lie in the redder ($G-J > 3.5$ mag) lobe of the data, which corresponds to non-overestimated $G_{RP}$ flux. %photometry. 
To remove the spurious stars, we empirically fit a line to the lower envelope of UltraCoolSheet objects and include an additional offset of $\Delta(G-J)$ of $-$0.1 mag to account for the photometric uncertainties of 2MASS magnitudes at $J$-band signal-to-noise (SNR) $\gtrsim$ 10. This resulting cutoff, indicated by a red line in Figure \ref{fig:gj_grp}, is $G - J \geq 1.37(G - G_{RP}) + 1.50$ mag, above which lie all 37,148 late-M dwarfs and brown dwarfs in the GCNS with reliable Gaia photometry.

\subsection{TESS Sensitivity Limitations}
\label{sec:tess_lims}
As we are only interested in objects with TESS light curves, we limit our sample based on TESS magnitude. The expected TESS magnitude was calculated from Gaia photometry using the formula
\begin{eqnarray}
    T = G - 0.00522555(G_{BP} - G_{RP})^3 \nonumber \\+ 0.0891337(G_{BP} - G_{RP})^2  \\ - 0.633923(G_{BP} - G_{RP}) + 0.0324473 \nonumber
\end{eqnarray}
as determined in \citet{stassun2019}. While the relationship deteriorates in the M-dwarf regime ($G_{BP} - G_{RP} > 2$ mag), it is still accurate to $\pm$0.1~mag at $G_{BP} - G_{RP} \sim 4$ mag (spectral type $\sim$L0). We set a limit of $T \leq 18$ mag. At this magnitude, the smallest light curve variation detectable in TESS is approximately 10\% \citep{miles2023}, which is already larger than the variation expected for periodic signals of stellar rotation. As a result, we expect to be complete to all small-amplitude periodic variations detectable by TESS. This results in a list of 21,405 candidate late-M dwarfs, which are listed in Table \ref{tab:alltargets}.

\begin{deluxetable*}{ccccccccc}
    \tabletypesize{\footnotesize}
    \digitalasset
    \tablecaption{All 21,405 candidate late-M dwarfs with $T\leq18$ mag \label{tab:alltargets}}
    \tablewidth{0pt}
    \tablehead{
        \colhead{Gaia ID} & 
        \colhead{$M_{RP}$} & 
        \colhead{$T$ mag} & 
        \colhead{$G - G_{RP}$} & 
        \colhead{$G - J$} & 
        \colhead{Not Binary} & 
        \colhead{Uncrowded} & 
        \colhead{Has PDCSAP Data} & 
        \colhead{Periodic}
    }
    \startdata
    90022514620544 & 11.80 & 16.49 & 1.53 & 3.87 & No & No &   &   \\
689569884943232 & 11.99 & 16.33 & 1.40 & 3.62 & Yes & Yes & No &   \\
756571374829440 & 12.28 & 15.22 & 1.42 & 3.63 & Yes & No &   &   \\
2781562554898432 & 13.84 & 17.04 & 1.55 & 4.30 & Yes & No &   &   \\
2838320548071296 & 12.25 & 17.43 & 1.42 & 3.65 & Yes & No &   &   \\
3036954195695616 & 13.26 & 16.93 & 1.51 & 4.13 & Yes & No &   &   \\
3124060427493120 & 13.55 & 16.34 & 1.53 & 4.26 & Yes & Yes & No &   \\
4719932835786752 & 12.43 & 16.72 & 1.42 & 3.64 & Yes & Yes & No &   \\
5780304426212864 & 12.41 & 16.88 & 1.41 & 3.66 & Yes & No &   &   \\
5797037618821888 & 13.12 & 17.69 & 1.49 & 4.10 & Yes & Yes & No &   \\
    ... & ... & ...& ...& ...& ...& ...& ... & ... \\
    \enddata
    \tablecomments{The full table is available in machine-readable form.}

%\textcolor{red}{(SM: Verify that this refers to the 1826 candidate single uncontaminated UCDs, or the 399 ones with PDCSAP data. If the latter amend the table heading to reflect this.)}

\end{deluxetable*}

%  \subsection{Cross-matching with other catalogues}

\subsection{Removing Unresolved Binaries in Gaia}
\label{sec:binaries}
 
This work aims to focus on the rotation of single stars, as binarity can impact a star's angular momentum evolution. The first criterion to remove binaries is that the Gaia Renormalized Unit Weight Error (RUWE) needs to be less than or equal to 1.4 \citep{lindegren2018}. This is the standard value used in the literature, as there is a strong delineation between well- and poorly-behaved solutions for single stars at RUWE = 1.4 \citep{lindegren2018}. As such, we use RUWE as one indicator of binarity and remove stars with RUWE $\geq$ 1.4. It is important to note that some binaries 
%particularly equal-flux binaries, 
can have values below this cut-off as seen in recent works \citep[e.g.][]{stassun2021, penoyre2022, castro2024}, so this should not be used in isolation.

Equal-flux binary stars that are unresolved in Gaia can be seen in the catalogue when objects are plotted on an M$_{RP}$ vs. $G-G_{RP}$ color-magnitude diagram. These objects lie clustered 0.753 magnitudes above the main sequence in what is known as the binary main sequence \citep{hurley1998}. 

%Thus, these objects could be removed by fitting a curve to the main sequence and then applying an offset of approximately 0.38 magnitudes, halfway to the binary main sequence. 

\begin{figure}
	\centering
	\includegraphics[width=1.0\linewidth]{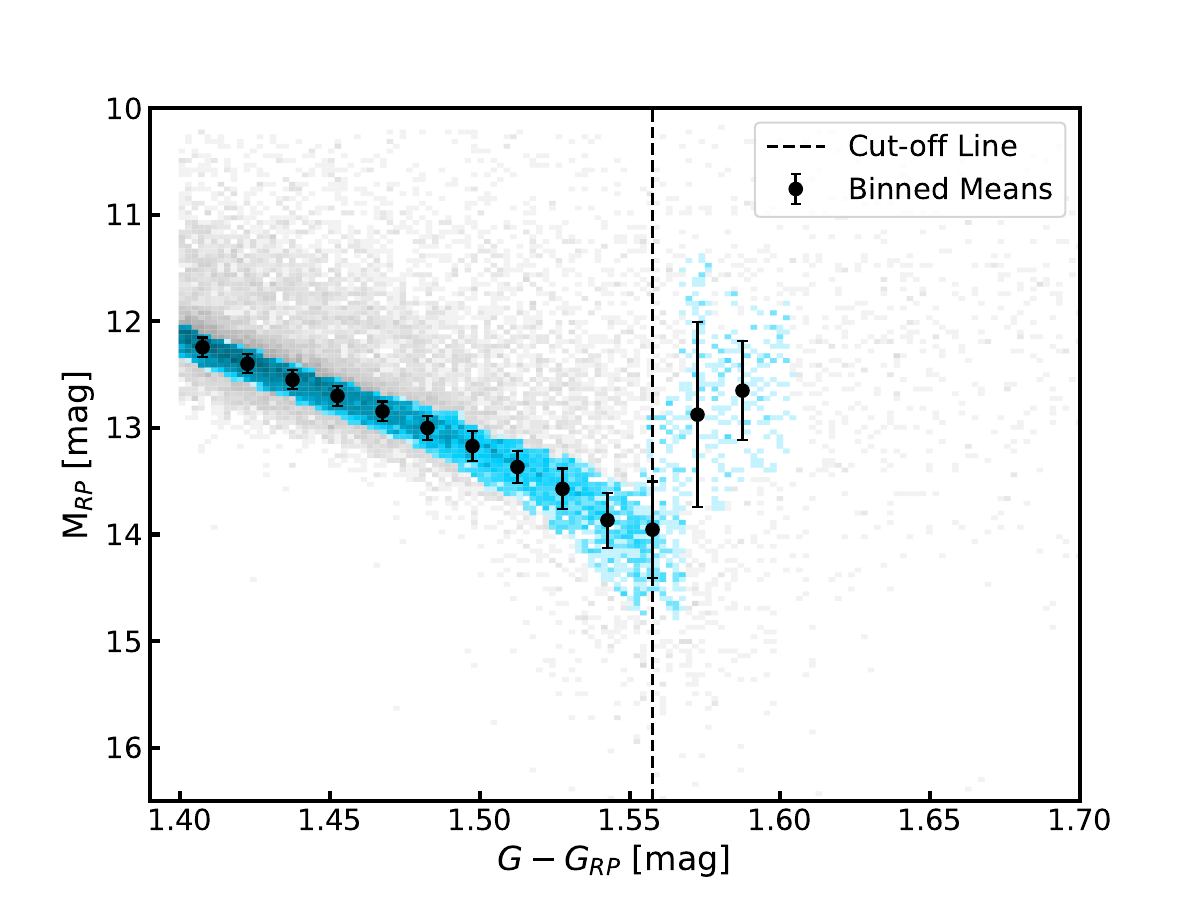}
	\caption{Empirical determination of the Gaia main sequence for $G - G_{RP} >$ 1.40 mag late-M dwarfs. This Gaia $G - G_{RP}$ vs. $G_{RP}$ color-magnitude diagram shows the 21,405 $T \leq 18$ mag targets in the input sample. The 50\% of the objects removed above and below the main sequence are in gray, with the remaining objects in blue. The mean and standard deviation for each 0.015 mag-wide bin are plotted in black. %\textcolor{red}{(SM: I suggest making this a 2D histogram, like you've done for Figure 3. Otherwise it is hard to tell that there is an over-density at the locus of the main sequence.)}
    }
	\label{fig:cmd_ms_fit}
\end{figure}

\begin{figure}
	\centering
	\includegraphics[width=1.0\linewidth]{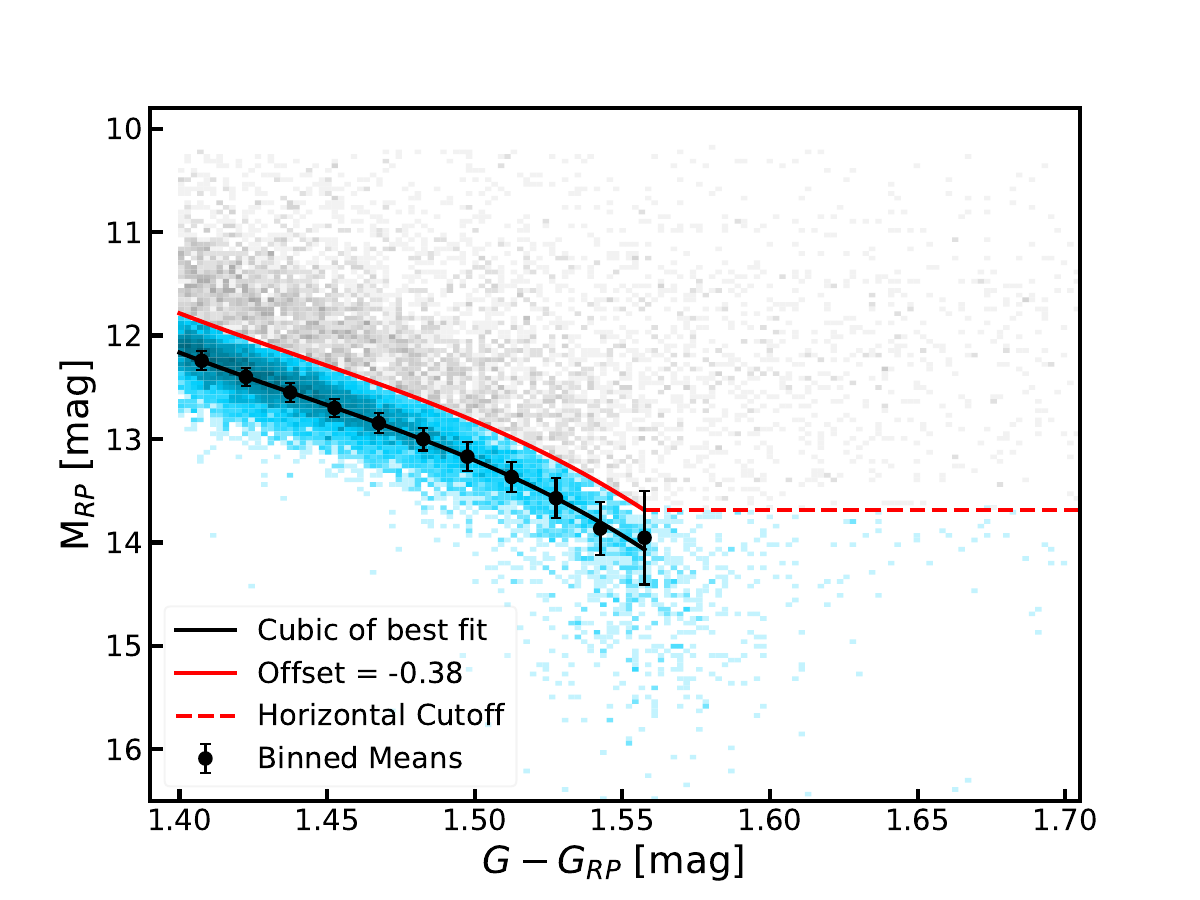}
	\caption{Selection of candidate single late-M dwarfs. The same color-magnitude diagram seen in Figure \ref{fig:cmd_ms_fit} is displayed as a two-dimensional histogram to show the density of stars with candidate single stars in blue and candidate binaries in gray.  The black line shows the best fit to the main sequence (binned black data points). Plotted in red is the dividing line 0.38 mag above the empirical main sequence (Section~\ref{sec:binaries}), used to remove candidate equal-flux binary stars. The horizontal dashed line at $M_{RP}=13.69$ mag is an extension of this dividing line for $G-G_{RP}>1.56$ mag colors. }
	\label{fig:cmd_binary}
\end{figure}

To first determine an empirical main sequence, we used a sliding bin with a width of 0.015 mag and a step size of 0.005 mag. For each step, we used a cumulative distribution function to keep 50\% of the data with the narrowest $G - G_{RP}$ range, i.e. the 50\% of the data with the highest $M_{RP}$ density per $G - G_{RP}$ bin.  The remaining data were rebinned with a bin size of 0.015 mag to perform least-squares fitting on the means and standard deviations. This can be seen in Figure \ref{fig:cmd_ms_fit}.

The curve best fitting for the late-M dwarf main sequence is 
% \begin{eqnarray}
\begin{eqnarray}
M_{RP} = 212.63(G - G_{RP})^3 - 919.02(G - G_{RP})^2 \nonumber \\
+ 1333.87(G - G_{RP}) - 637.44.    
\end{eqnarray}

We apply a $-0.38$ mag vertical offset to this empirical main sequence to the halfway locus between the main sequence and the binary main sequence (blue line parallel to the main sequence fit in Figure~\ref{fig:cmd_binary}). All candidate single late-M dwarfs are fainter than this locus.

As there are fewer data points and a larger scatter at the red end of our sample, the empirical main sequence becomes less well-defined, and the division between the main sequence and binary main sequence becomes unclear. Therefore, we do not fit the entire main sequence. Instead, we stop fitting the main sequence at the point where the mean of the remaining data deviates from the trend. As seen in Figure \ref{fig:cmd_ms_fit}, this occurs at $G - G_{RP} = 1.56$ mag. For redder Gaia colors, we extend the line dividing candidate single and candidate binary late-M dwarfs at a constant $M_{RP} = 13.69$ mag. All stars fainter than this combined cutoff are considered single and are kept in the sample. A visual representation of this cutoff can be seen in Figure \ref{fig:cmd_binary}.

Of the 21,405 objects identified above, 15,478 remain after binary removal; those with RUWE $\geq 1.4$ and candidate unresolved equal-flux binaries. We note, however, that these are called ``candidates'' because despite the combined RUWE and photometric selection criteria, some unresolved binaries may still remain in the sample.

\subsection{Removing Targets in Crowded TESS Fields}
\label{sec:crowded}

\begin{figure*}
	\centering
	
        \includegraphics[width=0.42\linewidth]{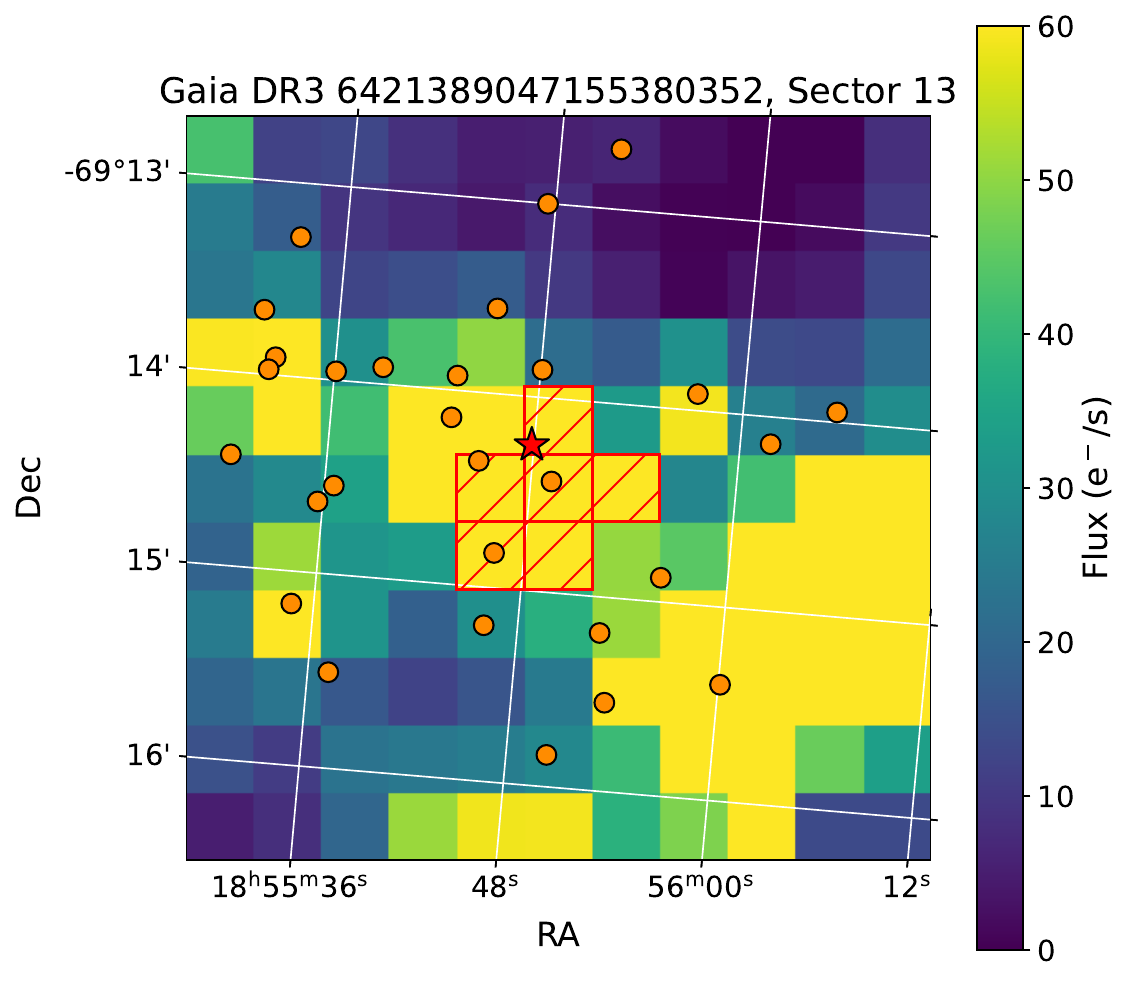}
        \includegraphics[width=0.42\linewidth]{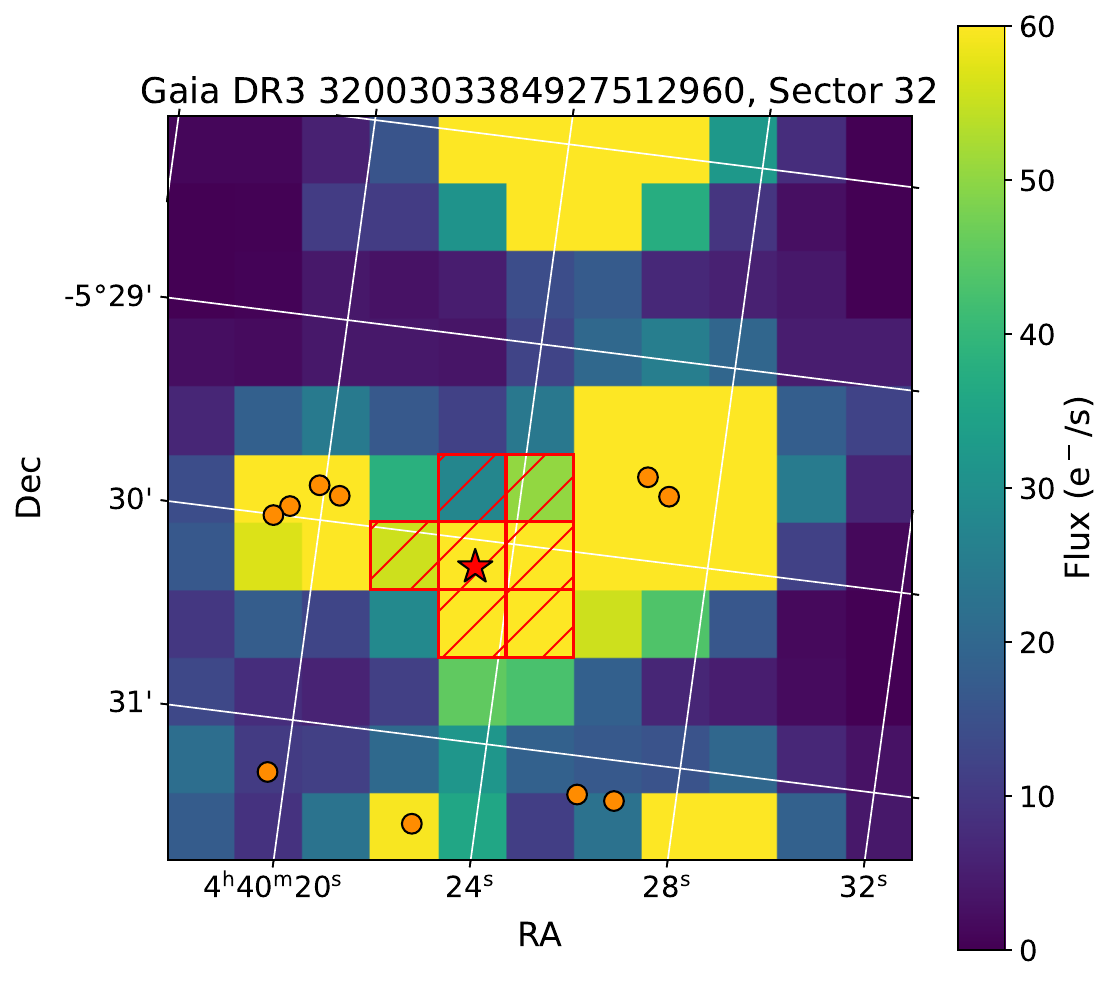}
        \includegraphics[width=0.42\linewidth]{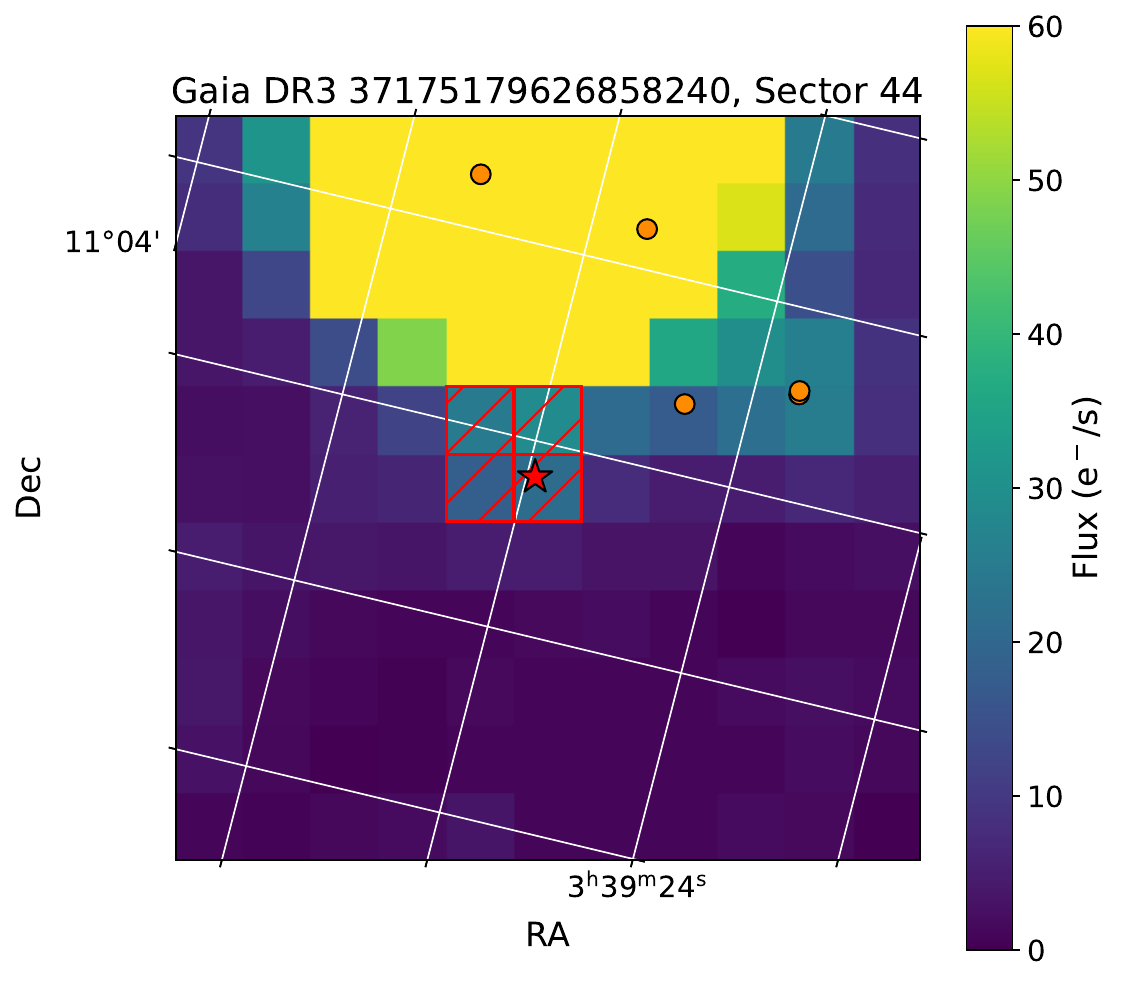}
        \includegraphics[width=0.42\linewidth]{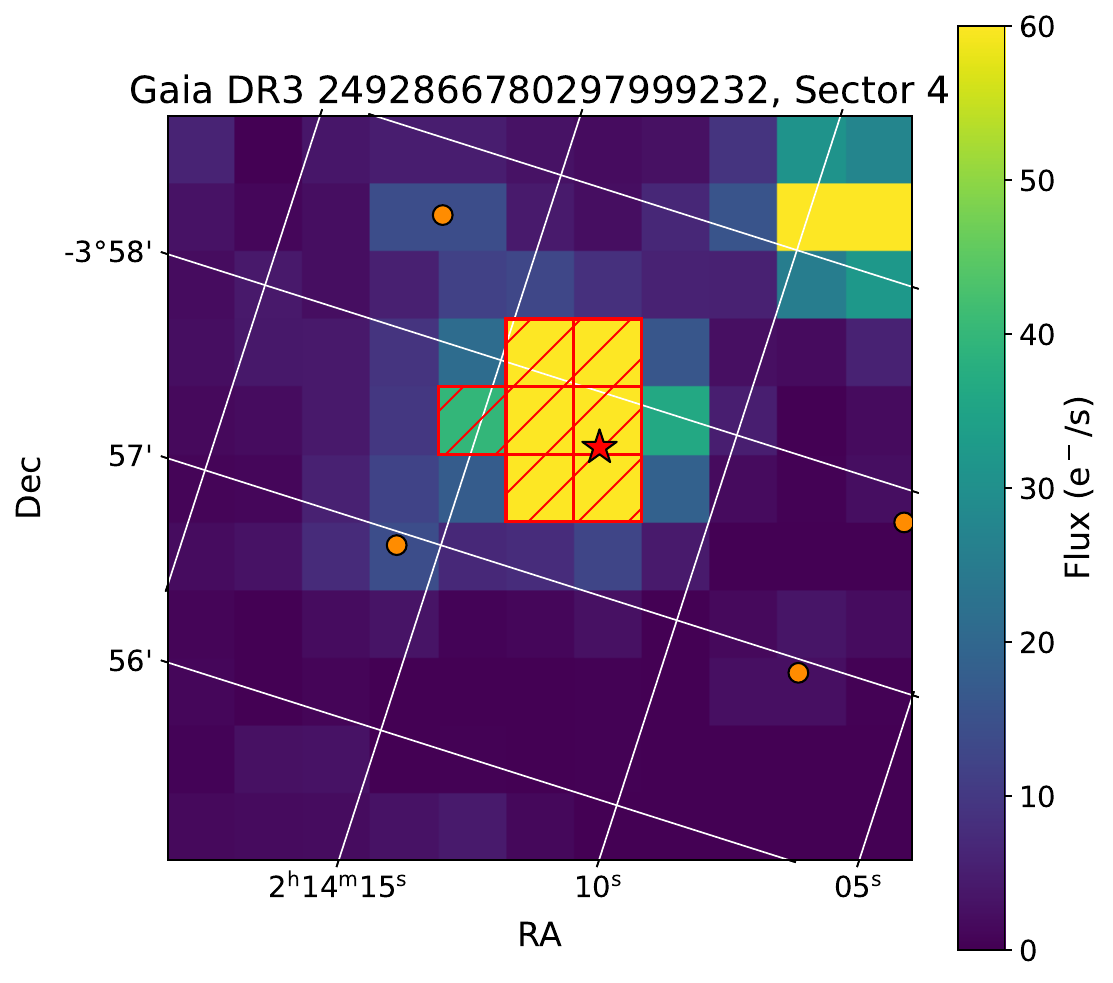}
	\caption{Target pixel files displaying three stars which are in crowded fields based on criteria i (top left), ii (top right), iii (bottom left) and one which is isolated (bottom right). The target (listed in the title of each plot) is plotted as a red star, while nearby stars with $G < 20$ mag in Gaia are plotted as orange circles. The red-hatched boxes denote pixels within the SPOC PDCSAP aperture.}
	\label{fig:tpfs}
\end{figure*}

Due to the large pixel size ($21^{\prime\prime}$) of TESS, contamination is a major concern in determining the source of periodicity in a light curve. Multiple stars can be contained within a single pixel. In addition, it has been found that stars more than three pixels away from the aperture can contaminate light curves \citep{higgins2023}. We designed a set of criteria that eliminate stars contaminated by crowding in TESS (Figure \ref{fig:tpfs}). Stars affected by crowding have:
\renewcommand{\theenumi}{\roman{enumi}.}%
\begin{enumerate}
    \item one or more $G\leq20$ mag stars within a one-pixel ($21^{\prime\prime}$) radius, or
    \item one or more stars of equal brightness or brighter than the target within a three-pixel ($63^{\prime\prime}$) radius, or
    \item one or more $G\leq12$ mag stars within a five-pixel ($1.75^{\prime}$) radius.
\end{enumerate}

Contamination by brighter stars at further distances could still be present for a very small number of stars at this step. However, as this should only affect a handful of targets, we allow these targets to be removed in the final stages of analysis by \texttt{TESS-Localize} \citep[see Section \ref{sec:tl}]{higgins2023}. At this stage, only 12\% of this list (1816 targets) were uncrowded and considered for the next steps.

\subsection{Availability of TESS SPOC PDCSAP Data}
\label{sec:spoc_pdcsap}
The final list of targets to be analyzed was determined by the availability of a high-level processed light curve from TESS. Specifically, we required that targets be processed with the TESS Science Processing Operations Center (SPOC) pipeline using Pre-search Data Conditioning Simple Aperture Photometry \citep[PDCSAP;][]{jenkins2016}. We did not require 2-minute cadence data, although we preferentially used it when available. When unavailable, we used the longer 1800s (Sectors 1-26), 600s (Sectors 27-55), or 200s (Sectors 56+) cadences. The 20s cadences were not used for the sake of computation time. Data were downloaded on 2025 January 16th, with data available up to Sector 85. 

\begin{figure}
	\centering
	\includegraphics[width=1.0\linewidth]{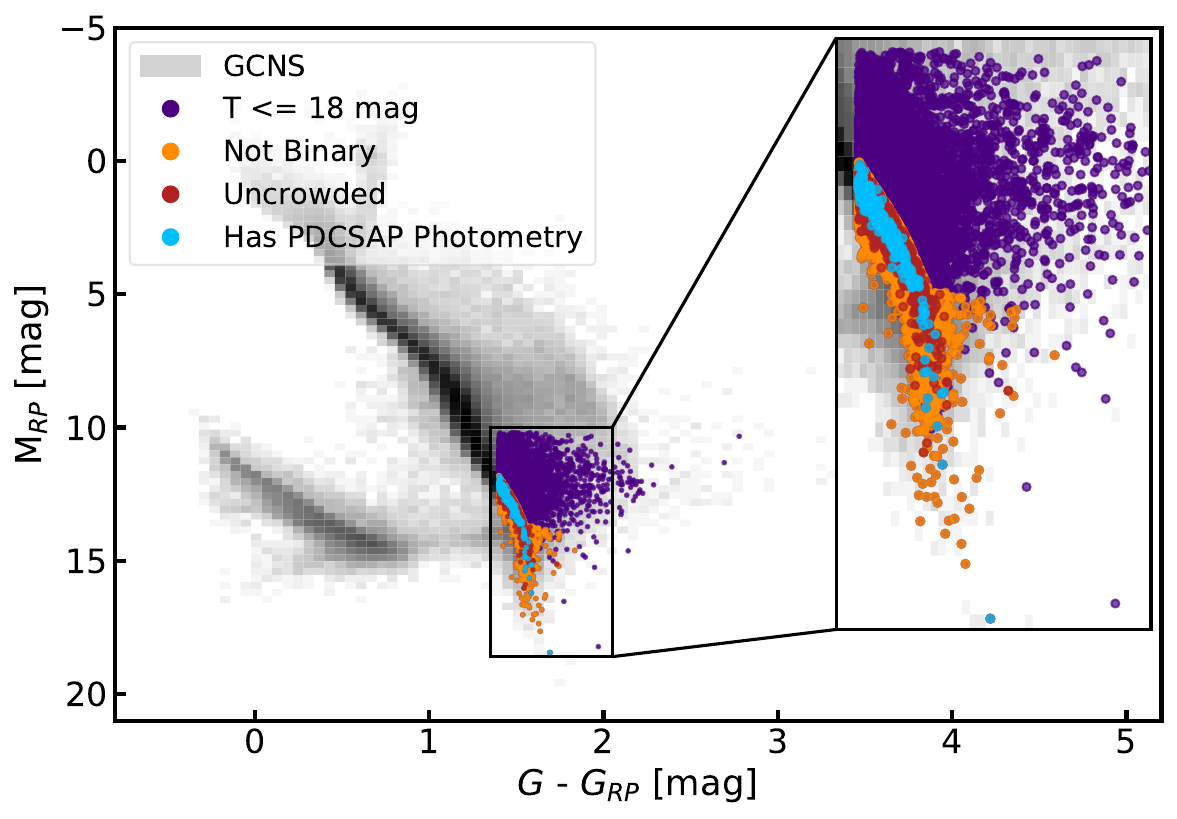}
	\caption{A color-magnitude diagram of the Gaia Catalogue of Nearby Stars, overlaid with late-M dwarf candidates. The colors indicate progressive selection for stars that: are brighter than TESS $T$ magnitude 18 (purple), are likely single (orange), are in uncrowded fields (red), and have PDCSAP photometry (blue; see Section~\ref{sec:data_filtering}).}
	\label{fig:cmd_all}
\end{figure}

\begin{figure}
	\centering
	\includegraphics[width=0.9\linewidth]{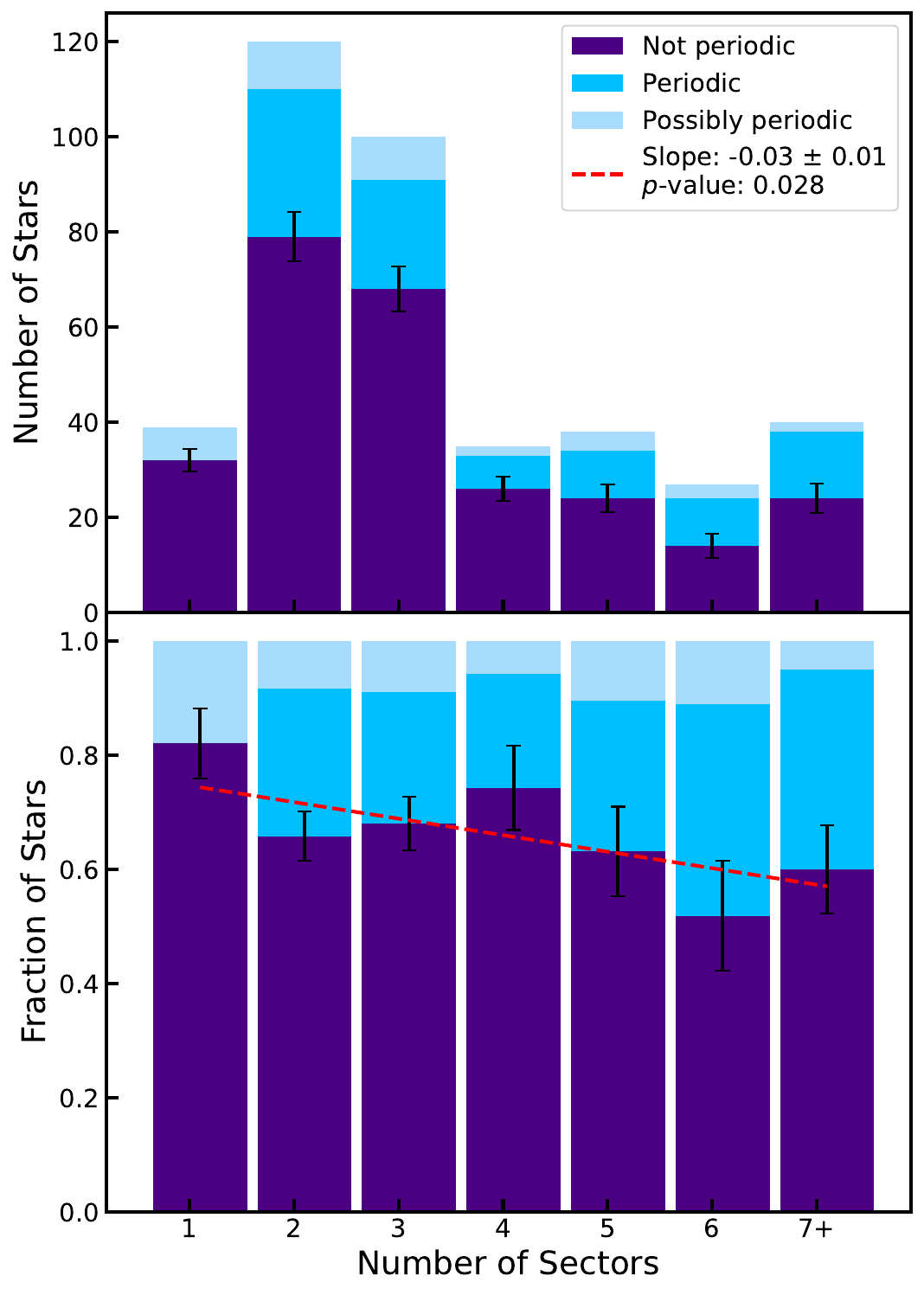}
	\caption{\textit{Top:} A histogram of the maximum number of available sectors of TESS data with either PDCSAP or eleanor reductions for the 399 likely single late-M dwarfs in this study. The histogram is divided into non-periodic, possibly periodic, and periodic stars as defined in Section~\ref{sec:GP}. \textit{Bottom:} The fraction of stars in each bin that are not periodic, possibly periodic, and periodic. Overplotted in both panels is the binomial uncertainty of the non-periodic fraction, and the red dashed line in the bottom panel is the line of best fit to these uncertainties. The implications of these plots are discussed in Section~\ref{sec:varfrac}.}
	\label{fig:hist_sectors}
\end{figure}

There were 399 targets with PDCSAP data in our sample of 1826 candidate single late-M dwarfs in TESS. These 399 stars comprise the sample studied in this work. The final sample can be seen in Figure \ref{fig:cmd_all}, which also shows the uncrowded $T \leq 18$ mag sample (Sections \ref{sec:excess} and \ref{sec:tess_lims}) and the sample after binary removal (Section \ref{sec:crowded}). The top panel of Figure~\ref{fig:hist_sectors} presents a histogram of the 399 stars, showing the distribution of targets with different numbers of TESS sectors available. Most stars have two or three sectors of data. The largest number of available sectors for a target is 42 in the northern continuous viewing zone.

\subsection{Assigning Spectral Types}
\label{sec:spt}

To better understand the final distribution of targets in the sample, we examined their spectral types. However, as many of the targets have not been previously studied, published spectroscopically determined spectral types do not exist for the full sample and must be approximated.

The SIMBAD astronomical database \citep{wenger2000} yielded published spectral types for 153 of the 399 targets. For the remaining 246 objects, we estimated spectral types from their $G - G_{RP}$ colors using the standard values determined by \citet{pecaut2013}. We assigned photometric spectral types using a simple thresholding scheme: values halfway between the color standards were used as dividing lines. For example, since M5.5 and M6 correspond to $G - G_{RP}$ = 1.38 and 1.43 mag, respectively, objects with $G - G_{RP} > 1.405$ mag were assigned M6, and those with $G - G_{RP} < 1.405$ mag were assigned M5.5.

\begin{figure}
    \centering
	\includegraphics[width=1.0\linewidth]{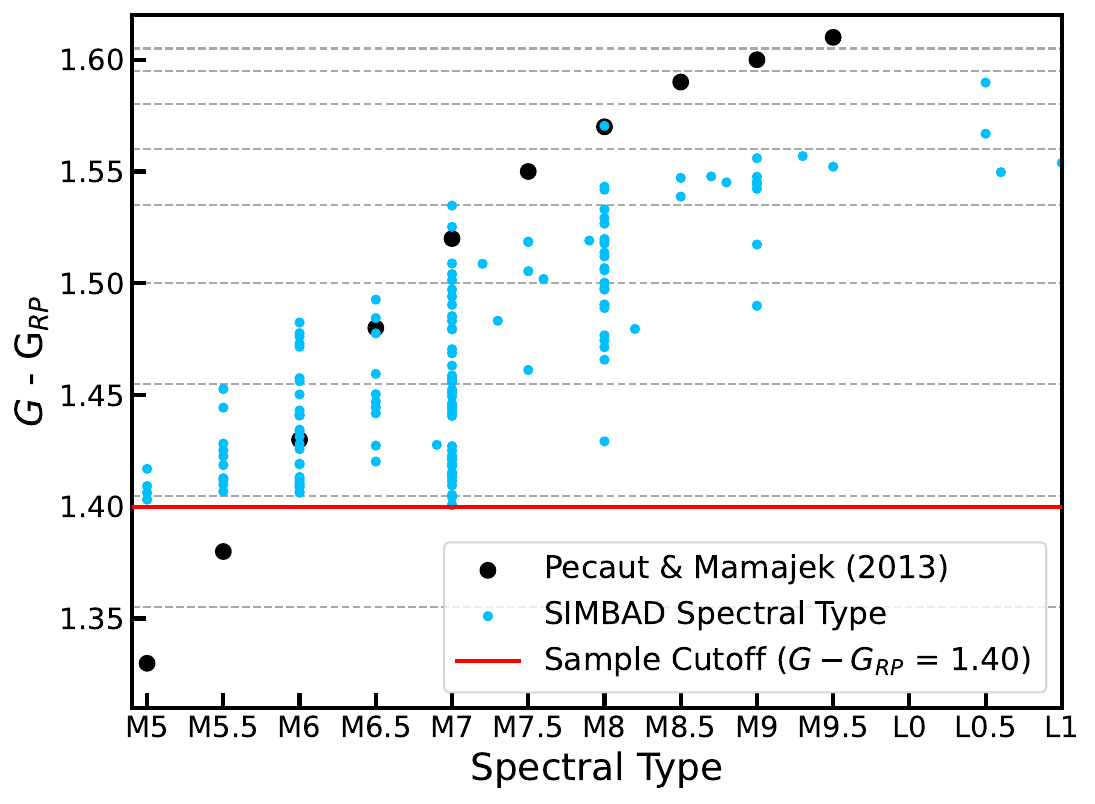}
	\caption{A comparison of Gaia $G - G_{RP}$ color and spectral type of stars with spectral types from the literature taken from the SIMBAD astronomical database \citep{wenger2000} compared to standard values selected by \citet{pecaut2013}. Gray dashed lines indicate the halfway point between the \citet{pecaut2013} values, marking a cutoff between approximated spectral types.}
	\label{fig:grp_spt}
\end{figure}

Figure \ref{fig:grp_spt} shows a comparison between the published spectral types and the photometric spectral types following \citet{pecaut2013}, with gray dashed lines indicating the boundaries used for assigning photometric spectral types. We note that while the trends generally agree, stars of the same spectral type have a wide scatter in color, and the photometrically assigned spectral types may be too early compared to the spectroscopic spectral type determinations for $\geq$M7 dwarfs. We defer an updated calibration of Gaia photometric spectral types for late-M dwarfs to later work. For now, we used them, but note that our photometric spectral types should not be used as a precise classification.

\begin{figure}
	\centering
	\includegraphics[width=1.0\linewidth]{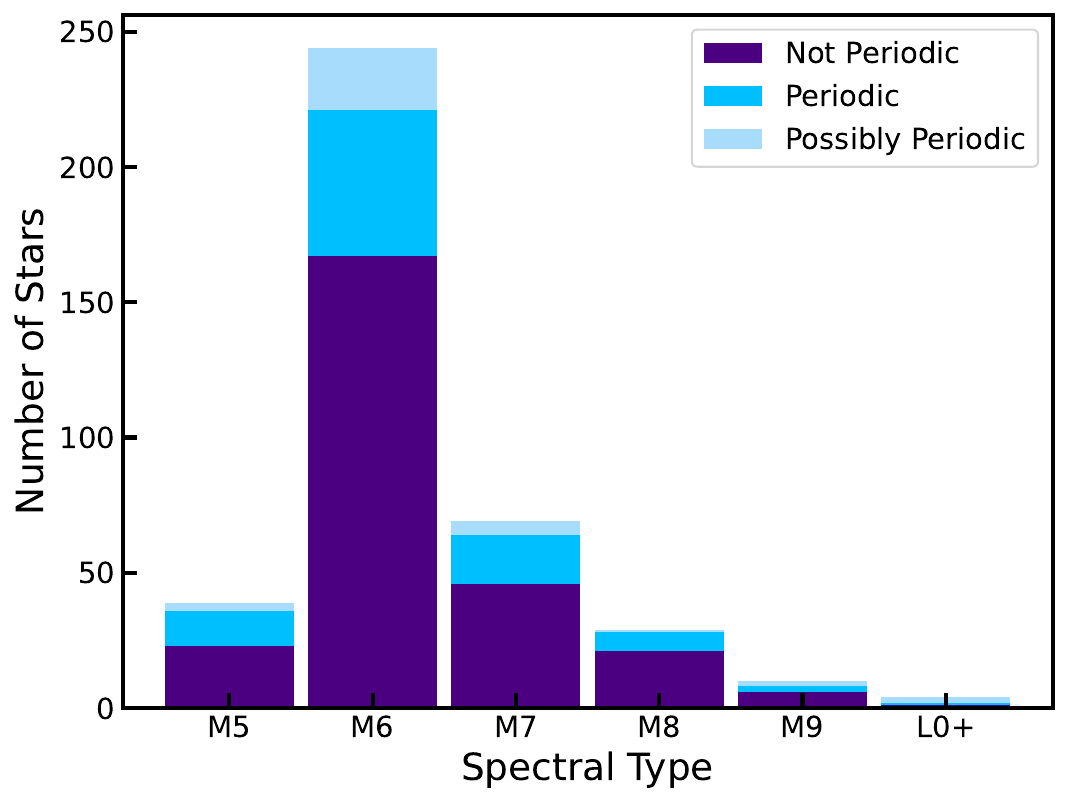}
	\caption{A histogram of spectral types of stars in the 399 likely single late-M dwarfs in this study. The histogram is divided into non-periodic, possibly periodic, and periodic stars as defined in Section~\ref{sec:GP}. The majority of stars have spectral types M6 or later. Spectral types were taken from the SIMBAD astronomical database \citep{wenger2000}, and where unavailable, they were calculated from $G - G_{RP}$.}
	\label{fig:spt_hist}
\end{figure}

The sample ranges from spectral types M5-L9, with the majority classified as M-dwarfs with a spectral type of M6 and later, as seen in Figure \ref{fig:spt_hist}. For simplicity, we refer to the entire sample as ``late-M dwarfs'', including the four M5s and 35 M5.5s whose color places them potentially within the boundary of the late-M dwarf classification, as well as eight L dwarfs that extend slightly beyond it.

\section{Identifying Periodically Variable Objects}
\label{sec:per_det}

% Perhaps this title should be more like 'Search for photometrically variable UCDs' or 'Identifying periodically variable objects'. While the detection of periods is the goal, I think you should first talk about 'identifying objects photometrically variable', and then determining periodicities since it will not be possible to measure a rotation period in all the 399 candidates--some of them will not be variable and some others will exhibit some kind of variability that might not allow to derive a period.

The first step in our analysis was to identify stars that exhibit photometric variability in TESS light curves. Only a subset of these objects are expected to show periodic signals suitable for rotation period measurement, as some may be non-variable or may exhibit irregular variability without a clear periodic signature. For those with detectable periodic modulation, we then attempted to determine robust rotation periods.

To do this, we analyzed both SPOC PDCSAP data and data reduced using the \texttt{eleanor} pipeline \citep[][referenced hereafter as ``eleanor light curves'' or ``eleanor data'']{feinstein2019}. \texttt{eleanor} offers different levels of accuracy in their data processing algorithm. We chose the 'corrected' method, which is based on Principal Component Analysis. While PDCSAP light curves are reduced from Target Pixel Files (TPFs), \texttt{eleanor} performs its light curve reduction on TESS Full Frame Images (FFIs). This means eleanor data are available for any target in our initial input catalogue, as long as it is in the majority of the sky that TESS has observed ($\sim$85\%). This is in contrast to the PDCSAP data, which is only available in certain sectors where TPFs exist. As of 2025 January 16th, when the data were downloaded, the latest sector for which eleanor data were available was 85. As eleanor data is processed from FFIs, the cadence of data is 1800 s, 600 s, or 200 s, depending on the sector (as described in Section \ref{sec:spoc_pdcsap}).

\begin{figure*}
	\centering
	\includegraphics[width=0.49\linewidth]{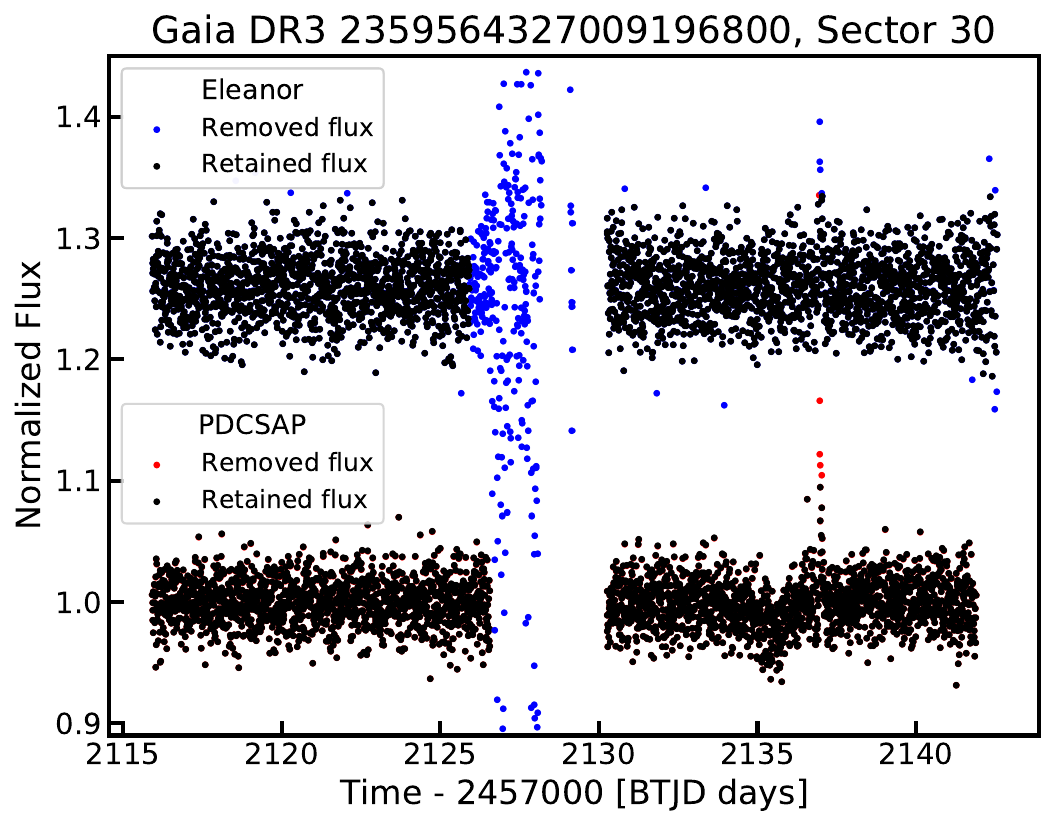}
        \includegraphics[width=0.49\linewidth]{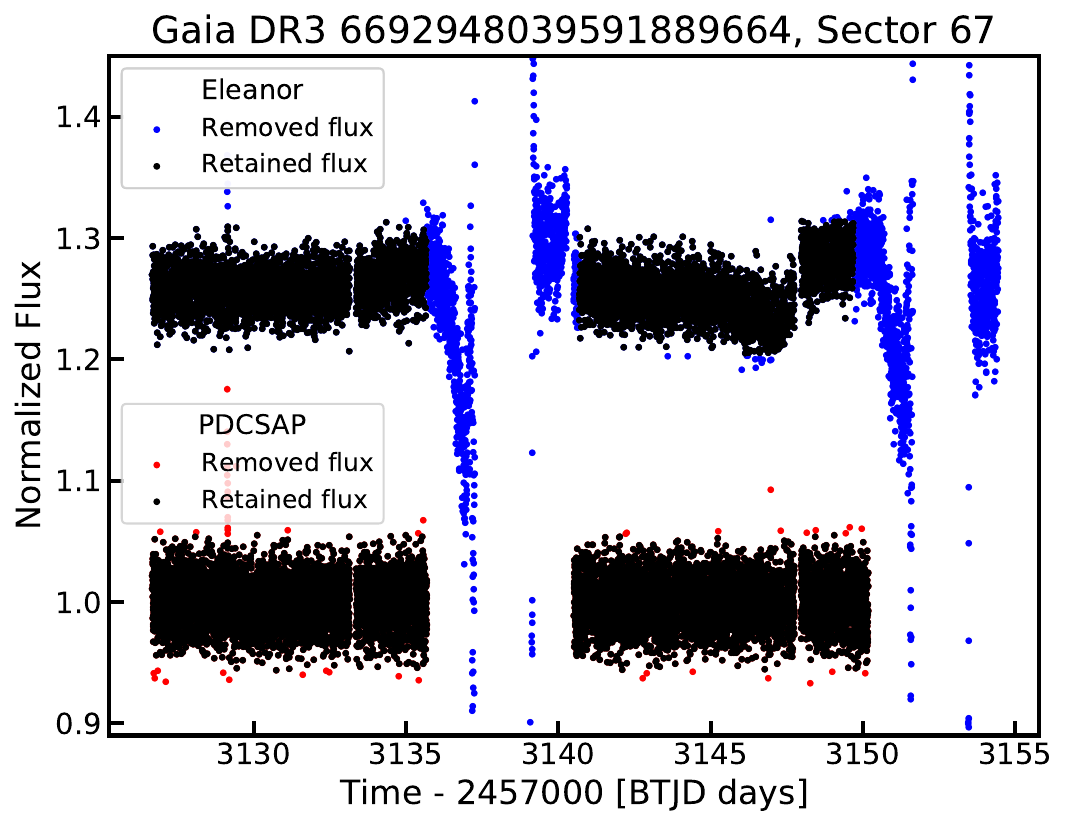}
	\caption{Normalized light curves of two targets in the sample. The left plot shows a target where the eleanor data (offset by +0.25 for clarity) after outlier rejection is cleaner than the PDCSAP flux, and the right plot shows the opposite. Colored points indicate data clipped before analysis (blue = eleanor, red = PDCSAP).}
	\label{fig:spocvseleanor}
\end{figure*}

It is entirely feasible to analyze the TESS late-M dwarf sample solely with \texttt{eleanor}. However, we find that both eleanor and PDCSAP light curves can contain non-astrophysical features. An example of this is shown in Figure \ref{fig:spocvseleanor}. In the left panel, one of our targets exhibits a dip in the PDCSAP light curve but not in the eleanor data. As the dip is not present in both methods, we cannot be certain of its origin. Conversely, the right panel shows a case where the PDCSAP reduction exhibits fewer systematics than the eleanor reduction. Therefore, we use either method as a validation of the other to improve the reliability of our findings on astrophysical variability. 

\subsection{Step 1: Data Filtering}
\label{sec:data_filtering}

Regardless of the reduction method, systematics can prevail in the light curves on timescales $\gtrsim$1 day. The main contributors are periodic variations in earthshine and moonlight due to TESS's 13.7-day 2:1 resonant orbit with the Earth-Moon system and momentum dumping at 1.5, 2, 2.5, 3, 5, and 13.7 days \citep{vanderspek2018}. We also observe a previously unreported periodic systematic around 0.73 days, which may be an overtone of the 1.5-day momentum dump. To decrease the effect of these periodic systematics, we apply a median filter with a sliding bin width of 2 days using \texttt{wotan} \citep{hippke2019}. This also significantly diminishes our sensitivity to periods longer than 2 days.

To retain some sensitivity to $>$2-day periods, we also analyze the data that have not been detrended. These require careful examination because of the above temporal systematics, and the maximum period we can expect in a single sector is around the 13.7-day mark \citep[e.g.][]{martins2020, fetherolf2023}. Non-detrended PDCSAP and eleanor light curves have different properties. 
As the PDCSAP light curves are designed for exoplanet detection, long stellar trends are removed or dampened. 
% This has the effect of removing long-scale systematic trends, but also reducing the number of rotation periods longer than a few days we can detect. 
On the other hand, the ``corrected'' eleanor reduction pipeline does not dampen long stellar trends. The result is that while eleanor, in theory, can retain longer periods, it is more susceptible to systematic variability and is often noisier (see the right panel of Figure \ref{fig:spocvseleanor}). As a result, while we only remove outliers greater than 3$\sigma$ from PDCSAP light curves, for eleanor light curves, we clip not only outliers greater than 3$\sigma$ but also entire 1-day bins where more than 5\% of the data is above the 3$\sigma$ threshold.

We utilized a Lomb-Scargle periodogram \citep[Section \ref{sec:LSP};][]{lomb1976, scargle1982} to determine the preliminary periods of our targets. Targets found to likely be periodic were analyzed more robustly with a Gaussian process-based approach to more accurately determine periods, amplitudes, and their uncertainties (see Section \ref{sec:GP}).

\begin{figure*}
	\centering
	\includegraphics[width=0.49\linewidth]{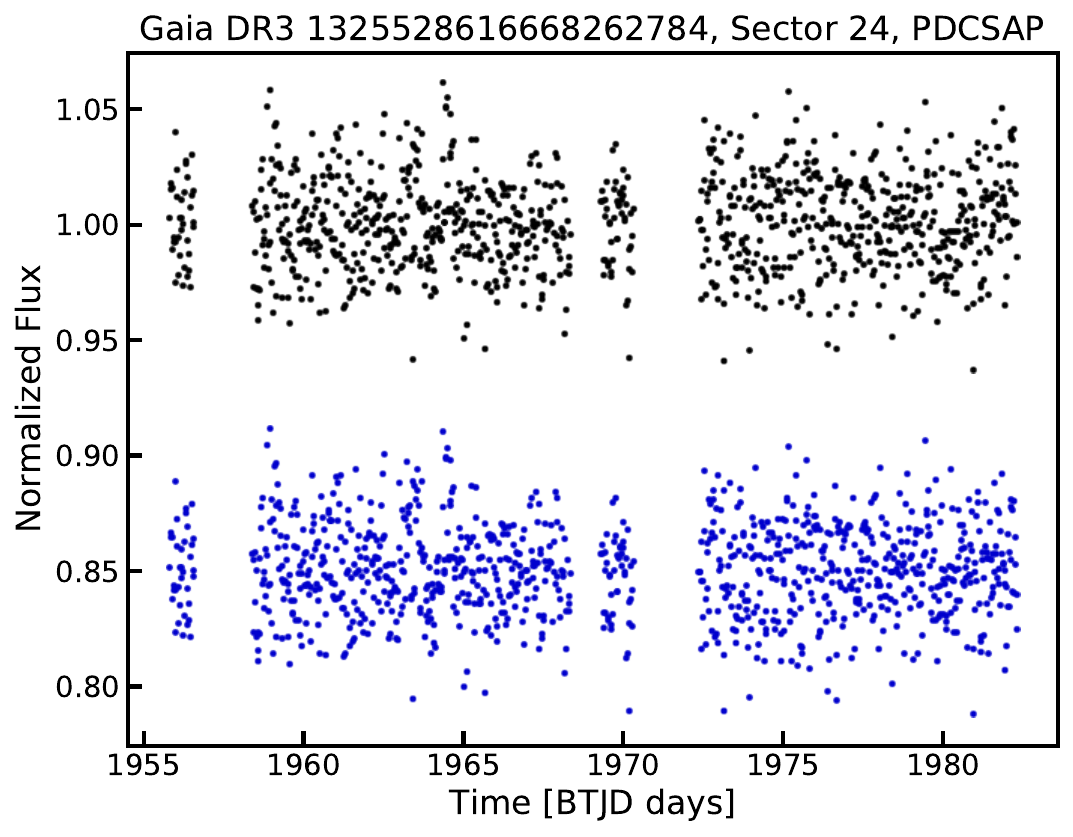}
        \includegraphics[width=0.49\linewidth]{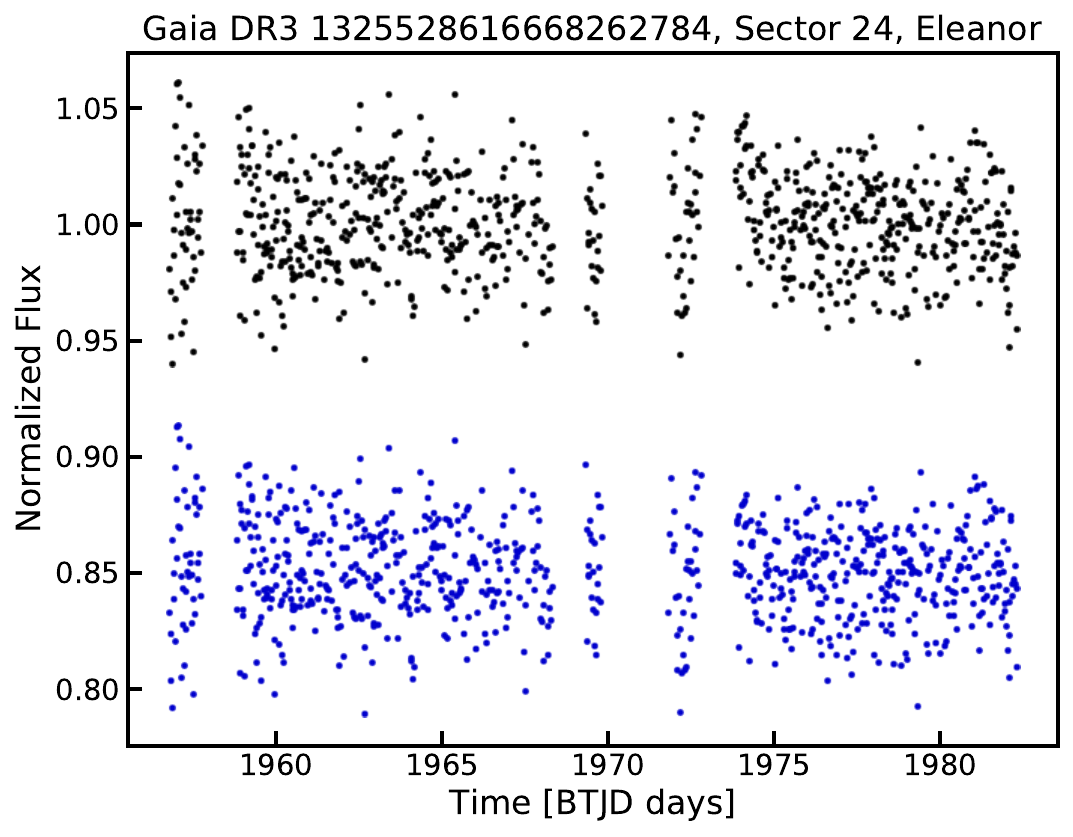}
        \includegraphics[width=0.49\linewidth]{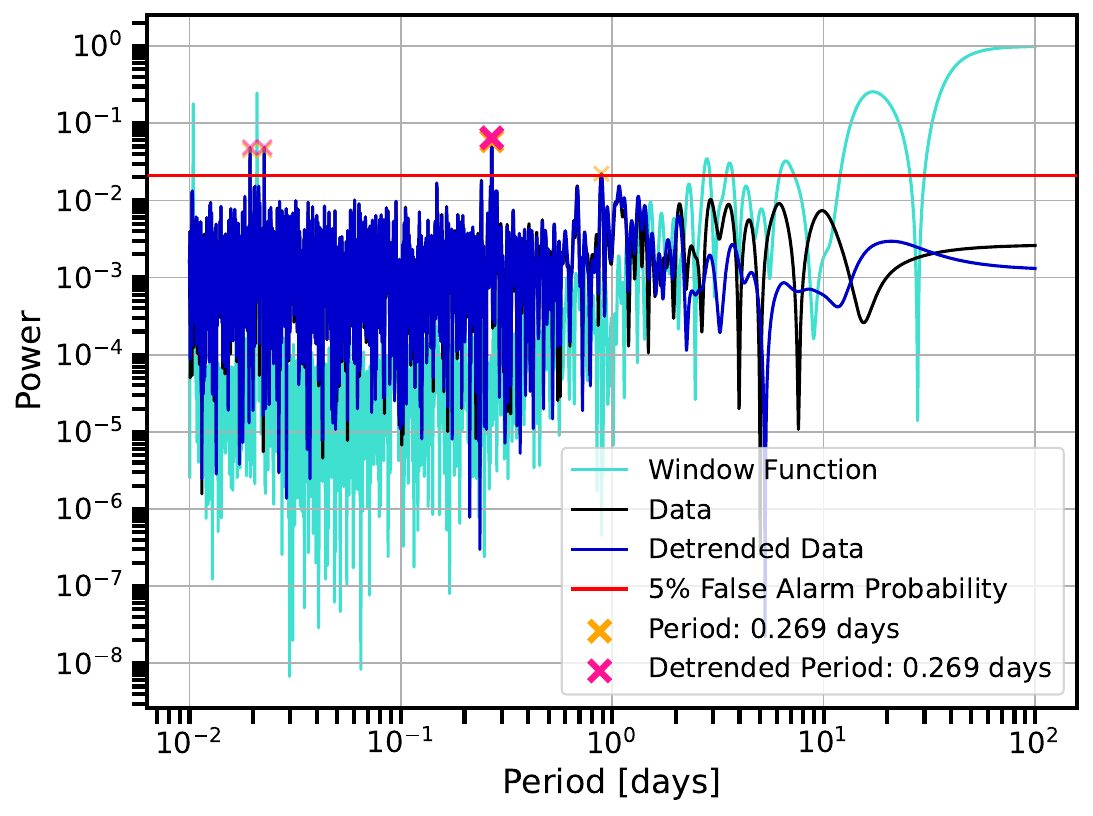}
        \includegraphics[width=0.49\linewidth]{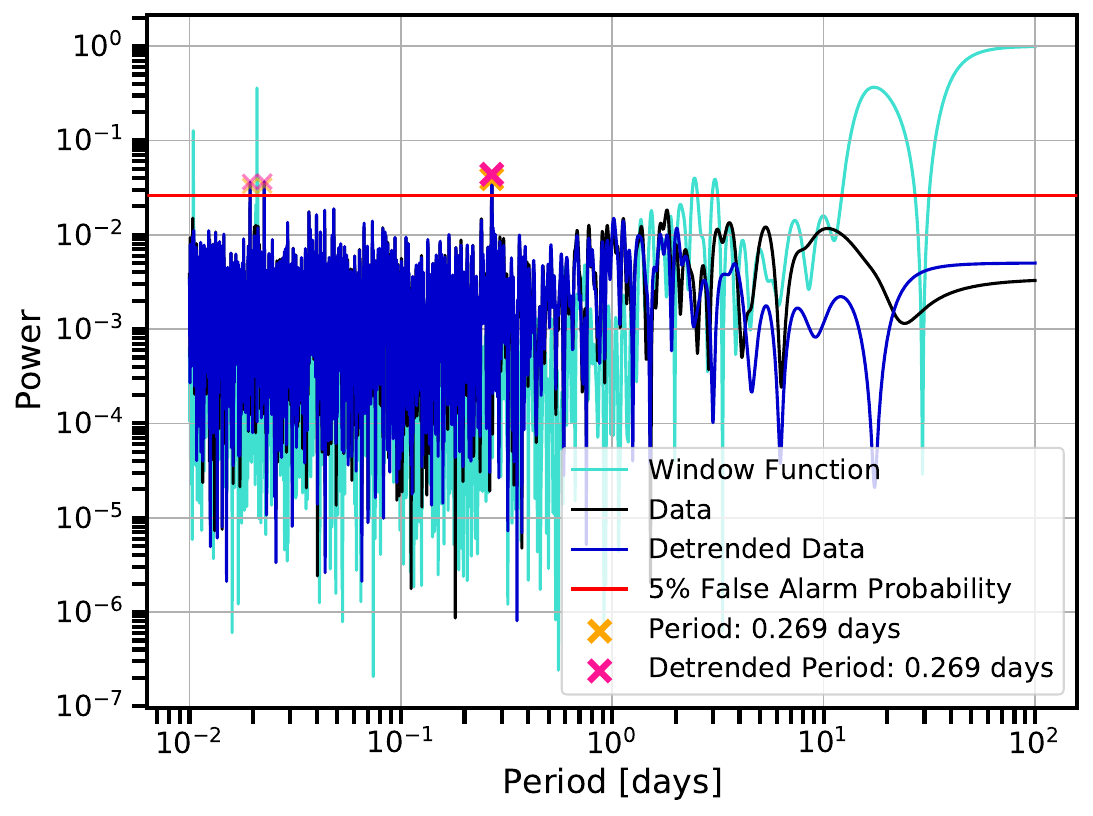}
        \includegraphics[width=0.49\linewidth]{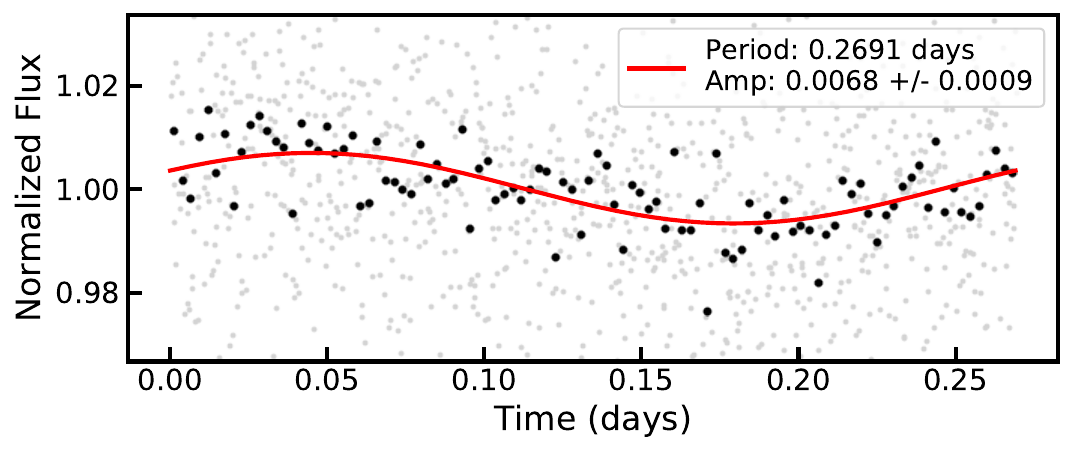}
        \includegraphics[width=0.49\linewidth]{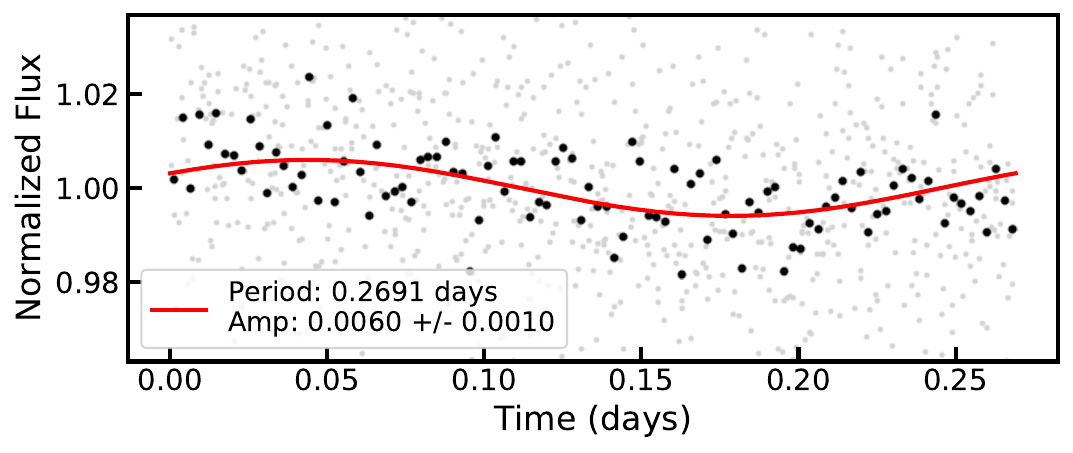}
        \includegraphics[width=0.49\linewidth]{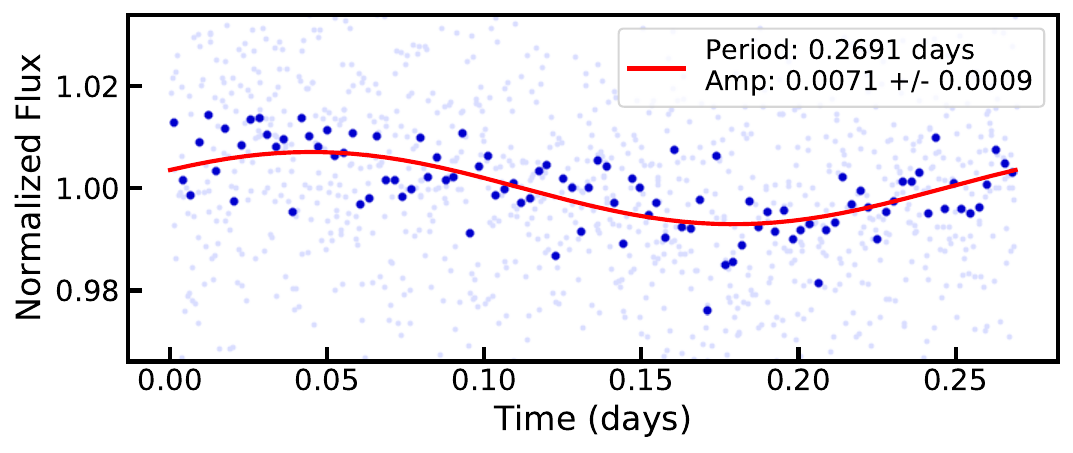}
        \includegraphics[width=0.49\linewidth]{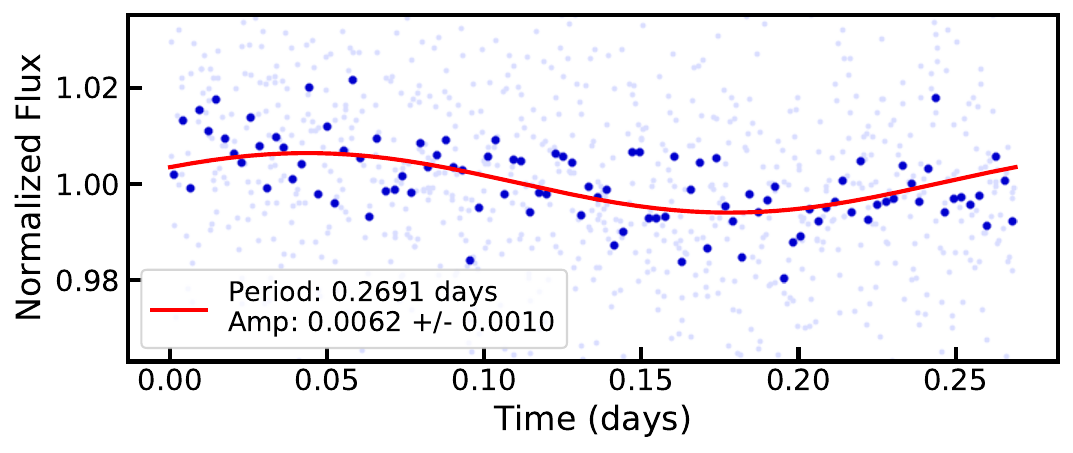}
	\caption{\textit{Top panels:} Normalized (black) and median detrended (blue) light curves for Sector 24 of TESS data on the M7 target Gaia DR3 1325528616668262784 with 30-min cadence. PDCSAP is on the left, and eleanor photometry is on the right. \textit{Middle panels:} LS periodograms for the respective light curves. Large pink and orange X's mark the highest peak, while small pink and orange x's mark lower power peaks still above the FAP but not above the window function. 
    \textit{Bottom panels:} light curves folded on the period with the highest peak in the periodogram, for both before (black) and after (blue) 2-day median detrending. The solid colored points are the data binned to 1/100 the phase, included only to visualize periodicity. The red curve is fit to the unbinned data. All four reduction methods produce the same 0.269-day period that is passed to the Gaussian Process (\S\ref{sec:GP}) for further refinement.
    %astrophysical, whereas the 3.07-day period seen in the non-detrended eleanor data is due to red noise and a peak in the window function.
    }
	\label{fig:lspmethodsfast}
\end{figure*}

 \begin{figure*}
	\centering
	\includegraphics[width=0.49\linewidth]{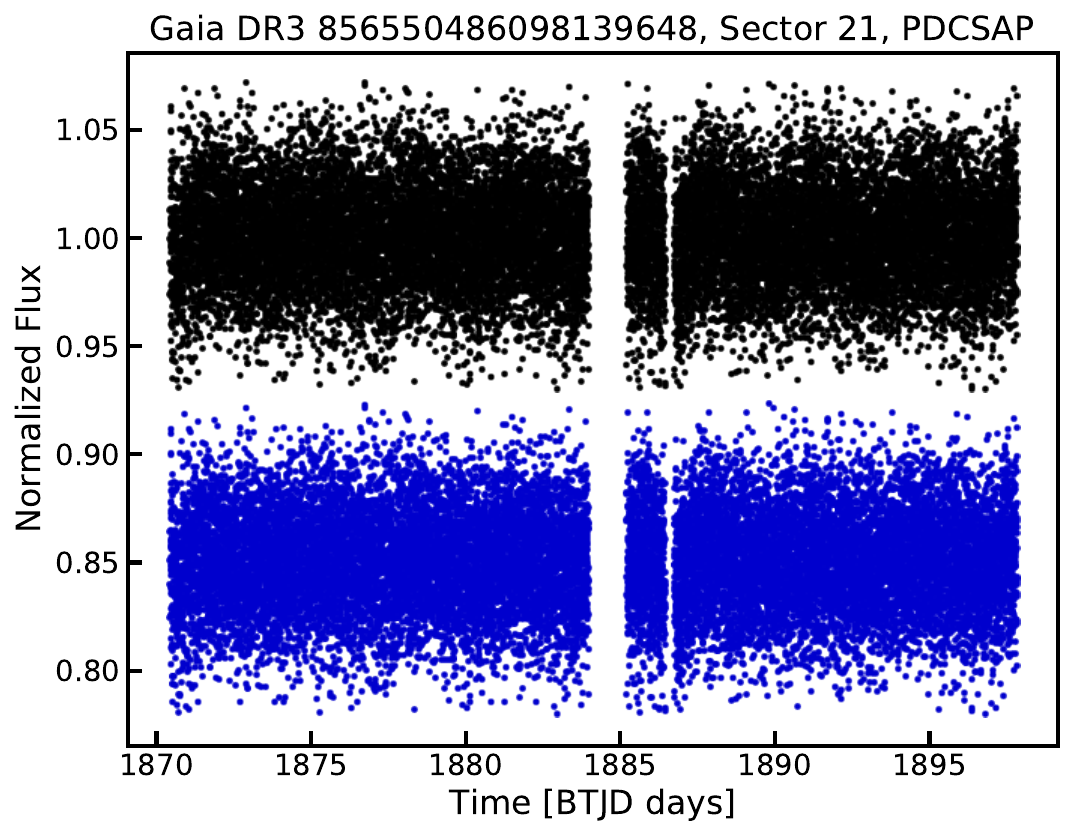}
        \includegraphics[width=0.49\linewidth]{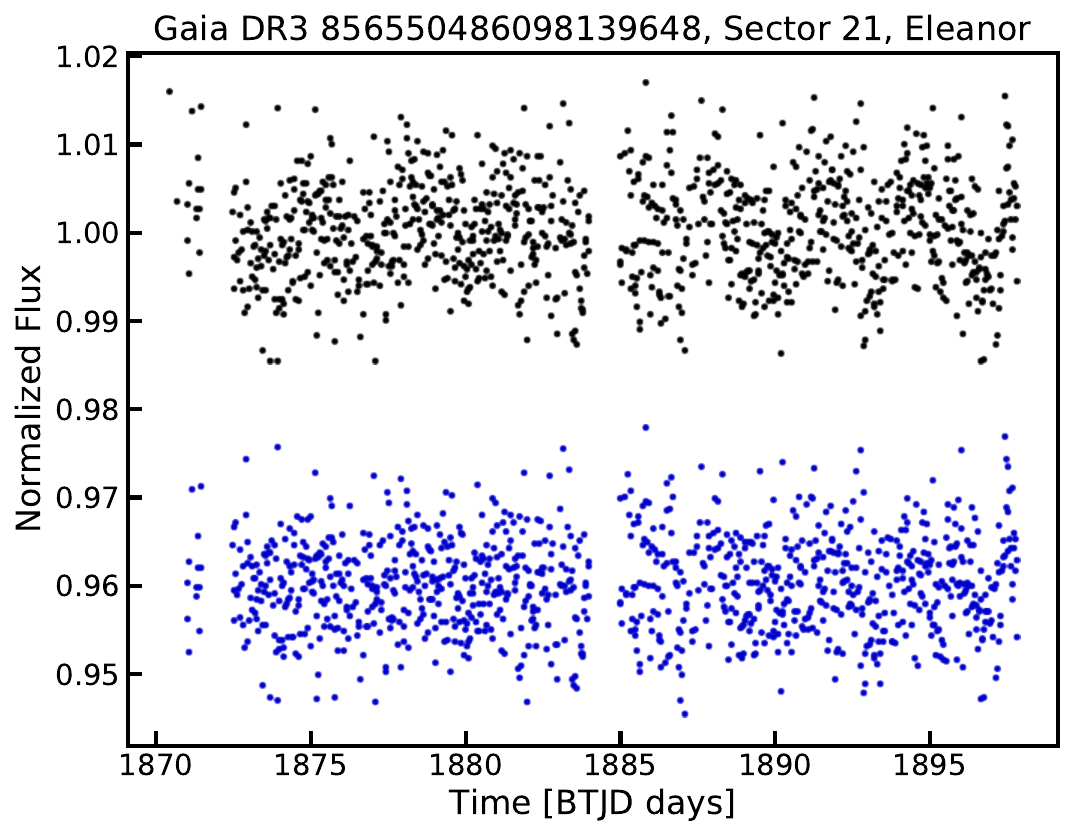}
        \includegraphics[width=0.49\linewidth]{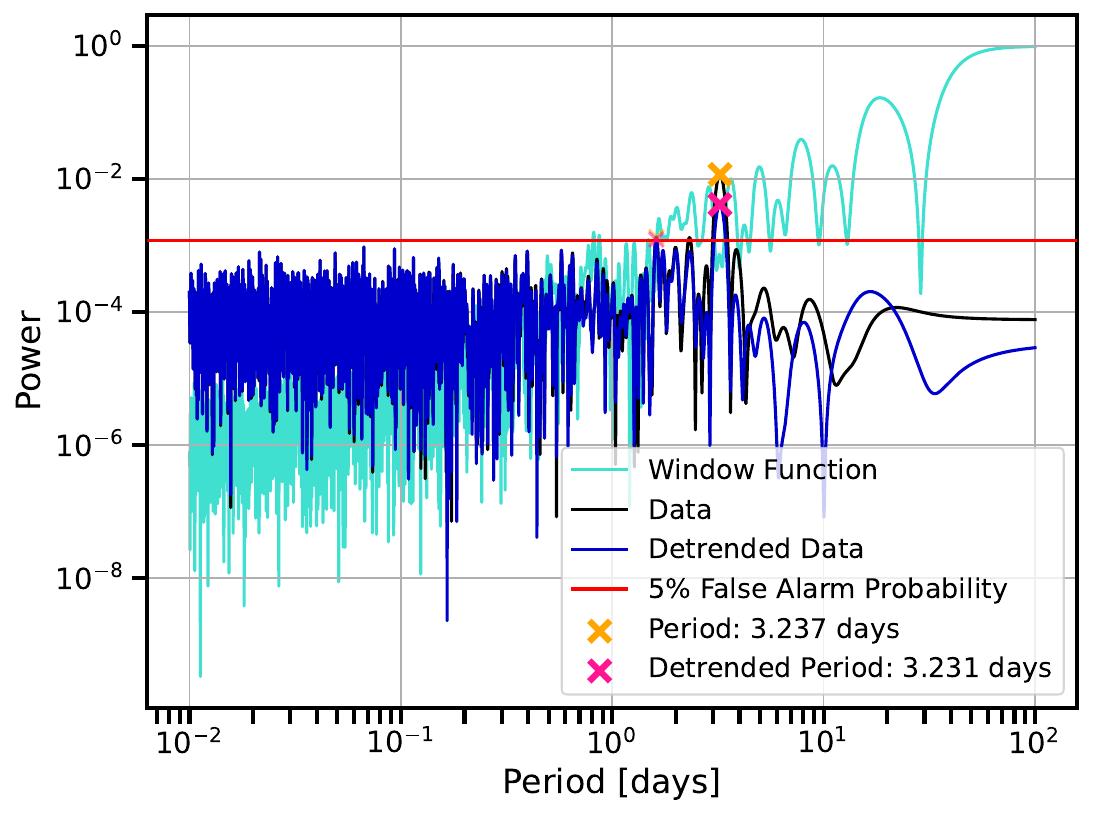}
        \includegraphics[width=0.49\linewidth]{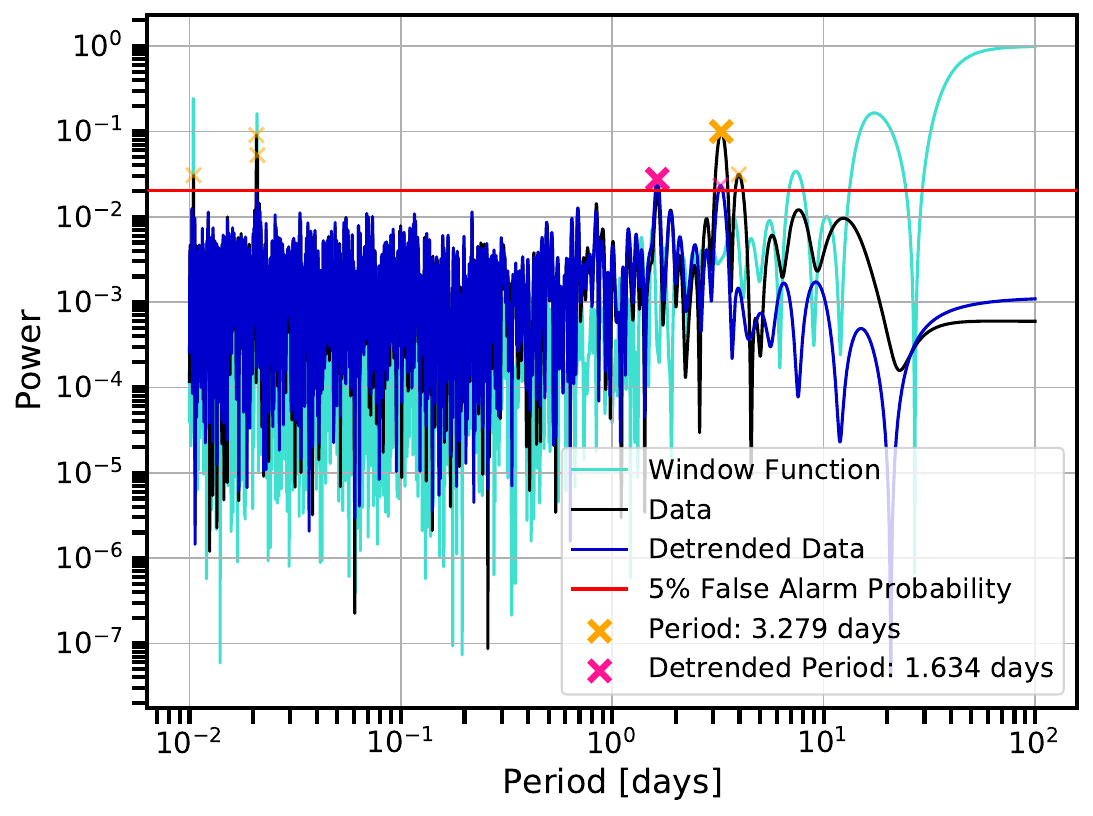}
        \includegraphics[width=0.49\linewidth]{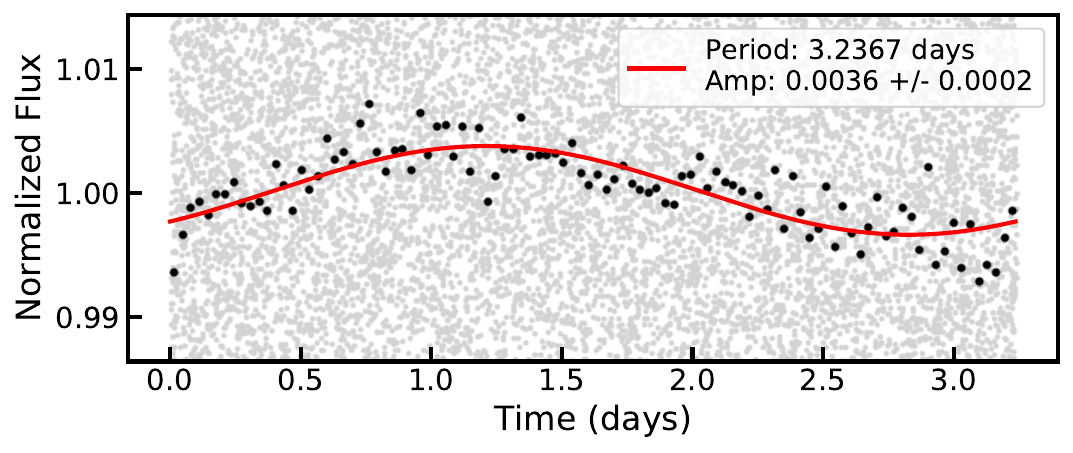}
        \includegraphics[width=0.49\linewidth]{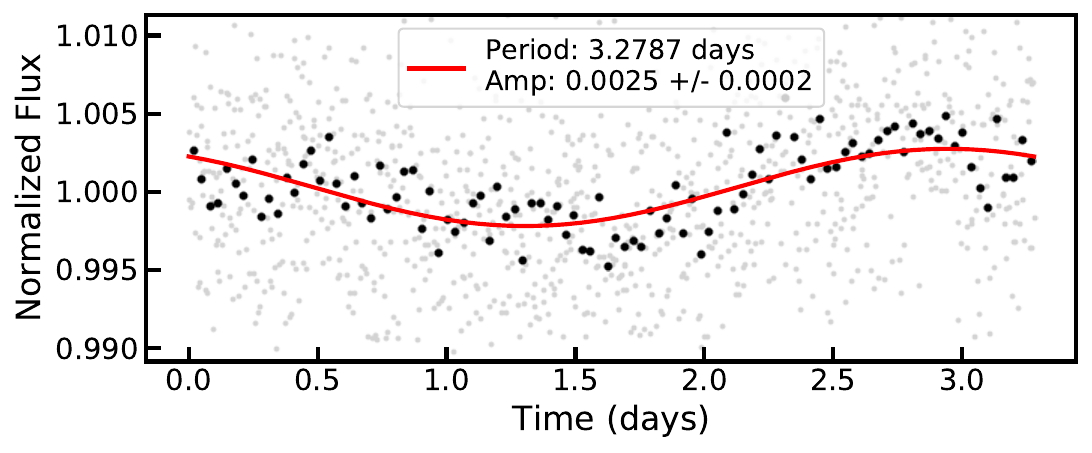}
        \includegraphics[width=0.49\linewidth]{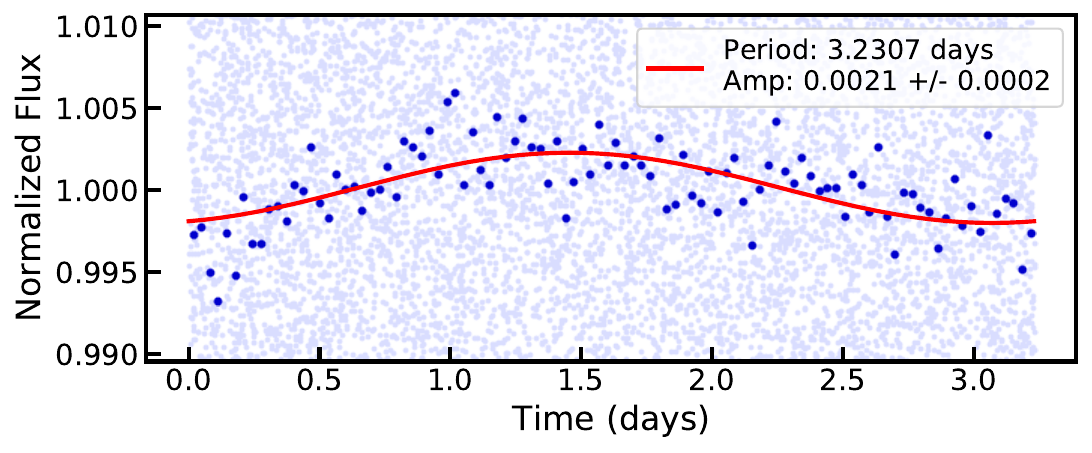}
        \includegraphics[width=0.49\linewidth]{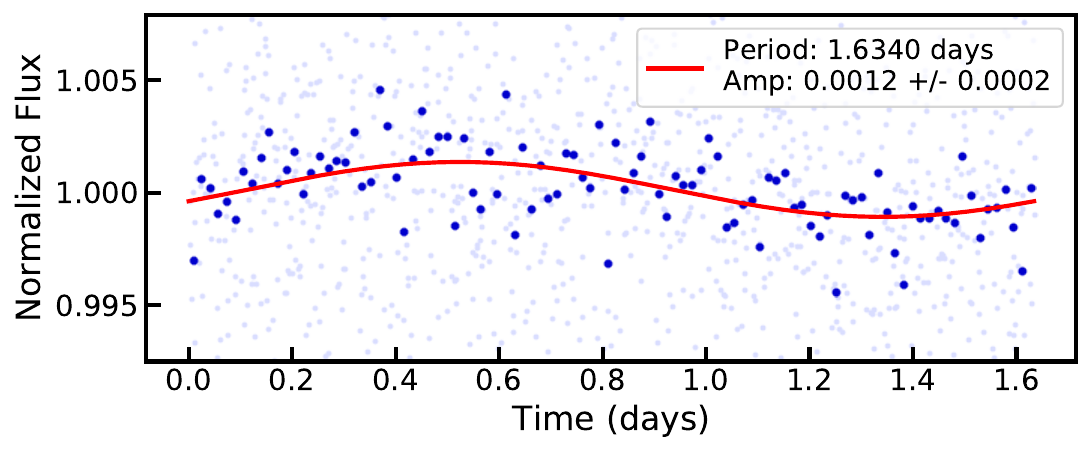}
	\caption{The same as Figure \ref{fig:lspmethodsfast}, but for the M6 dwarf Gaia DR3 856550486098139648 with an astrophysical $\approx$3.2-day period, recovered by three of the four reduction methods. The 1.6-day period seen in the eleanor detrended data (bottom right panel) is a 1/2 harmonic of the true period. Note that the 3.2-day period falls in a trough of the window function, so it is not caused by a spacecraft systematic.
    }
	\label{fig:lspmethodsslow}
\end{figure*}

\subsection{Step 2: Candidate Period Determination with Lomb-Scargle Periodograms}
\label{sec:LSP}
%  (e.g. Anthony et al (2022)) and red-noise can result in peaks in the LSP which technically have a FAP<<0.01 (e.g. Dorn-Wallenstein et al. (2019)).
 A Lomb-Scargle (LS) periodogram is an extension of a Fourier transform designed to determine the periodicity of a dataset with unequally spaced points with a sinusoidal pattern \citep{lomb1976, scargle1982}. As such, it is often used to determine the rotation periods of stars \citep[e.g.,][]{reinhold2013, miles2023, petrucci2024}.

 To determine periodicity, we first used the Generalized LS implemented by the \texttt{gls} class from the \texttt{PyAstronomy.pyTiming.pyPeriod} module \citep{pya}. A generalized LS is a modification to a LS which incorporates uncertainties and includes a constant term \citep{zechmeister2009}. We used a logarithmic frequency grid spanning from 0.01 days$^{-1}$ to 100 days$^{-1}$.

 We required that peaks had a false-alarm probability (FAP) less than 5\%. We set the initial FAP level high in order to allow for low-amplitude signals to be captured, as independent confirmation of low-confidence periodicities among sectors increases the confidence of the overall result. We also required that the LS power of any period be stronger than any neighboring peaks in the window function (determined from the observing cadence). In cases where multiple period peaks appeared on the periodogram, the highest peak was assumed to be the most likely source of periodicity. 
 % Revisit when other peaks are considered

 We apply the LS period search mainly on data from the individual TESS sectors (Section~\ref{sec:LSP_individual}). Where data exist from two or more TESS sectors, we also perform an all-sector LS period search (Section~\ref{sec:LSP_all}) to help identify any potential low-amplitude periods that may be below our detection thresholds in individual sectors.

\subsubsection{Individual Sector Analysis} 
\label{sec:LSP_individual}

 For each target, we analyzed each available sector with PDCSAP and eleanor, detrended and not. These comprise four versions of the original TESS light curve that we analyze for periodicity, and we refer to these as our four light curve reduction methods. 
 
 %Additionally, data from each sector were combined, and a Lomb-Scargle periodogram was run on the combined light curve. 

 Figures \ref{fig:lspmethodsfast} and \ref{fig:lspmethodsslow} show the differences between the four reduction methods for two stars, one with a rotation period of 0.26 days and another of approximately 3.2 days. It is worth noting that in many cases, there are multiple peaks with a higher power than the 5\% FAP. For periods longer than about 2 days, these are generally either related to peaks in the window function or associated with red noise, which can contribute to the power spectrum \citep{dorn2019}.
 Additionally, for 30 min-cadence data, we also commonly see two periodicity peaks around the 30-minute cadence peak in the window function (Figure \ref{fig:lspmethodsfast} or right set panels in Figure \ref{fig:lspmethodsslow}). We discard all such systematic periodicities as non-astrophysical.

\begin{figure}
	\centering
	\includegraphics[width=1.0\linewidth]{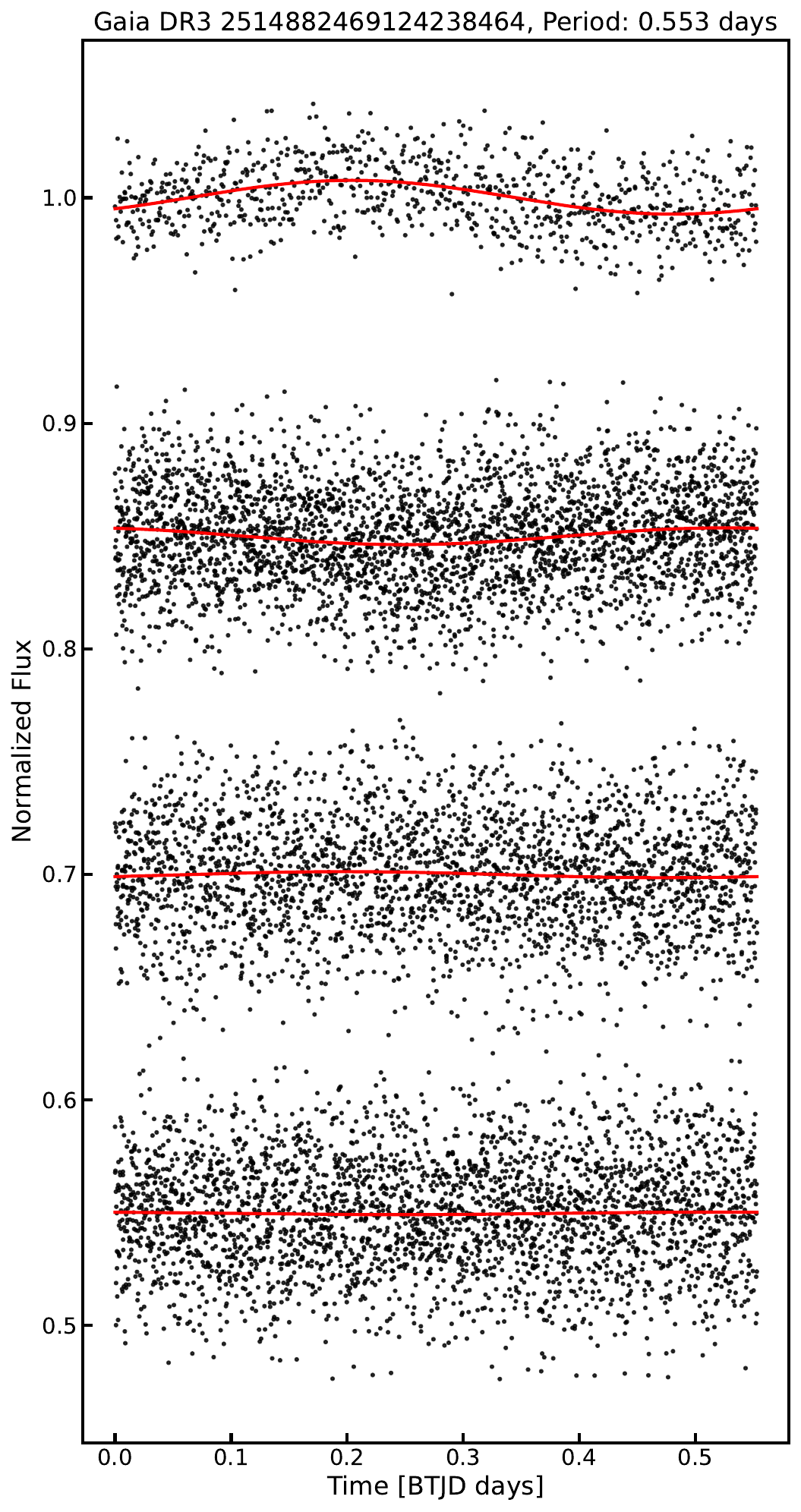}
	\caption{Normalized light curves of Gaia DR3 2514882469124238464 ($G - G_{RP} = 1.41$ mag, $\sim$M6) from Sectors 4, 31, 42, and 43, folded on the best period from the LS of 0.553 days. Individual sectors are offset from each other to display the changing amplitude of periodicity. }
	\label{fig:lcevo}
\end{figure}

\begin{figure*}
	\centering
	\includegraphics[width=\linewidth]{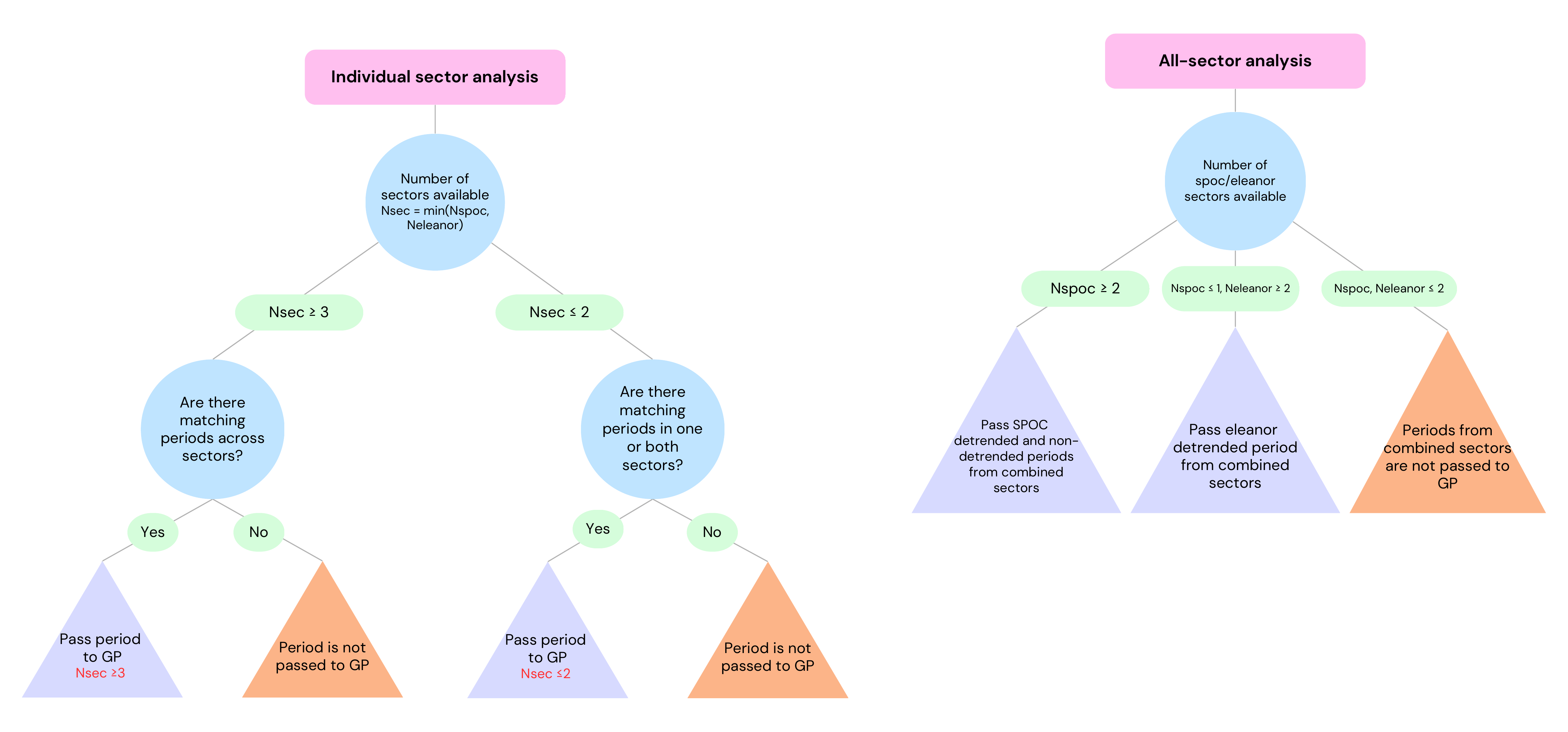}
	\caption{Decision tree to determine whether a candidate period from Step 2 (LS) should be passed for validation to Step 3 (Gaussian process). See Section~\ref{sec:LSP} for details.}
	\label{fig:decisiontreelsp}
\end{figure*}

 Due to the transient nature of spots on the star's surface, it is normal for periodicity present in one sector not to be seen in other sectors (see Figure \ref{fig:lcevo}). As such, we do not require the periodicity to appear in every sector of available data. 
 %Furthermore, it has been shown that astrophysical red noise can contribute to the power spectrum, causing longer-period peaks in the periodogram to exceed the 5\% FAP, despite not being true rotation periods \citep{dorn2019}. These periods, when present, should not repeat from one sector to the next. 
 As initial confirmation of any candidate period, we required that there be a match for the same period among at least two of the four reduction methods. The following algorithm was used to determine which periods were passed for subsequent analysis (Section~\ref{sec:GP}; see left panel of the decision tree in Figure \ref{fig:decisiontreelsp}):

 \begin{itemize}
     \item \textbf{Targets with 3+ sectors of TESS data (Nsec $\geq$ 3):} We required that the detected period matched across at least two sectors across any of the four reduction methods. 
     \item \textbf{Targets with 1 or 2 sectors of TESS data (Nsec $\leq$ 2):} We required that the detected period appeared at least twice, either in the same sector inferred with different reduction methods or in different sectors.  
 \end{itemize}

 We considered two periods $P_1$ and $P_2$ from different reduction methods ``matching'' if their fractional difference $f_{diff}\equiv\frac{|P_1-P_2|}{\bar{P}}$ is 
 \begin{equation}
     f_{diff} < \frac{1}{N_{cycles}} \approx \frac{\bar{P}}{27},
 \end{equation}   
where $N_{cycles}$ is the number of cycles present in a sector, equal to the average period $\bar{P}$ (in days) divided by the length of the sector (approximately 27 days). This value is a conservative underestimation of the precision with which we determine the period from the LS.

If there was an unequal number of sectors between the PDCSAP and eleanor data, we applied the criteria of the method with the least number of sectors. This gives the loosest constraint to pass to the next step. If multiple periods fulfilled the above criteria, we passed the period(s) with the most matching pairs.
%to the Gaussian Process. 

\subsubsection{All-Sector Analysis} 
\label{sec:LSP_all}
 
When two or more sectors of data were available, we combined the data from all individual sectors for each target and repeated the LS periodograms to help confirm candidate periods detected in individual sectors (Figure~\ref{fig:decisiontreelsp}, right panel). 
Data were normalized before combination, meaning that our sensitivity deteriorates for periods longer than %\textcolor{green}{that which could be \textcolor{orange}{determined} \sout{recovered} from} 
one sector.
 
The all-sector analysis showed that systematic trends were amplified in the eleanor data, predominantly when not detrended. As such, we did not use the non-detrended eleanor all-sector data. When there were two or more sectors of PDCSAP data available, any PDCSAP all-sector periods stemming from either the detrended PDCSAP, non-detrended PDCSAP, or the eleanor detrended data were added to the candidate periods for subsequent analysis if they were not already present. In the case where there were two or more eleanor sectors but only one or no PDCSAP sectors, only candidate periods from the detrended eleanor all-sector data were added.

All candidate periods that match the above criteria from the individual and all-sector analyses are combined to be passed to the subsequent step, which uses Gaussian processes to determine more accurate periods and uncertainties. If two candidate periods for the same target are within 20\% of each other, only the longer period is passed, as the Gaussian process samples a period space 40\% above and below the provided period. With the candidate period detection complete, 361 of 399 targets were passed to the next step.

 % Maybe add appendix, walking through pipeline for a few examples??

 % If more than 5 sectors were available, targets were passed on to the Gaussian Process pipeline if the period matched across more than 20\% of the sectors with any reduction method. In the case of five or fewer sectors, the requirement was that the same period appeared in more than 1 sector. In this step, this requirement only needed to be held with one of the four reduction methods (PDCSAP vs. eleanor, median and non-median filtered). The exception here is if only one or two sectors are available, we only require a period to appear in one sector, but it must match between eleanor and PDCSAP or median and non-median filtered data.

\subsection{Step 3: Period Validation with a Gaussian Process}
\label{sec:GP}

\begin{figure*}
	\centering
	\includegraphics[width=0.41\linewidth]{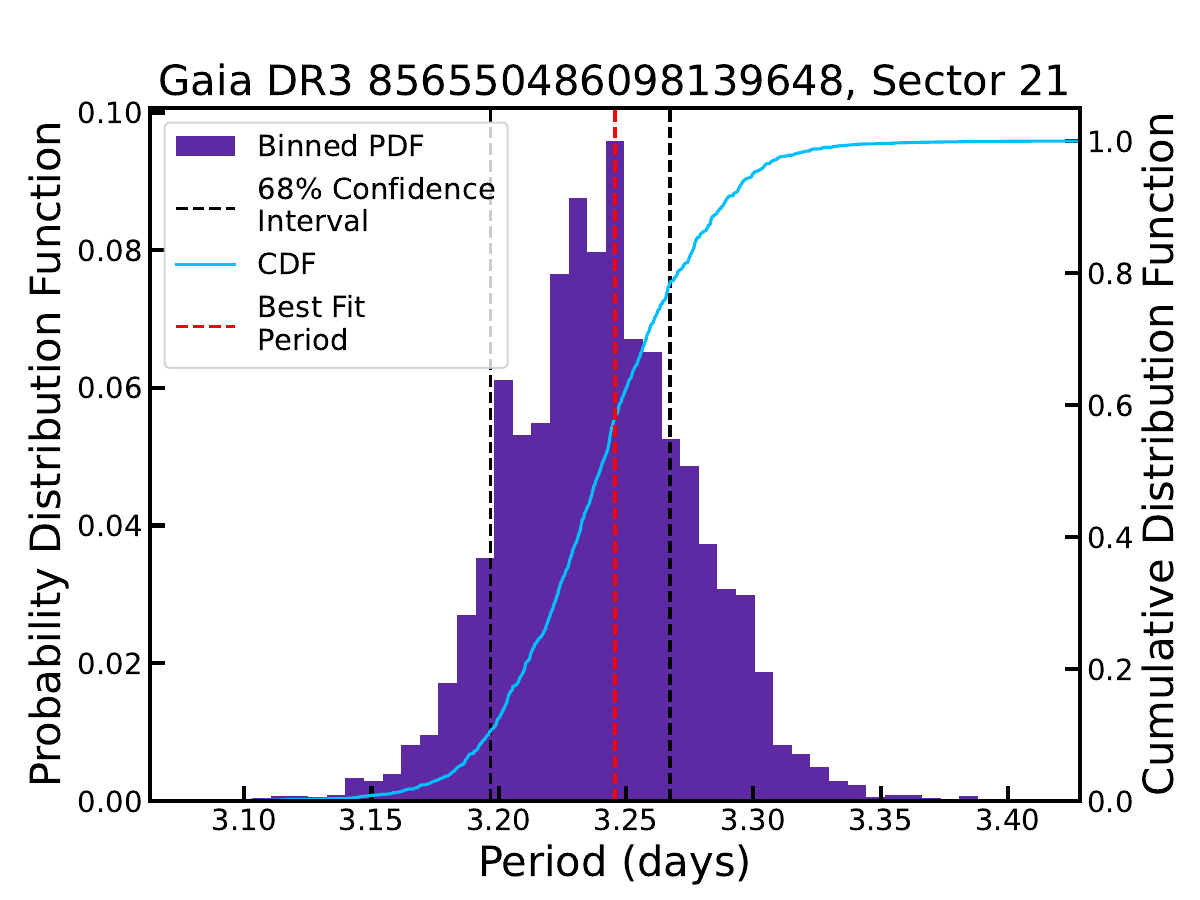}
        \includegraphics[width=0.41\linewidth]{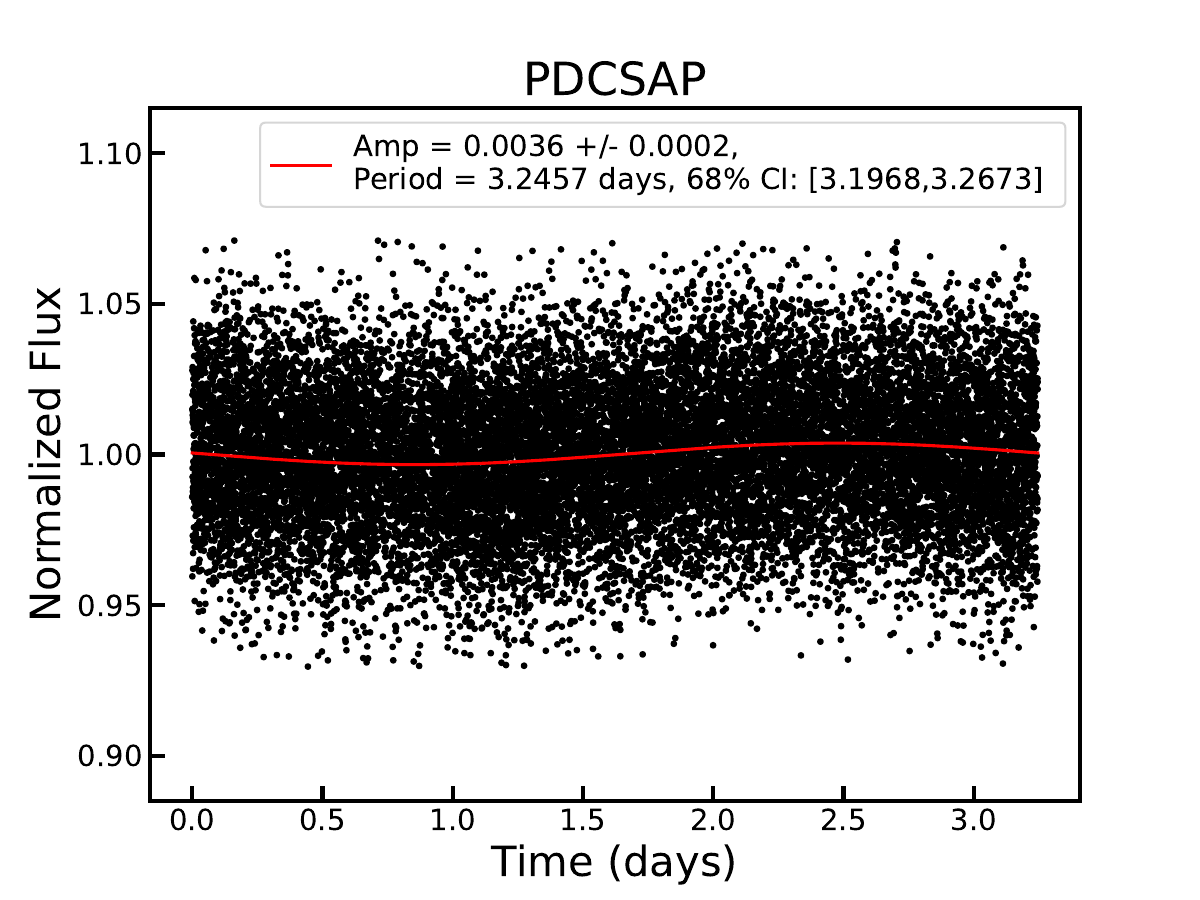}
        \includegraphics[width=0.41\linewidth]{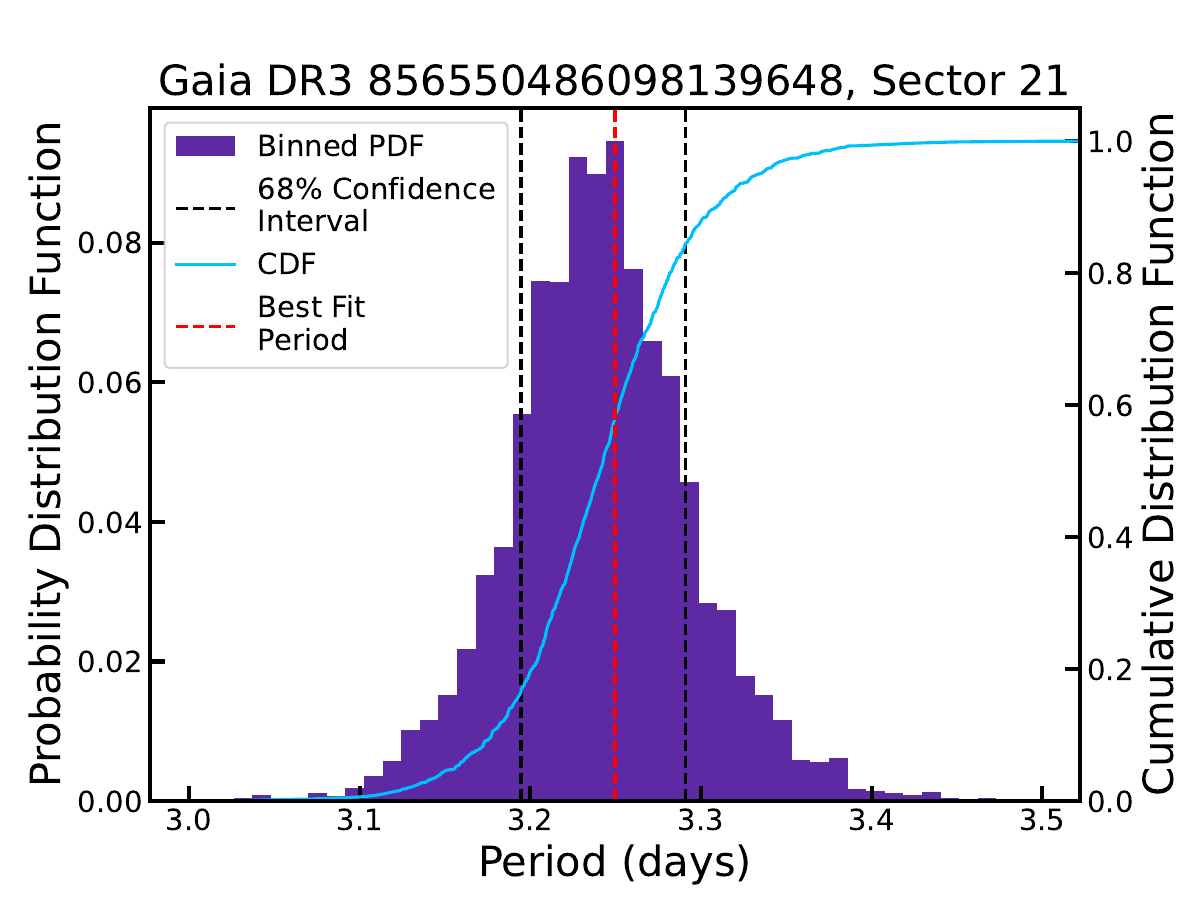}
        \includegraphics[width=0.41\linewidth]{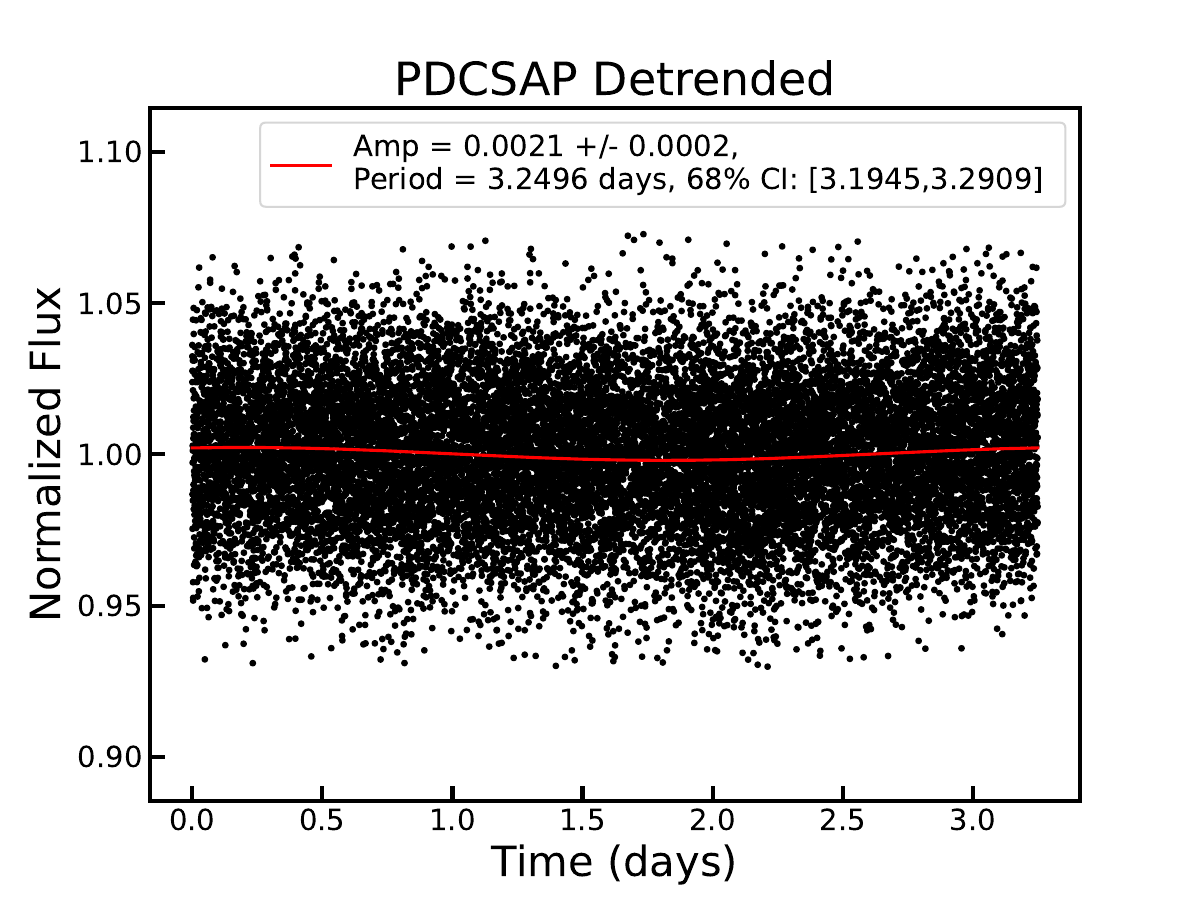}
        \includegraphics[width=0.41\linewidth]{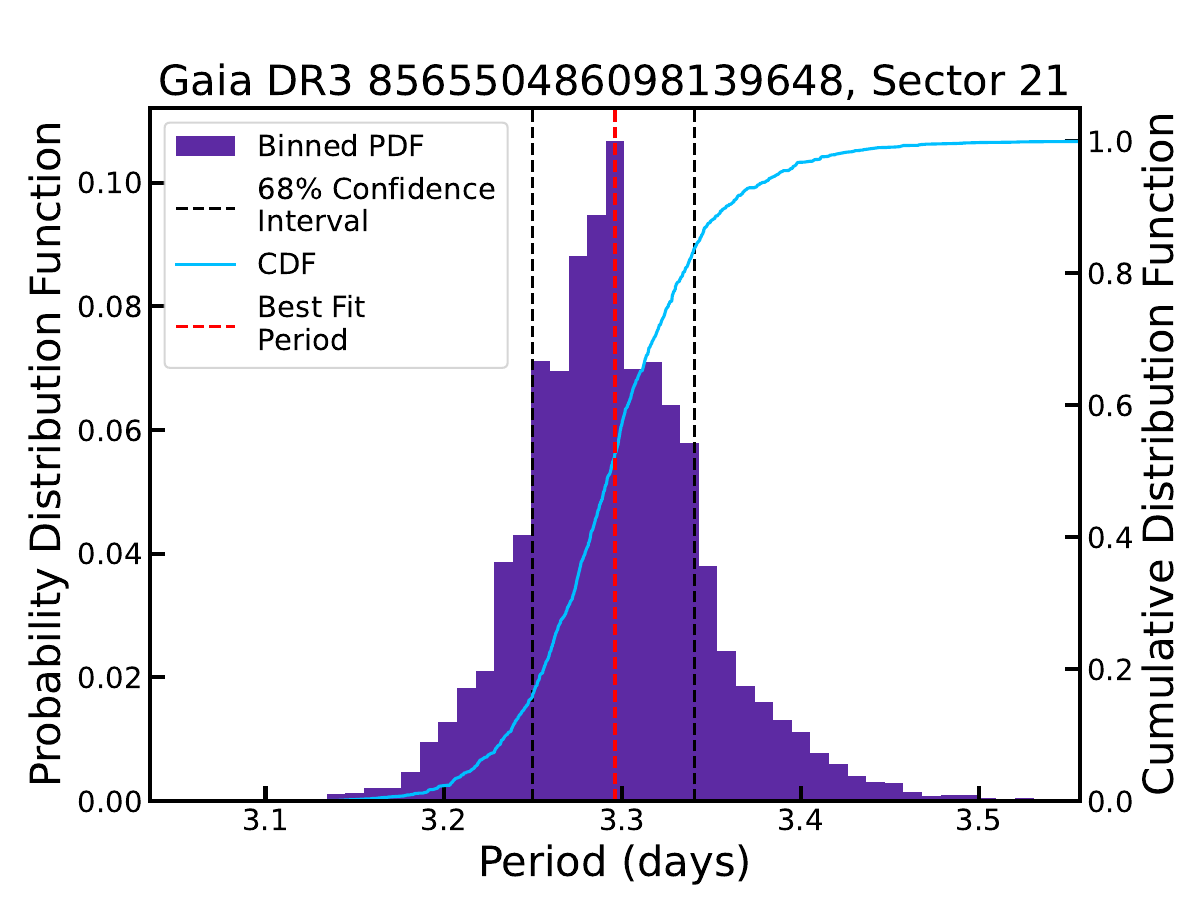}
        \includegraphics[width=0.41\linewidth]{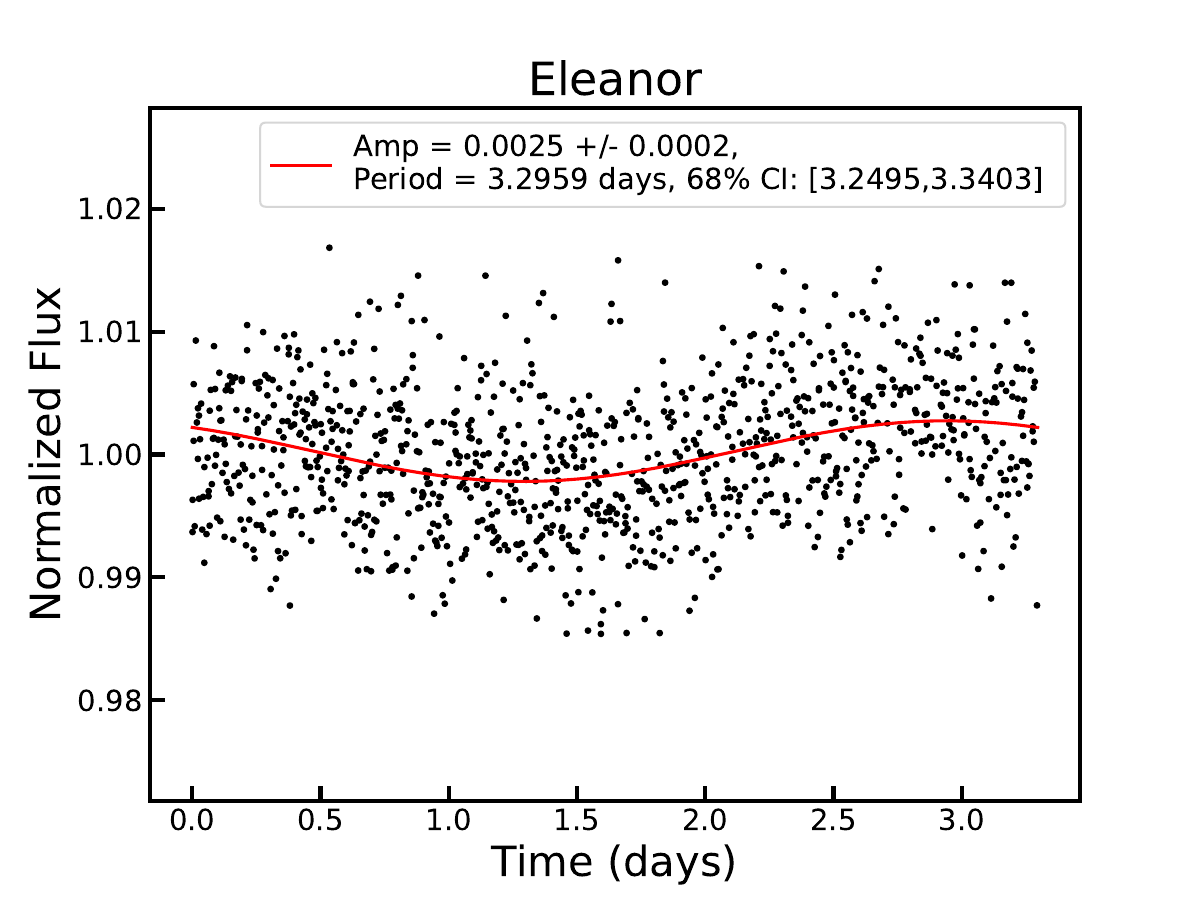}
        \includegraphics[width=0.41\linewidth]{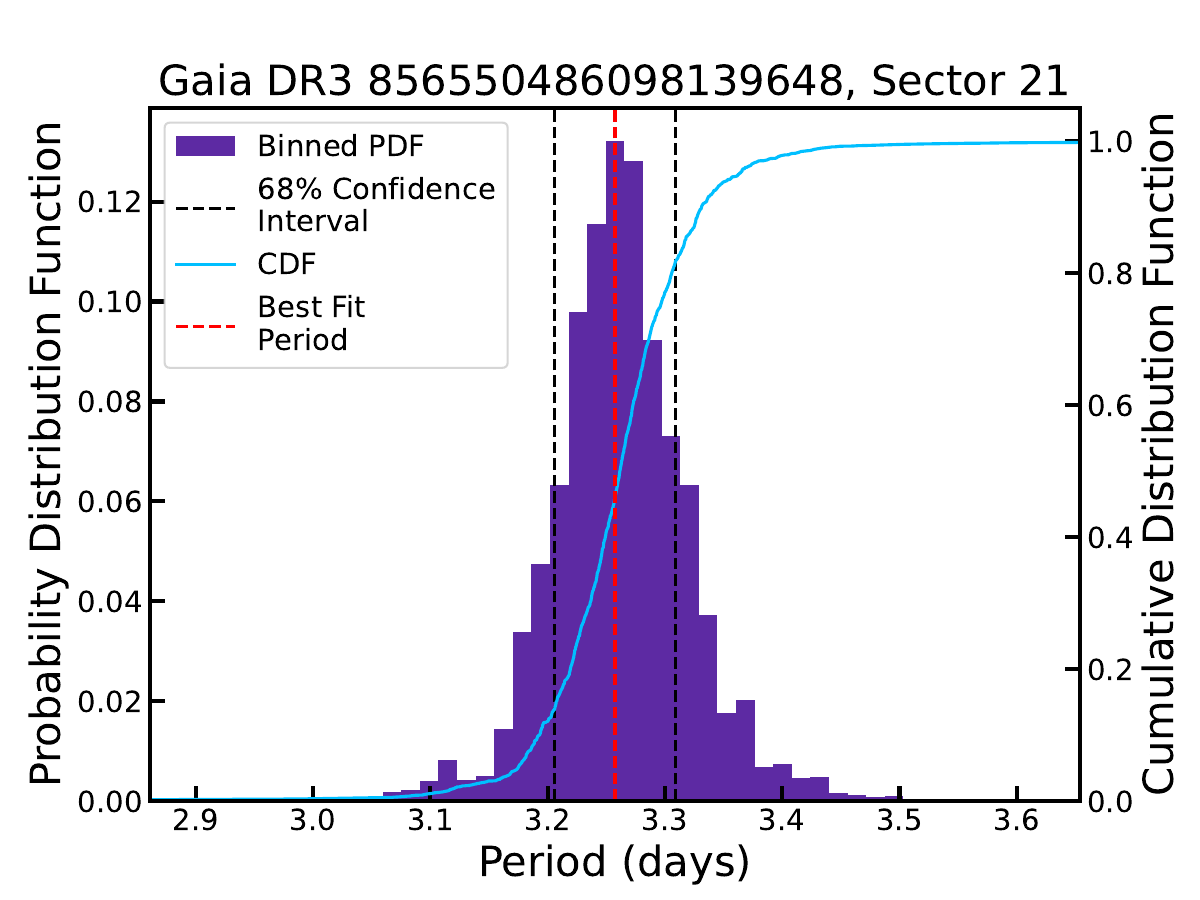}
        \includegraphics[width=0.41\linewidth]{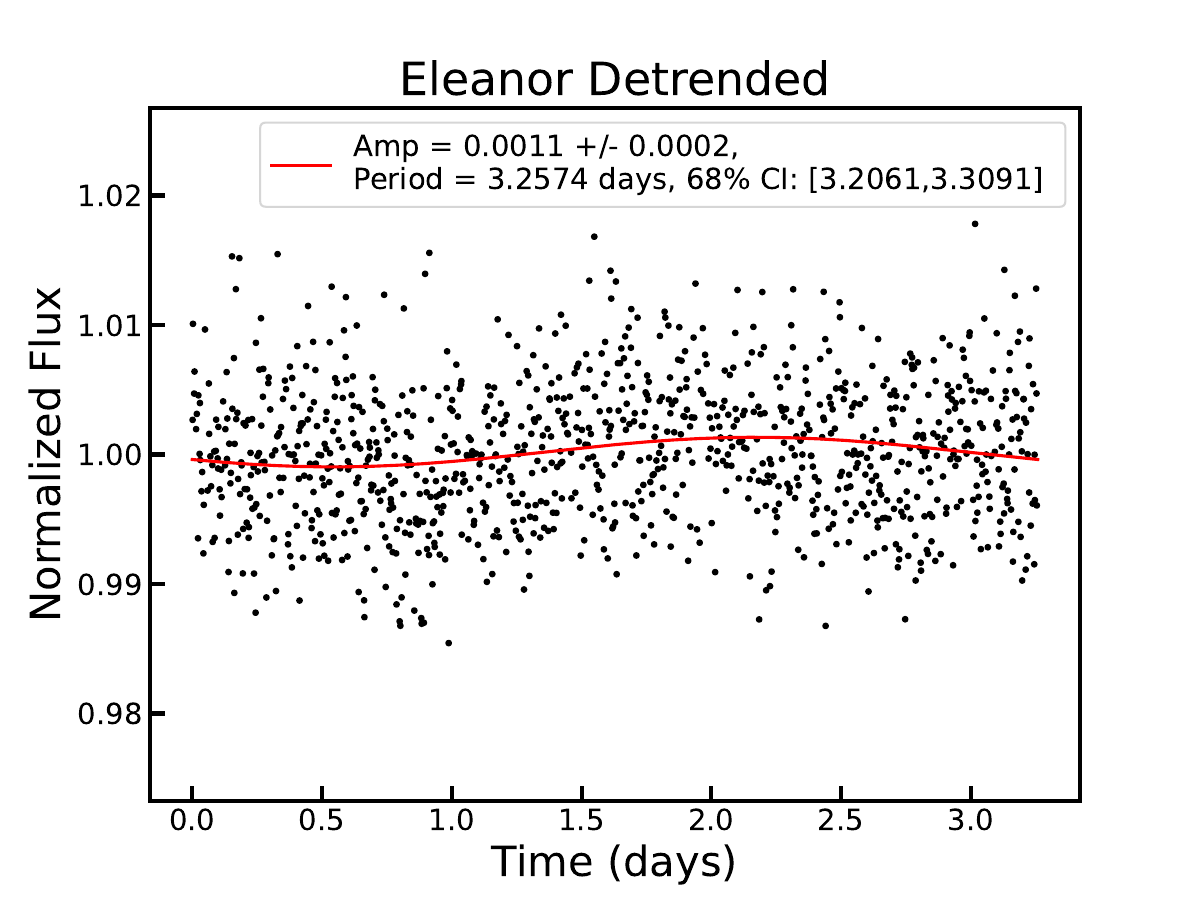}
	\caption{GP results for Sector 21 of the M8 target Gaia DR3 856550486098139648 shown in Figure \ref{fig:lspmethodsslow}. The candidate period determined from the LS analysis and passed onto the GP is 3.19 days, and the final period is 3.246$^{+0.022}_{-0.049}$ days from the Sector 21 PDCSAP non-detrended data. \textit{Left panels:} Binned PDFs displaying the distribution of periods %samples 
    from the nested sampling process %for the period parameter space, 
    for each of the four data reduction methods (see titles of right-hand panels). The most likely periods are marked with vertical red dashed lines. The CDFs are plotted in teal, from which the 68\% confidence intervals (black dashed lines) are calculated. \textit{Right panels:} Corresponding light curves folded on the best periods from the nested sampling process.}
	\label{fig:gpmethodsfast}
\end{figure*}

 \begin{figure*}
	\centering
	\includegraphics[width=0.41\linewidth]{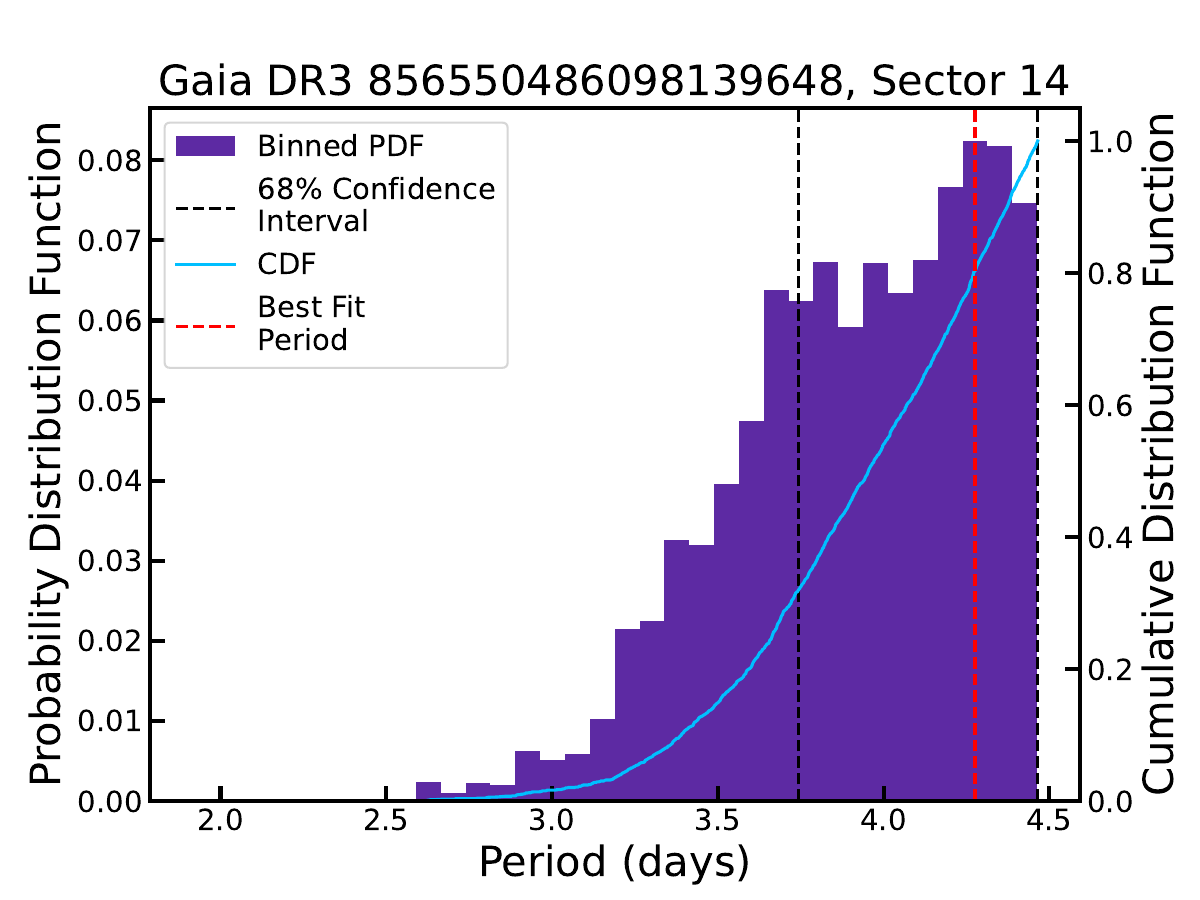}
        \includegraphics[width=0.41\linewidth]{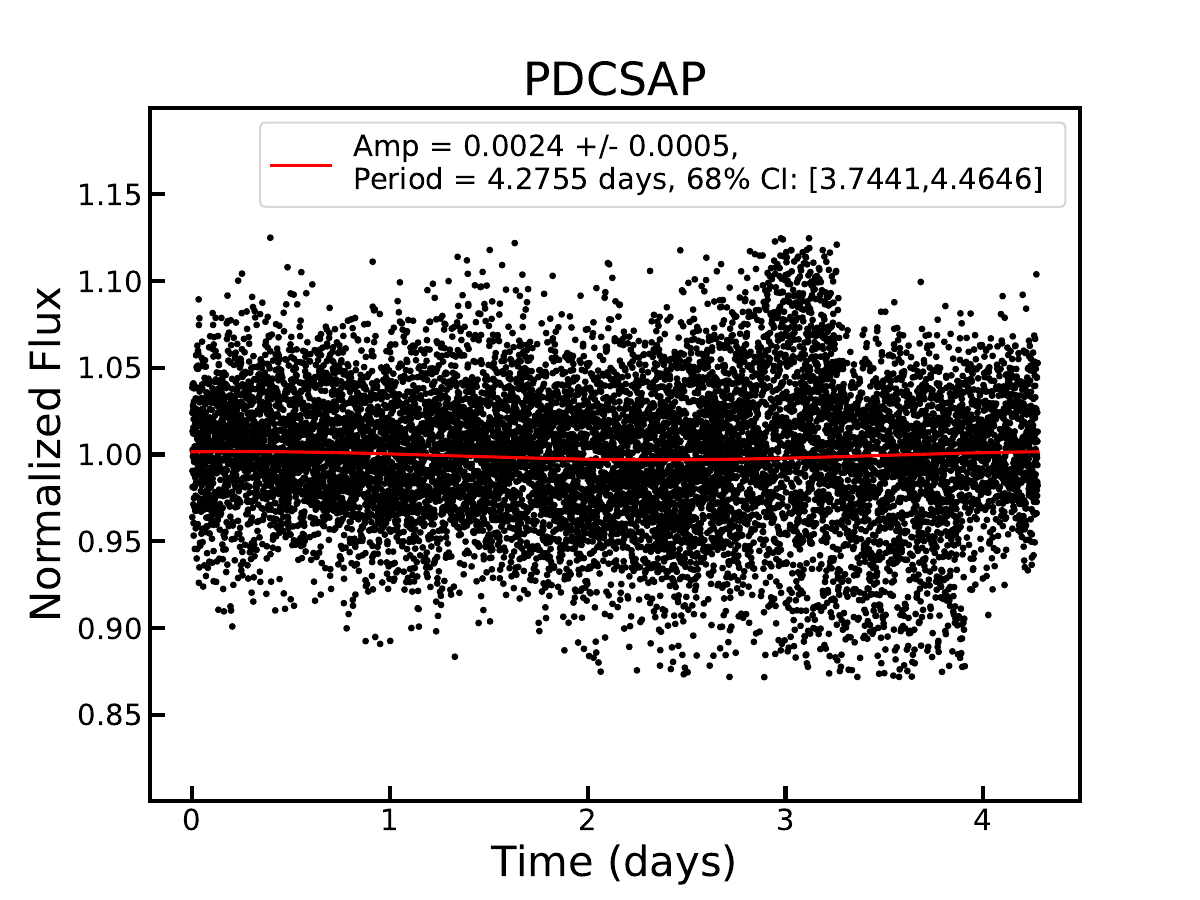}
        \includegraphics[width=0.41\linewidth]{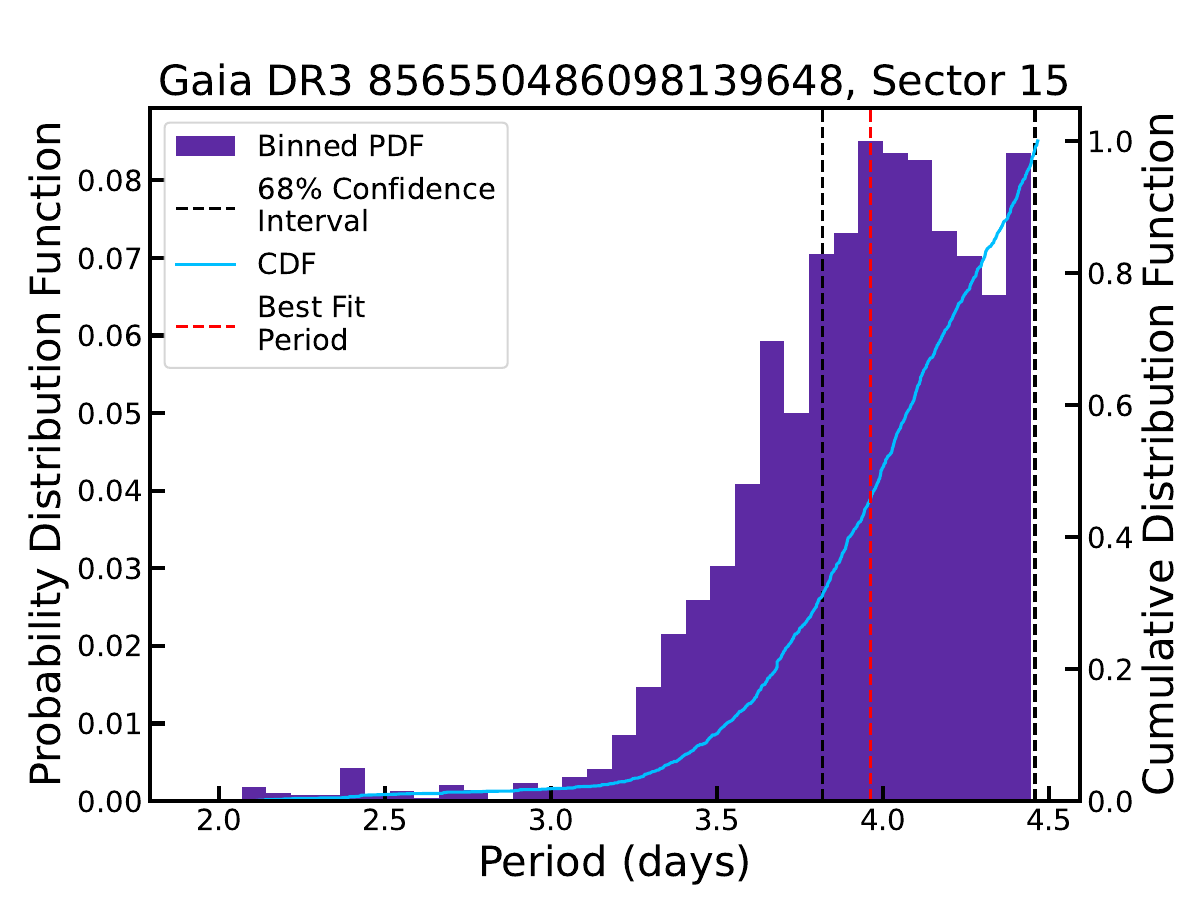}
        \includegraphics[width=0.41\linewidth]{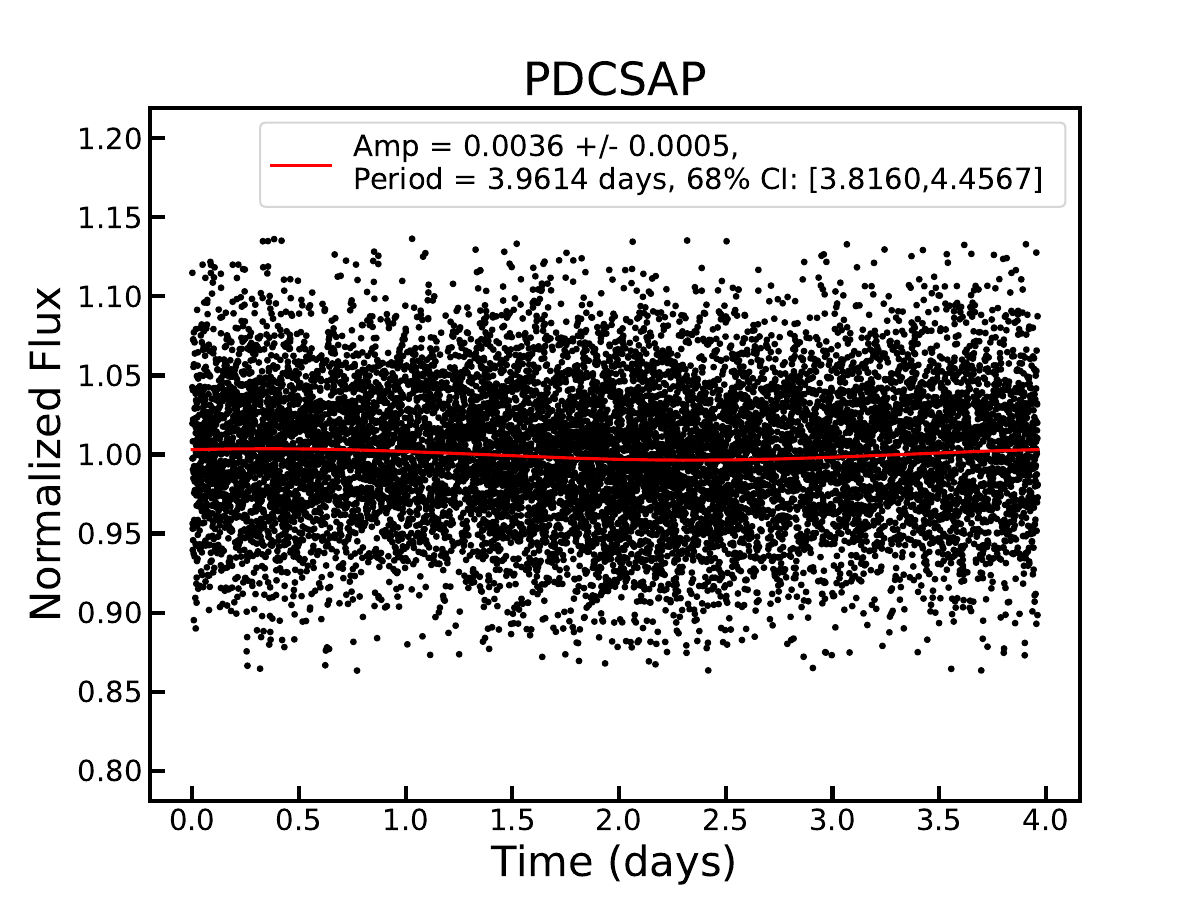}
        \includegraphics[width=0.41\linewidth]{spoc_GP_Period_days_CDF_zoom_s21_85655.pdf}
        \includegraphics[width=0.41\linewidth]{spoc_folded_lightcurve_s21_85655.pdf}
        \includegraphics[width=0.41\linewidth]{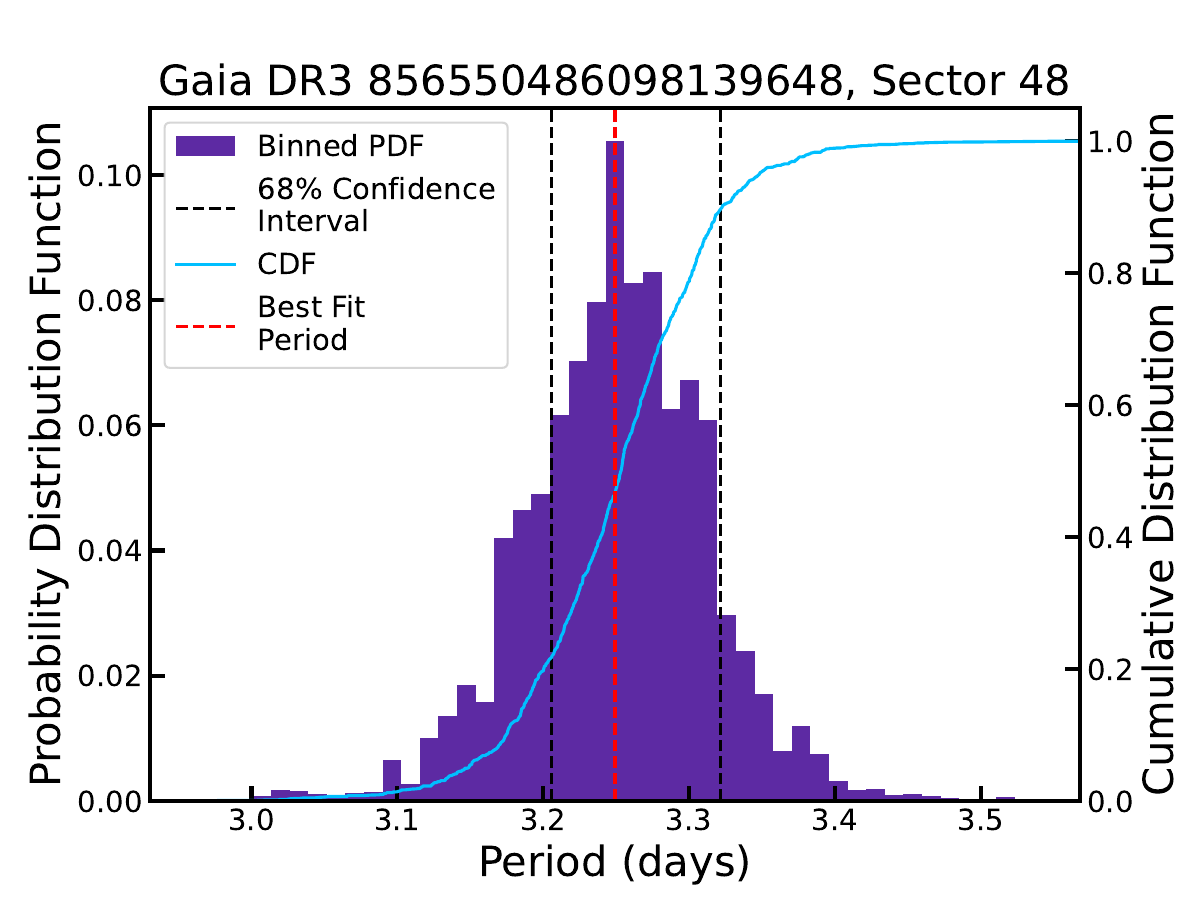}
        \includegraphics[width=0.41\linewidth]{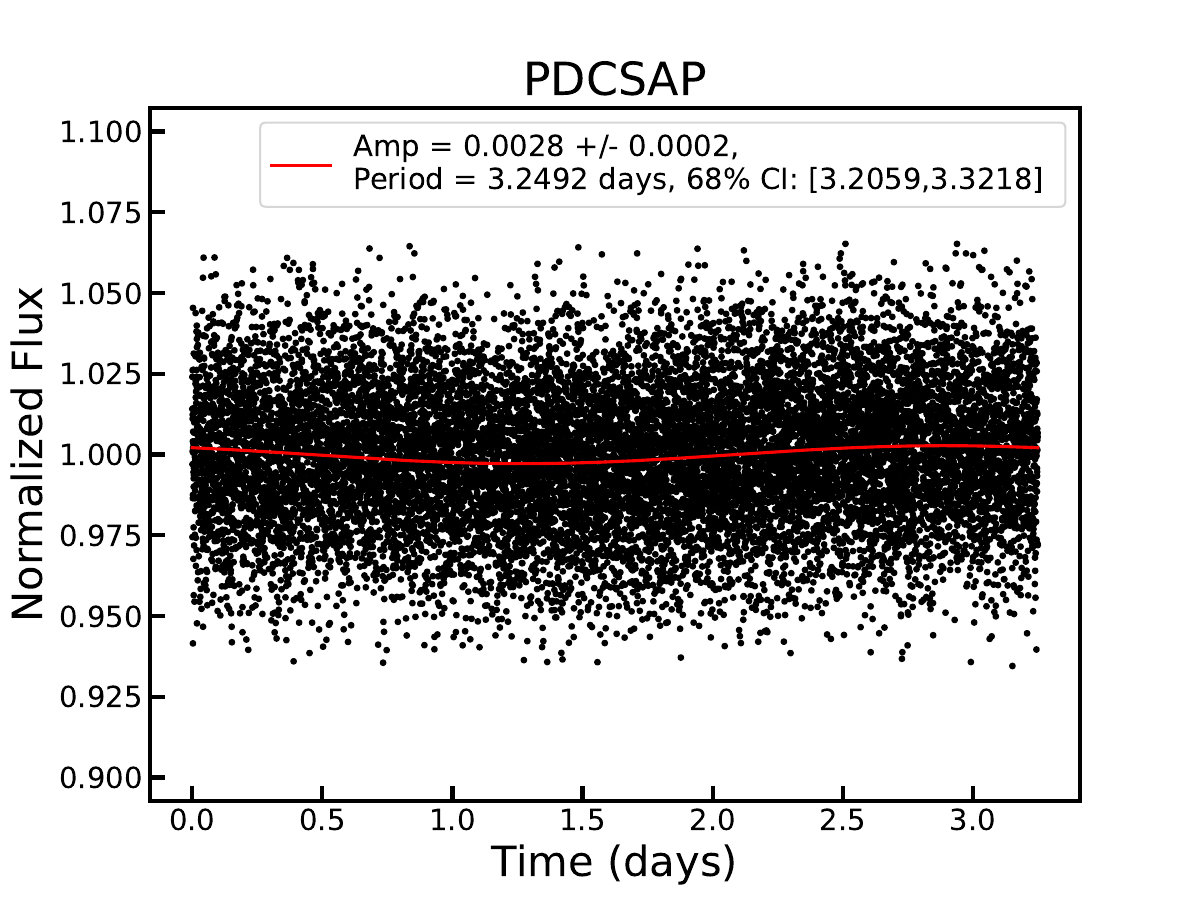}
	\caption{The same as Figure \ref{fig:gpmethodsfast}, but for PDCSAP non-detrended reductions for all sectors of available data. The candidate period determined from the LS analysis is confirmed in Sectors 21 and 48. Note: The 3.96 and 4.28-day periods determined in Sectors 14 and 15 are spurious, as the probability distribution function peaks at or very near the edge of a bin.}
	\label{fig:gpmethodsslow}
\end{figure*}

\begin{figure*}
	\centering
	\includegraphics[width=0.8\linewidth]{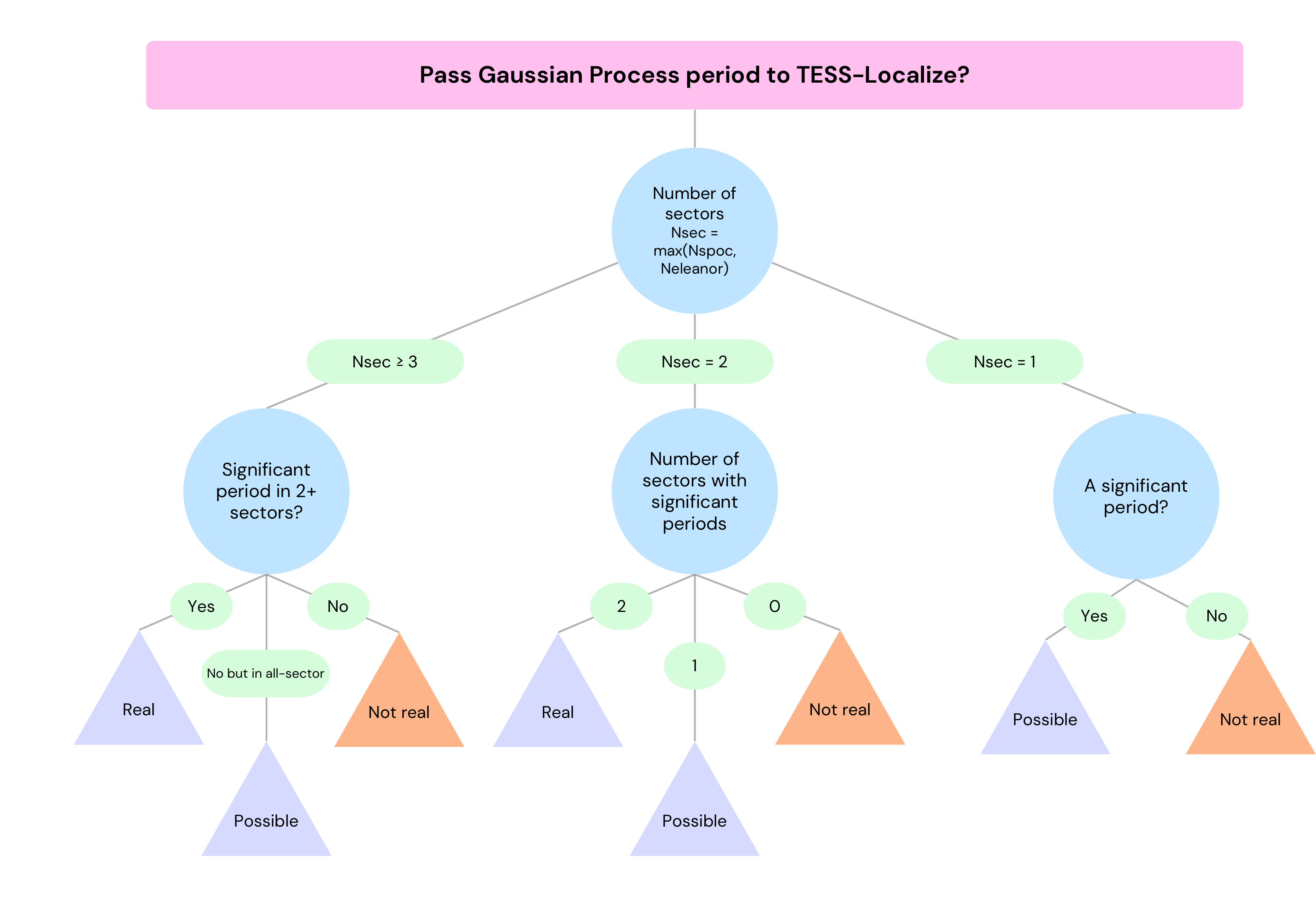}
	\caption{Decision tree to determine whether a target is periodic, based on the GP in Step 3 of the analysis.}
	\label{fig:decisiontreegp}
\end{figure*}

Gaussian processes (GPs) are stochastic models which use a kernel or ``covariance'' function to parameterize the covariance between data points \citep{angus2018}. While these ``hyperparameters'' are not physical in nature, an appropriate kernel choice can allow certain hyperparameters to be interpreted physically, such as rotation period. We utilized the \texttt{RotationTerm} kernel from \texttt{celerite2} \citep{foreman2023}. This kernel combines two stochastically driven damped harmonic oscillators (SHOs) with a power spectral density of the form:

\begin{equation}
    S(\omega) = \sqrt{\frac{2}{\pi}} \frac{S_0 \omega_0^4}{(\omega - \omega_0^2)^2 + \omega_0^2\omega^2/Q^2},
\end{equation}
where $\omega_0$ is the undamped angular frequency, and $Q$ is the quality factor. The rotation kernel specifically has modes at the period and 1/2 the period. The hyperparameters fit are the standard deviation $\sigma$, the primary period $P$, the quality factor of the secondary mode (minus 1/2) $Q_0$, the difference between the quality factors of the first and second modes $\delta Q$, and the fractional amplitude of the secondary mode to the first $f$. These set the hyperparameters for the two SHOs as follows:
\begin{eqnarray}
    Q_1 = \frac{1}{2} + Q_0 + \delta Q \nonumber \\
    \omega_1 = \frac{4\pi Q_1}{P\sqrt{4Q_1^2 - 1}} \\
    S_1 = \frac{\sigma^2}{(1 + f) \omega_1 Q_1} \nonumber 
\end{eqnarray}

\begin{eqnarray}
    Q_2 = \frac{1}{2} + Q_0 \nonumber \\
    \omega_2 = \frac{8\pi Q_2}{P\sqrt{4Q_2^2 - 1}} \\
    S_2 = \frac{f\sigma^2}{(1 + f) \omega_2 Q_2} \nonumber
\end{eqnarray}

To fit the GP model to the data, we used nested sampling with the package \texttt{dynesty} \citep{speagle2020}. We used 250 walkers and set bounds on the hyperparameters of [0, 1] for $\sigma$ and $f$, [0, 50] for $Q_0$ and $\delta Q$, and [$P_{LS}$ - 0.4*$P_{LS}$, $P_{LS}$ + 0.4*$P_{LS}$] for $P$, where $P_{LS}$ is most likely period as determined by the LS process. If multiple periods were passed, the GP was run for each provided period. 

From the nested sampling results, we construct a binned probability distribution function (PDF) with a bin width that is 1/5 the standard deviation of the samples. The reported period is chosen as the center of the bin of the highest peak in the PDF. The uncertainty on the period is the 68\% confidence interval with the narrowest period range as determined from the cumulative distribution function (CDF) of the samples. The amplitude of a fit is determined from the folded light curve in the same manner as it is in the LS step. An example of these PDFs and corresponding folded light curves for one star (the same one used as an example of the LS process in Figure \ref{fig:lspmethodsslow}) can be seen in Figure \ref{fig:gpmethodsfast} and \ref{fig:gpmethodsslow}.

Our approach returns a period regardless of whether the target displays variability or not. To control for this, we automatically removed periods if:
\begin{itemize}
    \item The signal-to-noise of the amplitude was less than 5, as a low signal-to-noise indicates a poor sinusoidal fit;
    \item The signal-to-noise of the period was less than 10 (i.e., $P_{rot} / \bar{\sigma} < 10$, where $\bar{\sigma}$ is the mean of the upper and lower uncertainties);
    \item The period was within 5\% of the edges of the sampled period space. A poor GP fit can get `stuck' in a local minimum on the edge of a parameter space. Examples of this edge case can be seen in the first two rows of Figure \ref{fig:gpmethodsslow};
    \item The period aligned with a peak in the window function.
\end{itemize}

\begin{figure*}
	\centering
        \includegraphics[width=0.31\linewidth]{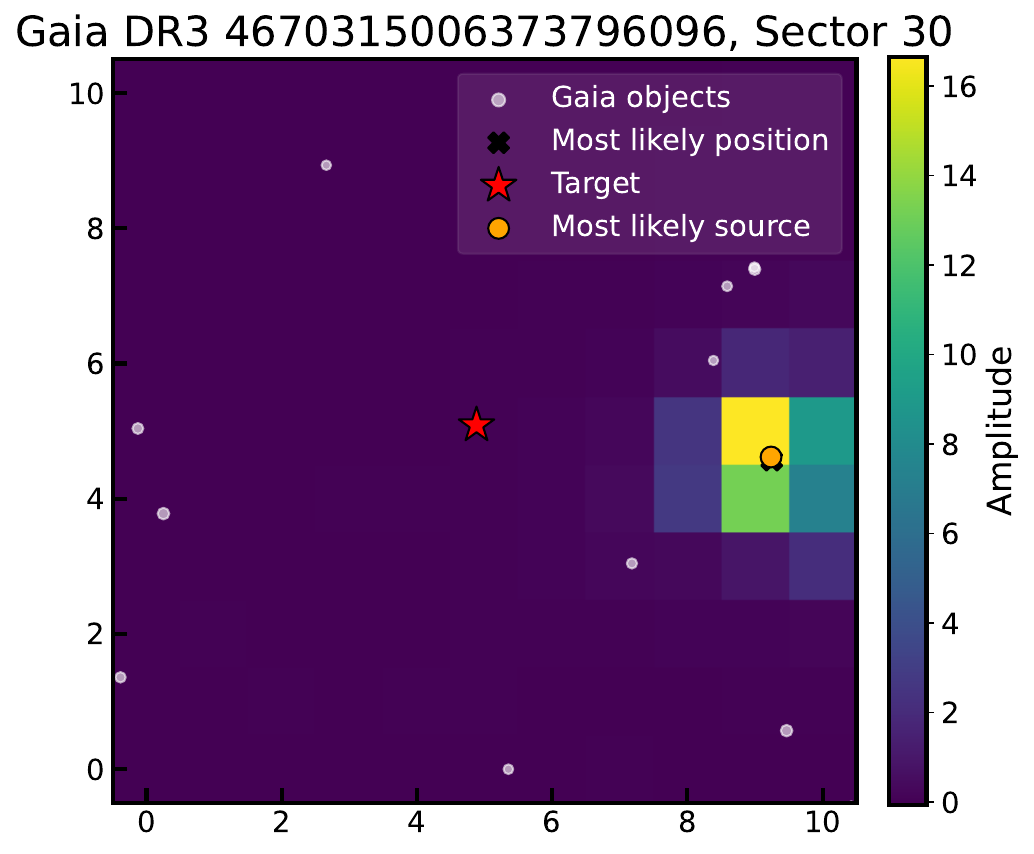}
        \includegraphics[width=0.31\linewidth]{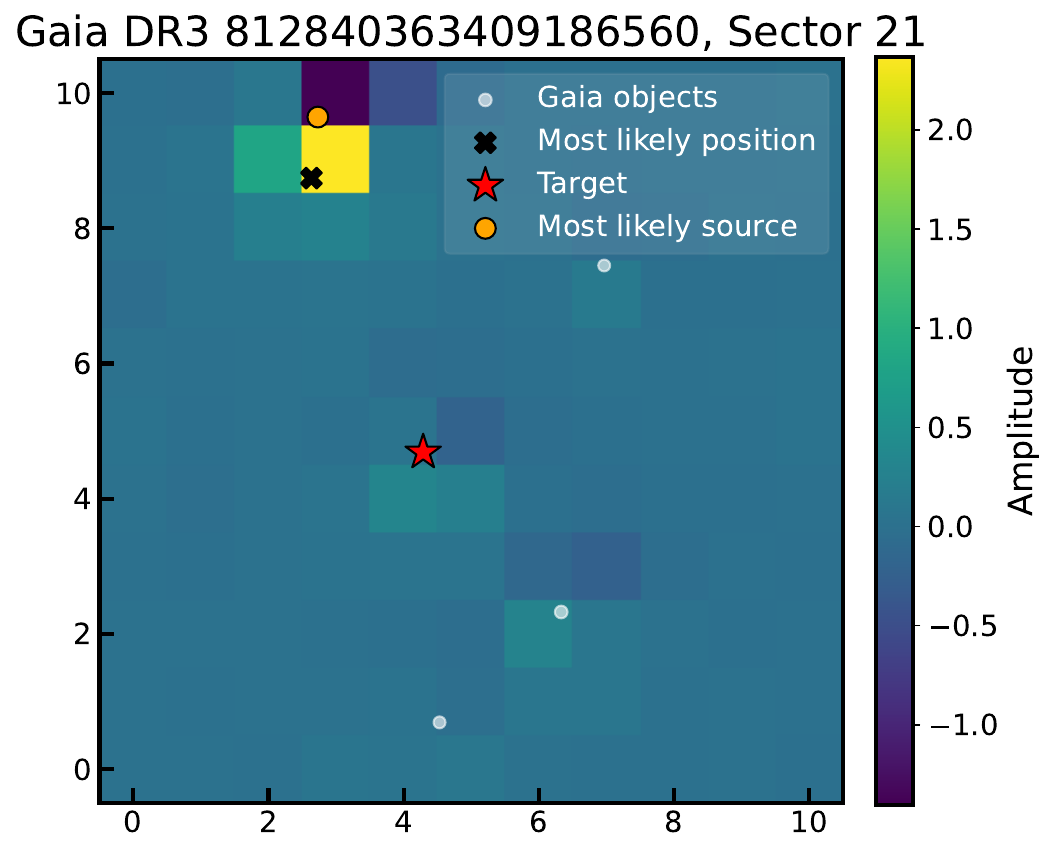}
        
        \includegraphics[width=0.32\linewidth]{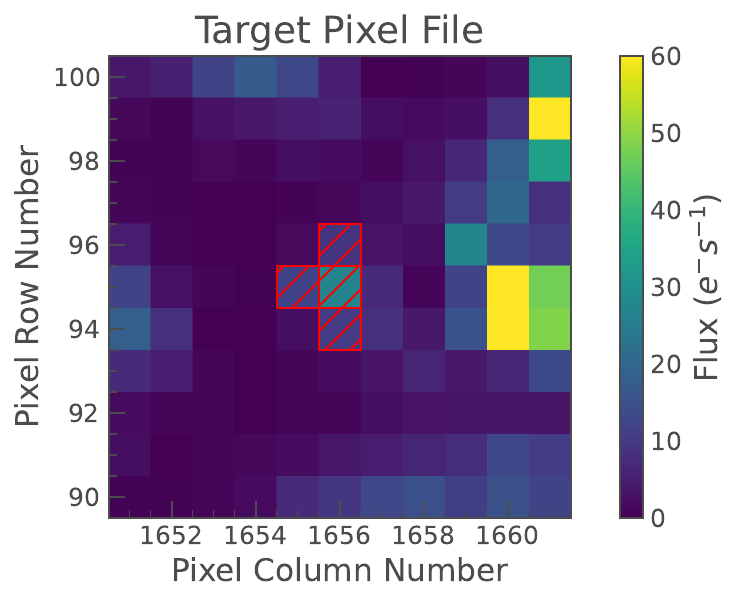}
        \includegraphics[width=0.32\linewidth]{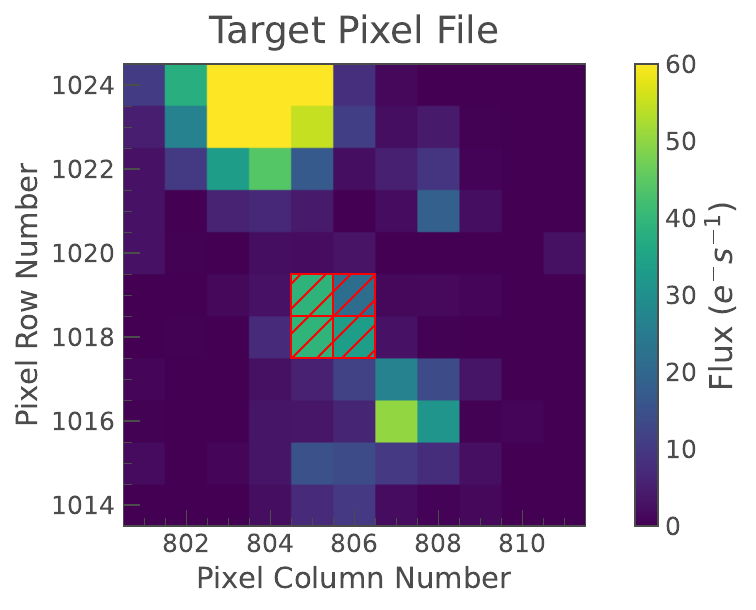}
        
        \includegraphics[width=0.31\linewidth]{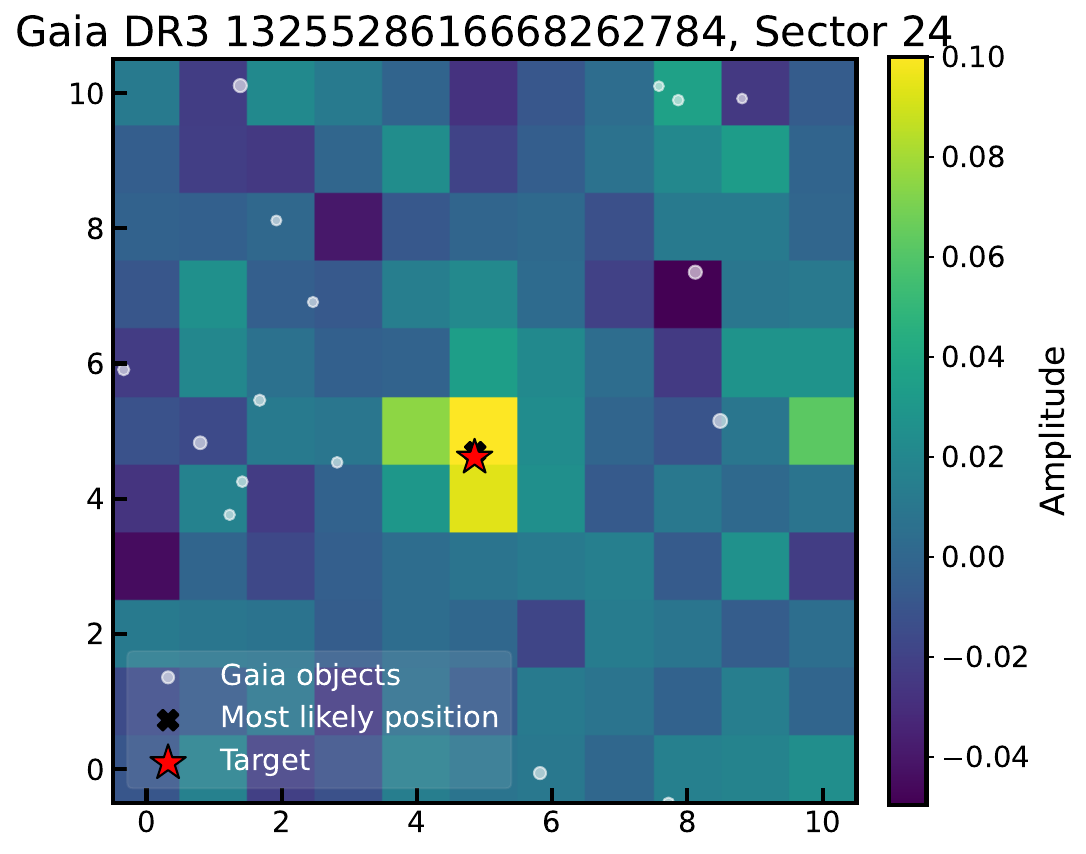}
        \includegraphics[width=0.31\linewidth]{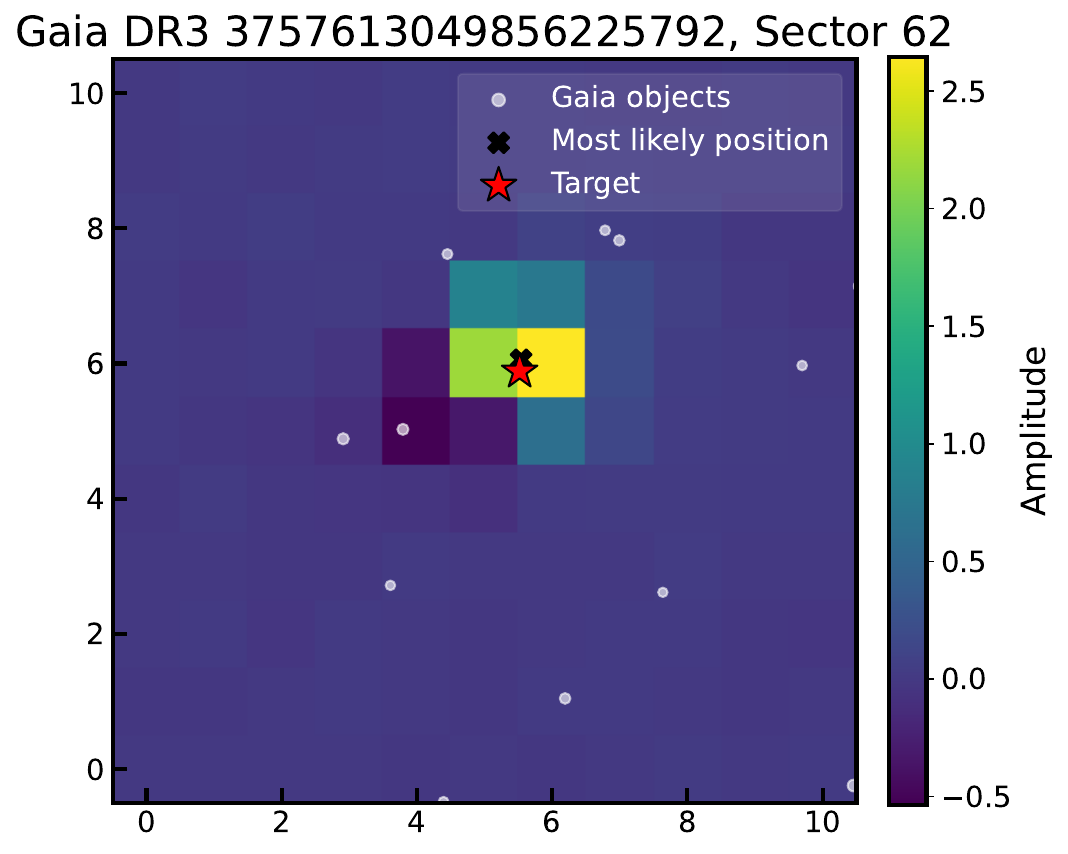}
        \includegraphics[width=0.31\linewidth]{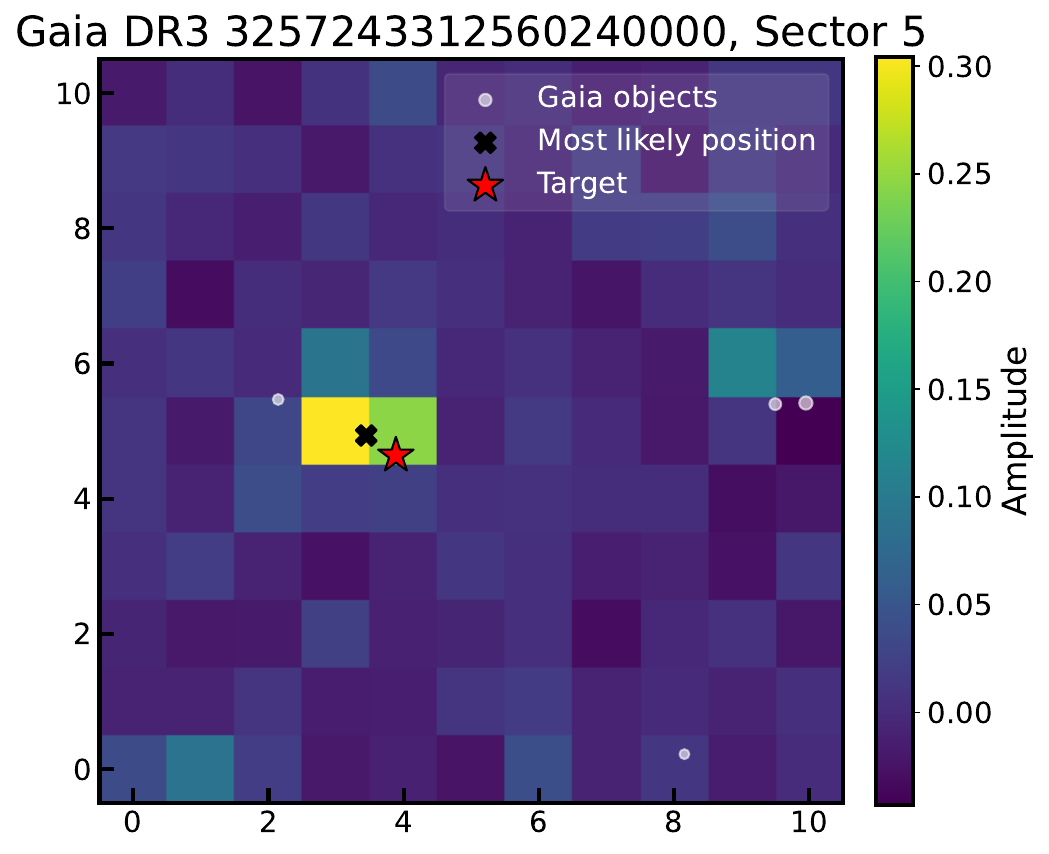}
        
        \includegraphics[width=0.32\linewidth]{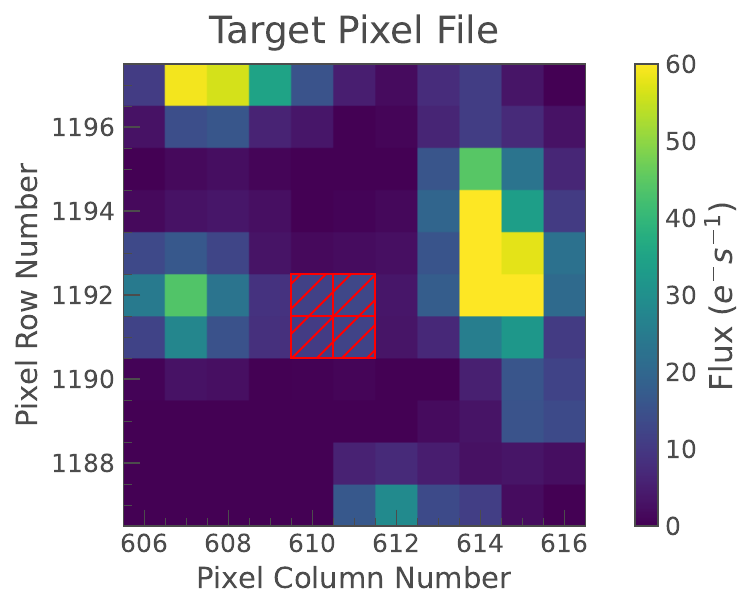}
        \includegraphics[width=0.32\linewidth]{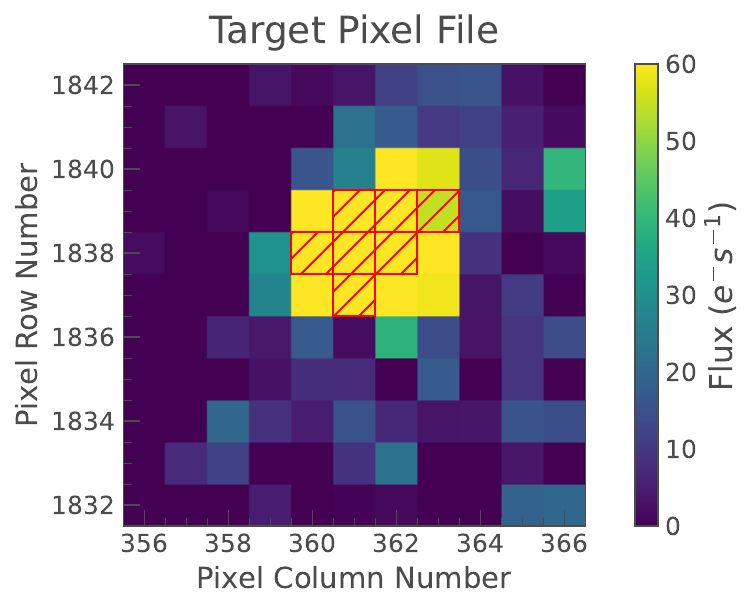}
        \includegraphics[width=0.32\linewidth]{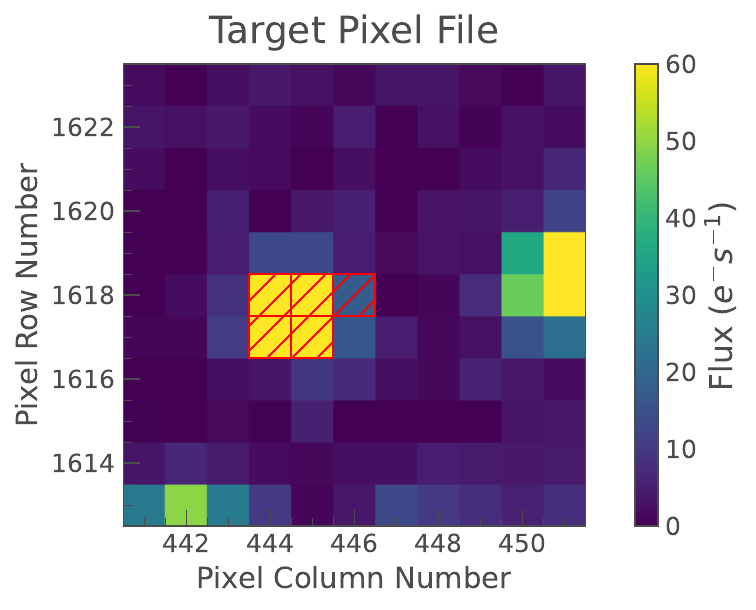}
        
	\caption{Examples of different scenarios in TL. \textit{Top two rows:} Amplitude heat maps (top row) and the corresponding target pixel files (second row) for TL FALSE results. The left panels show a case where a distant bright star located just off the TPF is the true periodic source, contaminating our star. The right panels show a case where the passed period is a TESS systematic, mapping the stars in the amplitude heat map. \textit{Bottom two rows:} The same as the top, for cases where TL returned TRUE. The left panels show a good result, in which the periodicity is confirmed to be co-located with the target star. The middle panels show a case where the $p$-value is low due to the high proper motion of the star. However, our visual inspection of this result retains it as a confirmed period. The right panels show an example where TL returns TRUE, but the period is not confirmed. This is a case of the same systematic as in the right panels of the top two rows, but here, our target is the brightest star in the field.} 
    %\textcolor{red}{(SM: A number of the panels don't render. Perhaps the links need correcting.)}}
	\label{fig:tl_heatmaps}
\end{figure*}

We can then classify a target as having ``real" periodicity if it has the same period (within uncertainties) in 2 or more sectors, independent of the reduction method. If it only has a period in one sector, but that period is also seen in the combination of all sectors, the target is given a ``possible'' status. If a target only has data in one or two sectors, it can be classified as having ``possible'' periodicity if it has a period in a single sector. This process is laid out in a decision tree in Figure \ref{fig:decisiontreegp}.

% A period can be upgraded from ``possible'' to ``real'' if it matches values determined with other (non-TESS) studies in the literature.  The final reported value is the one from the method and sector with the highest amplitude signal-to-noise. 

At this stage, there were 126 real and 46 possible periodic targets to be passed to \texttt{TESS-Localize} to confirm their source.

\subsection{Step 4: TESS-Localize}
\label{sec:tl}

\texttt{TESS-Localize} (TL) is a Python package designed by \citet{higgins2023} to determine the source of periodicity of a target with sub-pixel accuracy. It works on TESS data with TPFs. TL localizes periodicity by producing a sinusoidal fit for every pixel in the TPF, calculates amplitudes, and can automatically determine the appropriate TESS aperture for a star to gain accuracy within 1/5 of a pixel. By doing this sine fit, TL determines the most likely position for periodicity, independent of any knowledge of the actual stars in the TPF. This is the ``most likely position'' seen in Figure \ref{fig:tl_heatmaps}. It then matches that position with the closest Gaia star, giving the ``most likely source'' of variability. The full methodology of this package can be found in \citet{higgins2023}. We use the automatic aperture setting and input our best period from the GP. To visually examine the results from TL, we examine the provided amplitude heat maps, which give the amplitude of the input period in each pixel (Figure \ref{fig:tl_heatmaps}, top rows).

As our targets are selected to be uncrowded, there should not be many cases in which the source of periodicity is a star outside of our target's aperture. However, as mentioned in Section \ref{sec:crowded}, our check for crowding in TESS extended out only to a 5-pixel radius. If there is a much brighter (e.g., $G<10$ mag) star, contamination could still be present at larger distances.

TL uses simple aperture photometry to create light curves from the TPF. As such, a significant periodic signal found with one of the four methods in this work may not be significant in TL. In this case, TL may fail to locate the true source of periodicity in a TPF. Therefore, to confirm that the variability in TESS is co-located with our target, we seek a TRUE outcome from TL from at least one TESS sector. A FALSE outcome, indicating that the periodicity corresponds to a different star, can occur when:
\begin{itemize}
    \item The periodic star is not our target: e.g., in the case of a bright star more than 5 pixels away (Figure \ref{fig:tl_heatmaps}, first example); 
    \item The input period is the physical rotation period of the star, but is just not present in the sector run through TL (with a high enough amplitude SNR $\gtrsim$7.5 or amplitude $\gtrsim$0.4\%). In this case, TL should produce a TRUE result in the sectors where the period is present in TL, if such a sector exists.
    \item The input period is not a physical signal from the star, i.e., it is a systematic. In this case, the periodic signal is seen throughout the entire TPF and is directly proportional to the brightness of the stars in the pixels (Figure \ref{fig:tl_heatmaps}, second example);
\end{itemize}

In the case where there are all FALSE results, we visually examine the LS periodograms of the objects that were indicated to be the more likely source of variability to distinguish between these cases. This is because several of the above scenarios can occur simultaneously; the non-detrended PDCSAP data has an amplitude SNR less than 7.5 or amplitude less than 0.4\%, and the periodic signal is not physical or originates from a brighter nearby star. If a star only has too low amplitude or amplitude SNR for TL, it cannot be definitively confirmed, but there is also no reason to outright deny these periods. Therefore, they are included in the sample with a ``possible'' status. There are 12 such targets in our sample.

%As it is possible for a period to not be present in some of the sectors we are testing with TL, we only require that for a target to be confirmed as the true source of periodicity, it must be deemed TRUE in at least one sector.

\begin{figure}
	\centering
	\includegraphics[width=0.9\linewidth]{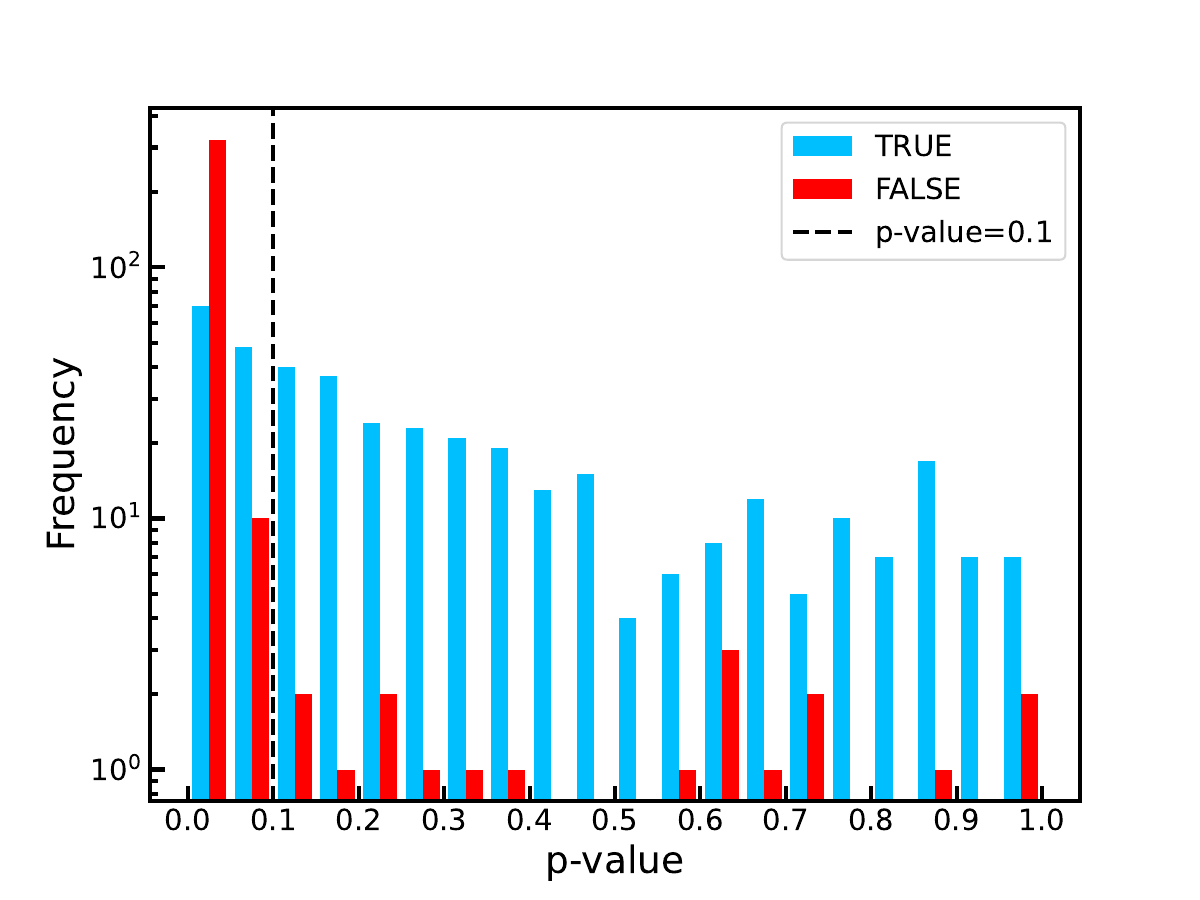}
	\caption{A comparison of the $p$-values for TESS-Localize results in which the target is the source of periodicity (TRUE) and where it is not (FALSE). The vertical line indicates a value of 0.1, below which all TRUE results are visually examined.
    %\textcolor{orange}{(SM: It would be good to have ``$p$-value'' written the same way in the text and on the Figure.)} IMPLEMENTED
    }
	\label{fig:pvaluehist}
\end{figure}

\begin{figure*}
	\centering
	\includegraphics[width=\linewidth]{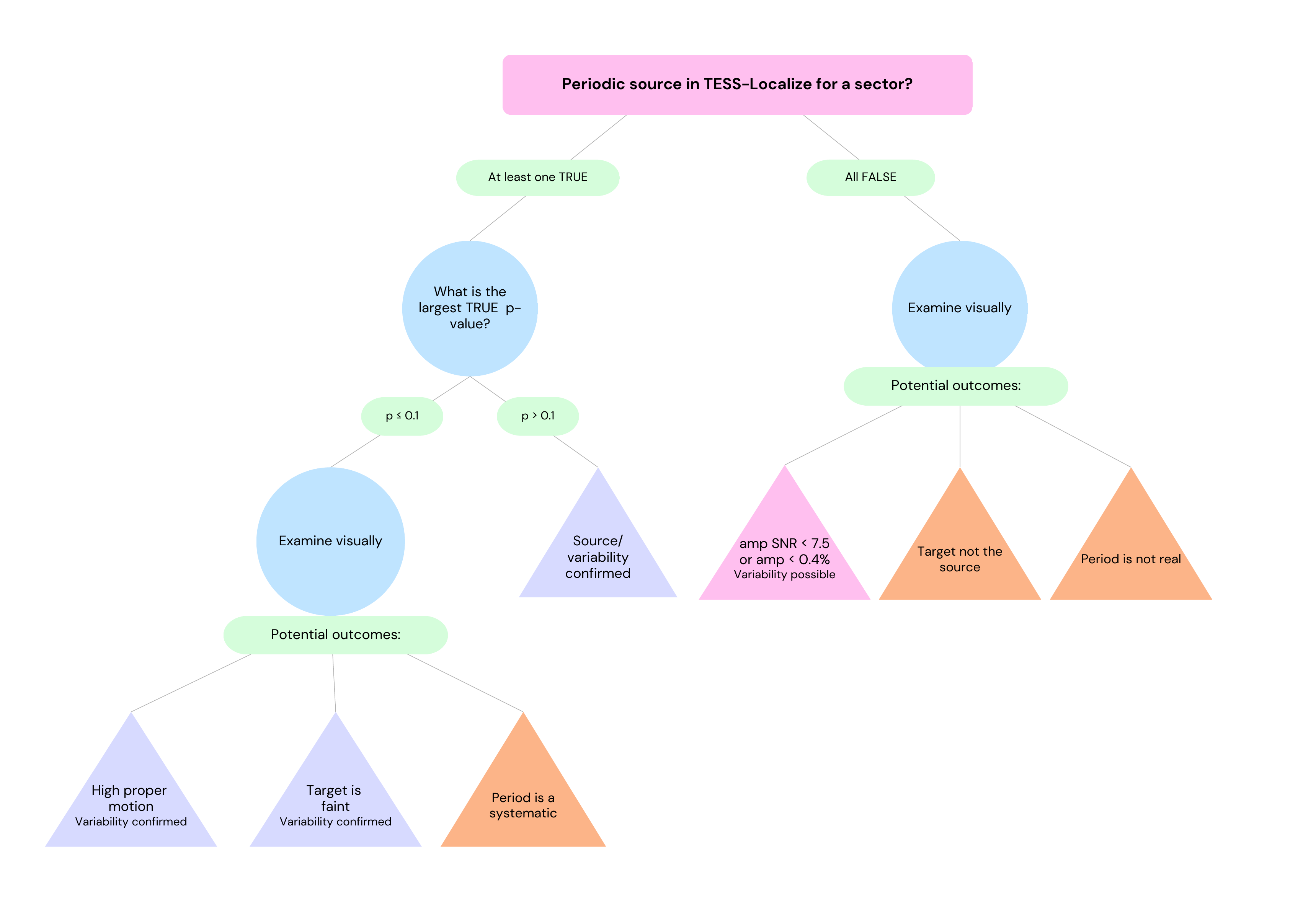}
	\caption{Decision tree detailing variability validation with TESS-Localize. }
	\label{fig:decisiontreetl}
\end{figure*}

TRUE results, suggesting that the periodicity is co-located with our target, also can not be blindly trusted. TL produces a metric called the `$p$-value' ($p$) for the co-location of the candidate periodicity with each Gaia source in the TPF. 
The $p$-value gives the fraction of fit locations with a lower probability of periodicity than the Gaia source. The higher the $p$-value, the more likely the target is the true source of periodicity. A low $p$-value could result, for example, from inputting a period which is a TESS systematic, but TL still determines our target as the source, as it is the brightest star in the TPF. An example of this can be seen in the bottom right panel of Figure \ref{fig:tl_heatmaps}. 

\citet{higgins2023} suggest that a lower threshold for the $p$-value should be set and gives the example that for their experiment, they expected 5\% of results with $p\leq 0.05$ to be false fits. We did not automatically discard all low $p$-value results as unreliable, as we found that in some cases, they reflected real variability.  
We found two such cases:
%when a low $p$ still properly identifies the source of periodicity. These are:
\begin{itemize}
    \item The target has a high proper motion. This results in the best-fit Gaia source being located in a different pixel than the source of periodicity, as the star has moved between the Gaia DR3 epoch (2016.5) and the epoch of observation in TESS (Figure \ref{fig:tl_heatmaps}, bottom middle panel).
    \item The star is faint ($<$30 e$^-$/s for the brightest TESS pixel of the star). This results in an overall lower amplitude to fit in TL.
\end{itemize}

Hence, instead of automatically discarding TRUE outcomes with low $p$-values, we visually examined the amplitude heat maps for all with TRUE outcomes with $p\leq 0.1$. We set the threshold based on our own analysis of our candidate variables, which revealed that 95\% of FALSE cases have $p$-values less than 0.1 (Figure \ref{fig:pvaluehist}).

The above process for validating the candidate periodicity of our targets with TL is detailed in Figure \ref{fig:decisiontreetl}. In our final list of periodic stars, we find 121 that TL confirms are the originators of the period we determined. These stars are left with the status determined for them in the GP analysis: ``real'' or ``possible''. An additional 12 could not be confirmed or refuted and are included with a ``possible'' status as described above.  Furthermore, we find that there are three cases where a contaminating bright star is the true source of periodicity and 37 where the period found is a systematic signal. These are given the status of ``not real''.  This leaves a final count of 95 ``real'' and 38 ``possible'' periods, out of 172 candidate periodicities passed to TL.

\setlength{\tabcolsep}{2pt}
\renewcommand{\arraystretch}{0.9}

\begin{deluxetable*}{cccccccc}
    \tabletypesize{\scriptsize}
    \digitalasset
    \tablecaption{The 133 candidate single periodic uncontaminated TESS late-M dwarfs found in this work \label{tab:periods}}
    \tablehead{
        \colhead{Gaia ID} & 
        \colhead{PDCSAP sectors} & 
        \colhead{Eleanor sectors} & 
        \colhead{$P_{rot}$ (days)} & 
        \colhead{Amp (\%)} & 
        \colhead{Best Sector} & 
        \colhead{Best Method} & 
        \colhead{Status} 
    }
    \startdata
    10147878144789504 & 4, 31, 44, 70, 71 & 4, 31, 42, 44, 70, 71 & 2.652$^{+0.060}_{-0.053}$ & 0.00394$ \pm0.00059$ & 31 & PDCSAP & Real \\
36685828234108032 & 5, 42, 43, 44, 70, 71 & 5, 42, 43, 44, 70, 71 & 5.99$^{+0.48}_{-0.66}$ & 0.00810$ \pm0.00049$ & 71 & Eleanor & Real \\
57083830512424448 & 42, 43, 44, 70, 71 & 42, 43, 44, 70, 71 & 0.6461$^{+0.0035}_{-0.0029}$ & 0.00704$ \pm0.00064$ & 70 & Eleanor Detrended & Real \\
65638443294980224 & 42, 43, 44, 70, 71 & 42, 43, 44, 70, 71 & 0.4907$^{+0.0039}_{-0.0025}$ & 0.00464$ \pm0.00076$ & 70 & PDCSAP Detrended & Possible \\
91196316201965184 & 42, 43, 70, 71 & 42, 43, 70, 71 & 1.528$^{+0.011}_{-0.018}$ & 0.00892$ \pm0.00017$ & 71 & Eleanor Detrended & Real \\
93316174620198656 & 17, 42, 43, 70, 71 & 17, 42, 43, 70, 71 & 0.5557$^{+0.0051}_{-0.0045}$ & 0.00414$ \pm0.00038$ & 70 & PDCSAP Detrended & Real \\
147258180719392512 & 43, 44, 70, 71 & 43, 44, 70, 71 & 0.6578$^{+0.0031}_{-0.0072}$ & 0.00392$ \pm0.00060$ & 71 & PDCSAP Detrended & Real \\
302952222668012032 & 57 & 57, 84 & 0.23071$^{+0.00023}_{-0.00023}$ & 0.02375$ \pm0.00028$ & 57 & PDCSAP & Real \\
576506489410890752 & 8, 34, 61 & 8, 34, 61 & 1.651$^{+0.021}_{-0.021}$ & 0.00583$ \pm0.00023$ & 61 & PDCSAP Detrended & Possible \\
577960765337401856 & 8, 34, 61 & 8, 34, 61 & 0.14266$^{+0.00022}_{-0.00026}$ & 0.01060$ \pm0.00043$ & 61 & Eleanor Detrended & Real \\
    ... & ... & ...& ...& ...& ...& ...& ... \\
    \enddata
    \tablecomments{The full table is available in machine-readable form.}
\end{deluxetable*}

\section{Results and Discussion}
\subsection{Rotation Period Statistics and Distribution}
\label{sec:rot_stats}

The final sample contains 133 (95 real, 38 possible) periodic variables among the 399 single, uncontaminated targets with PDCSAP data. The detected periods range from approximately 1.97 hours (0.0819 days) to approximately 6 days, with amplitudes between 0.08\% and 2.71\%. This is 33.3\%$\pm$2.4\% of the sample, divided into 23.8\%$\pm$2.1\% real periods and 9.5\%$\pm$1.5\% possible periods. The periodic objects are listed in Table \ref{tab:periods}. Here, we also recorded which sector and method produced the result with the highest amplitude SNR. In general, the method that produced the highest amplitude SNR was the PDCSAP pipeline data (101 periods, 75.9\%), either detrended (49 periods) or not (52 periods). \texttt{eleanor} was only responsible for the higher amplitude SNR 24.1\% of the time, also split roughly equally between detrended and not (18 detrended, 14 not). This shows that the PDCSAP pipeline is better for finding rotation periods; however, \texttt{eleanor} was still able to identify most periods.

\begin{figure}
	\centering
        \includegraphics[width=0.9\linewidth]{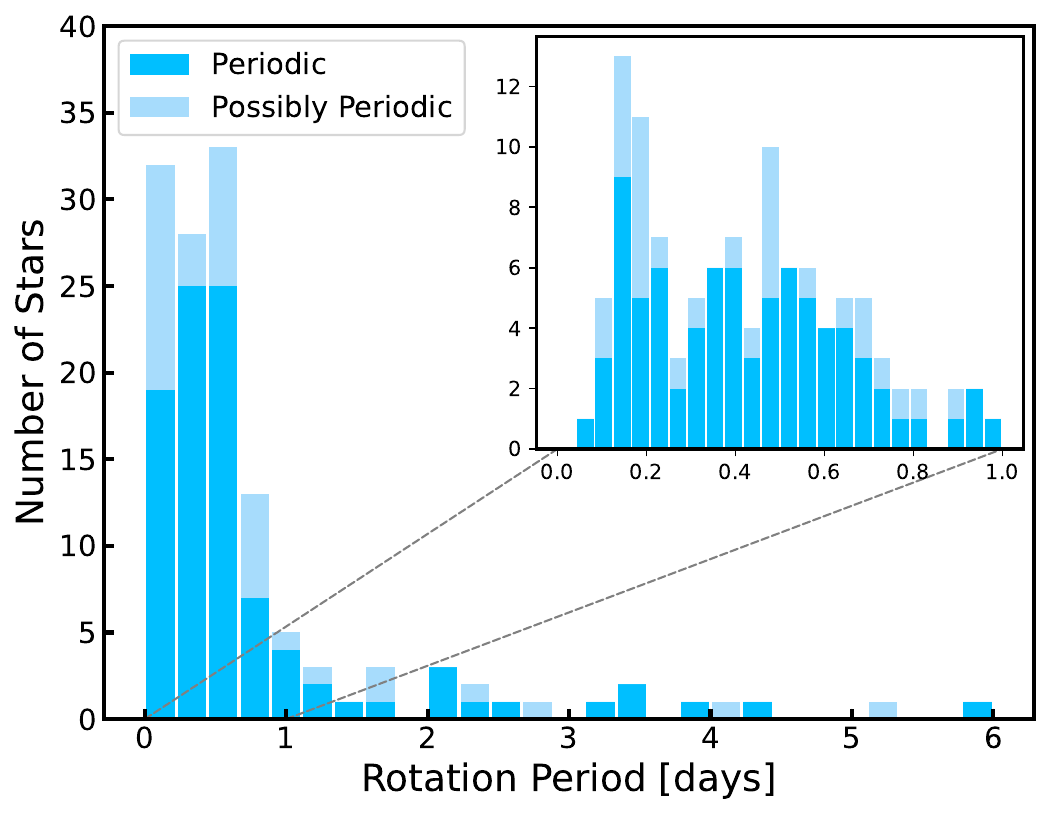}
	\caption{A histogram of rotation periods in our sample. An inset is included to see the distribution of periods under 1 day, which comprises the majority of the rotation periods identified.}
	\label{fig:periodhistinset}
\end{figure}

The period distribution can be seen in Figure \ref{fig:periodhistinset}. Most targets have periods under 24 hours (110 stars; 83\%). This is primarily due to bias in the period search process: an effect of systematics and peaks in the window function dominating at longer times as discussed in Section \ref{sec:per_det}, as well as our use of a 2-day window for detrending in two of four reduction methods. 

\begin{figure*}
	\centering
        \includegraphics[width=0.46\linewidth]{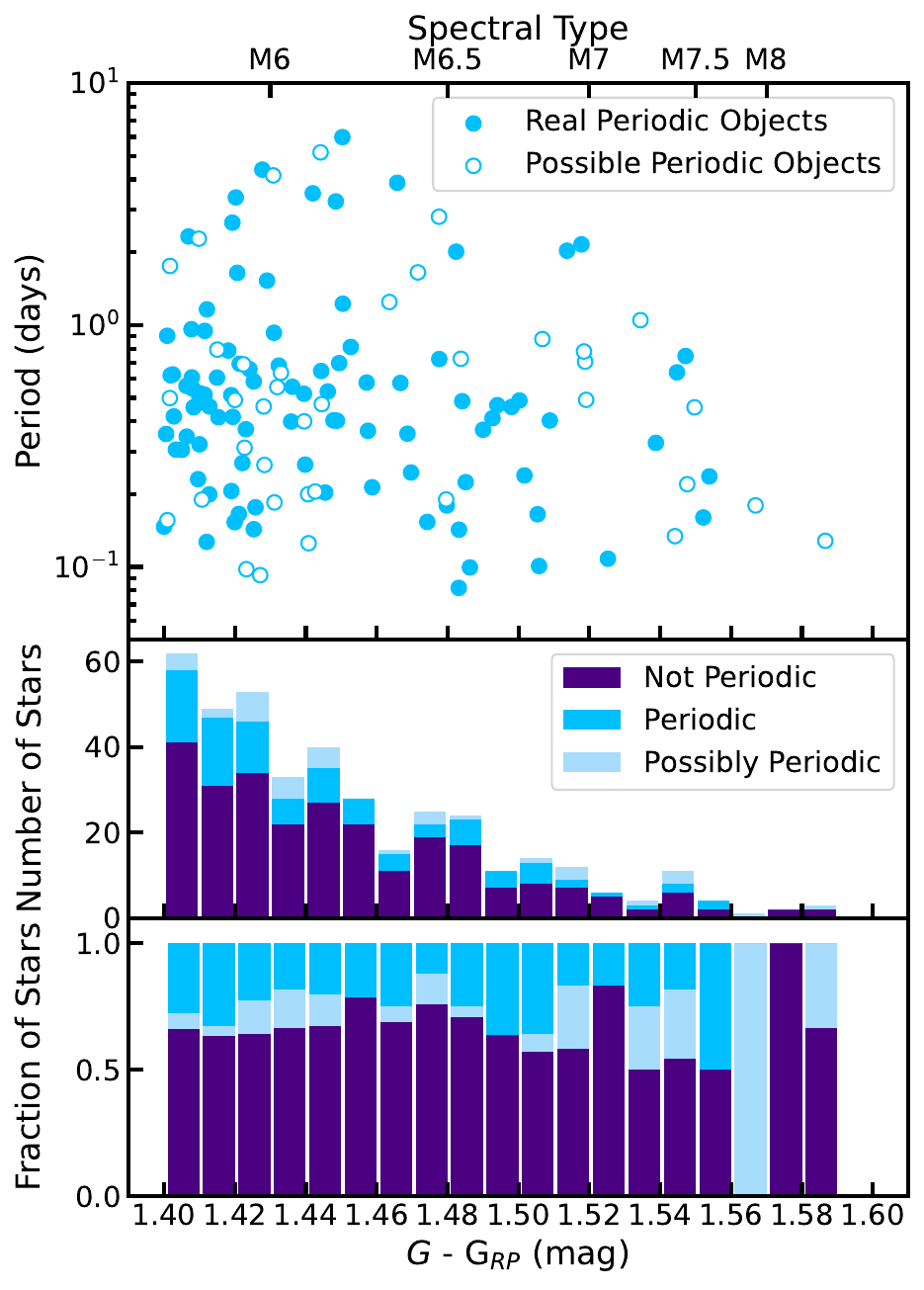}
        \includegraphics[width=0.46\linewidth]{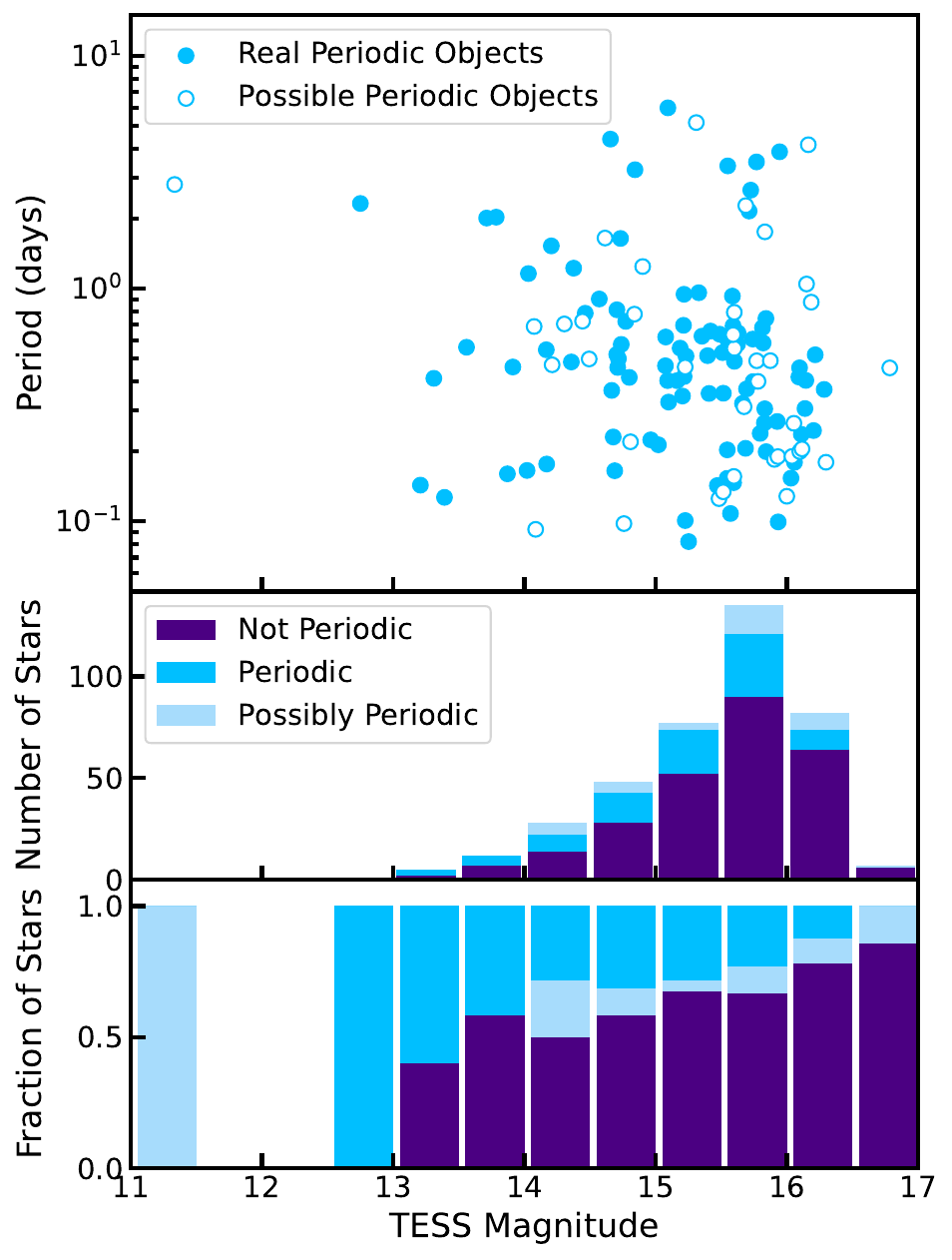}
	\caption{\textit{Top panels:} Rotation period vs.\ Gaia $G - G_{RP}$ color (and corresponding spectral type from \citet{pecaut2013}; left) and rotation period vs.\ TESS magnitude (right) of the targets in the sample. \textit{Middle panels:} Histograms showing the distribution of periodic objects in the sample. \textit{Bottom panels:} The fraction of stars in each bin that are not periodic, possibly periodic, or periodic. The left panel shows no correlation between period detections and $G-G_{RP}$ color, whereas there is a clear correlation between TESS magnitude and period detection in the right panel. }
	\label{fig:statplots}
\end{figure*}

\begin{figure}
	\centering
	\includegraphics[width=0.9\linewidth]{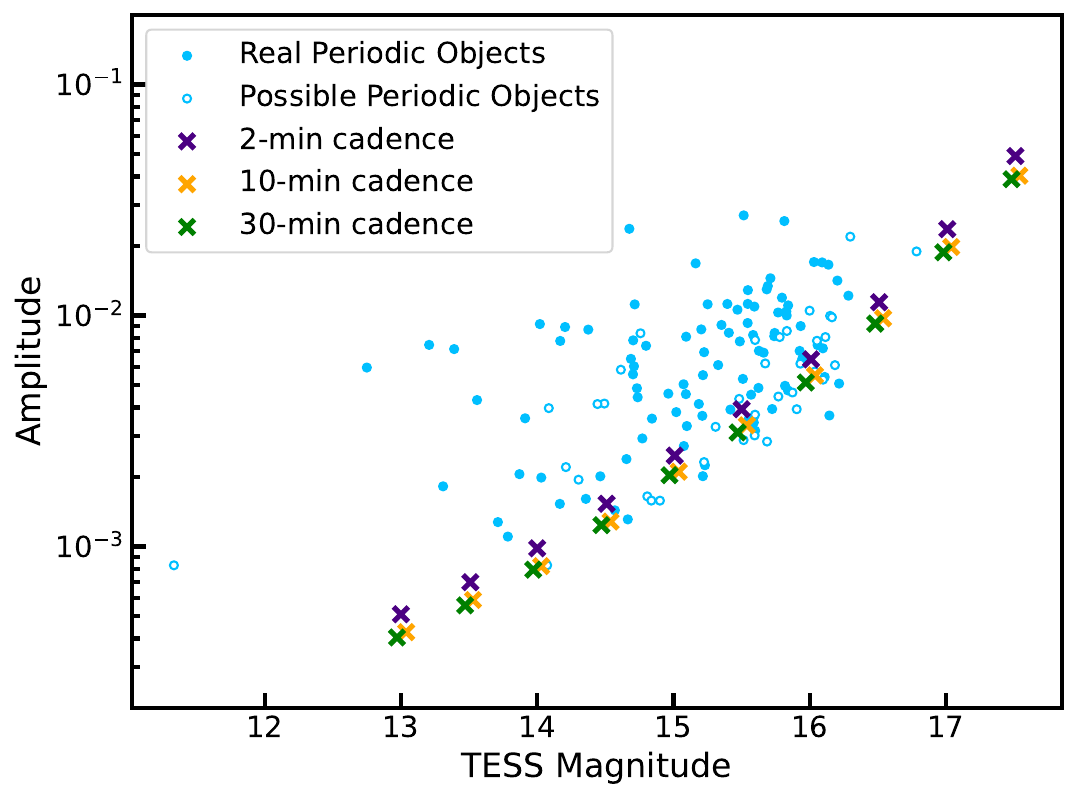}
	\caption{Amplitude vs.\ TESS magnitude of the sinusoidal variation of the folded light curve for the periodic (filled blue circles) and possibly periodic (empty blue circles) targets in this sample. Also plotted with crosses are the 2-min, 10-min, and 30-min cadence amplitude sensitivity limits to $<$ 1 day periodic variations of TESS as determined by \citet{miles2023}.}
	\label{fig:amptess}
\end{figure}

The distribution of period detections across $G-G_{RP}$ color and TESS magnitude can be seen in the top panels of Figure \ref{fig:statplots}. As seen in the bottom left panel of Figure \ref{fig:statplots}, there is little to no correlation between $G-G_{RP}$ color and period detection. The middle panel shows that the apparent overdensity of periodic objects at spectral types brighter than $\sim$M6.5 is due entirely to the underlying population, as most targets fall in this spectral type range. In contrast, the bottom right panel shows there is a correlation between TESS magnitude and detection. Such a brightness dependence is expected as it follows the decrease in signal-to-noise toward fainter magnitudes. This is in line with results from \citet{miles2023}, who calculated the TESS sensitivity for recovering periods and amplitudes of ultracool dwarfs in a single sector (with either 2-, 10-, or 30-minute cadence) using a 1-day median filter and a LS periodogram (Figure \ref{fig:amptess}).

We do not find any correlation between period length and color, as seen in the top panel of Figure \ref{fig:statplots}. However, the range of colors studied in this work is very narrow, and our bias to shorter ($\lesssim 1$ day) periods limits the range of periods 
 considered. It must be placed in the broader literature to get a full understanding (see Section \ref{sec:literature}). We also find no correlation between TESS magnitude and period duration (top panel of Figure \ref{fig:statplots}), indicating no brightness bias in detecting periods of specific durations, other than the general bias toward finding periods under one day.

\begin{figure*}
	\centering
	\includegraphics[width=0.48\linewidth]{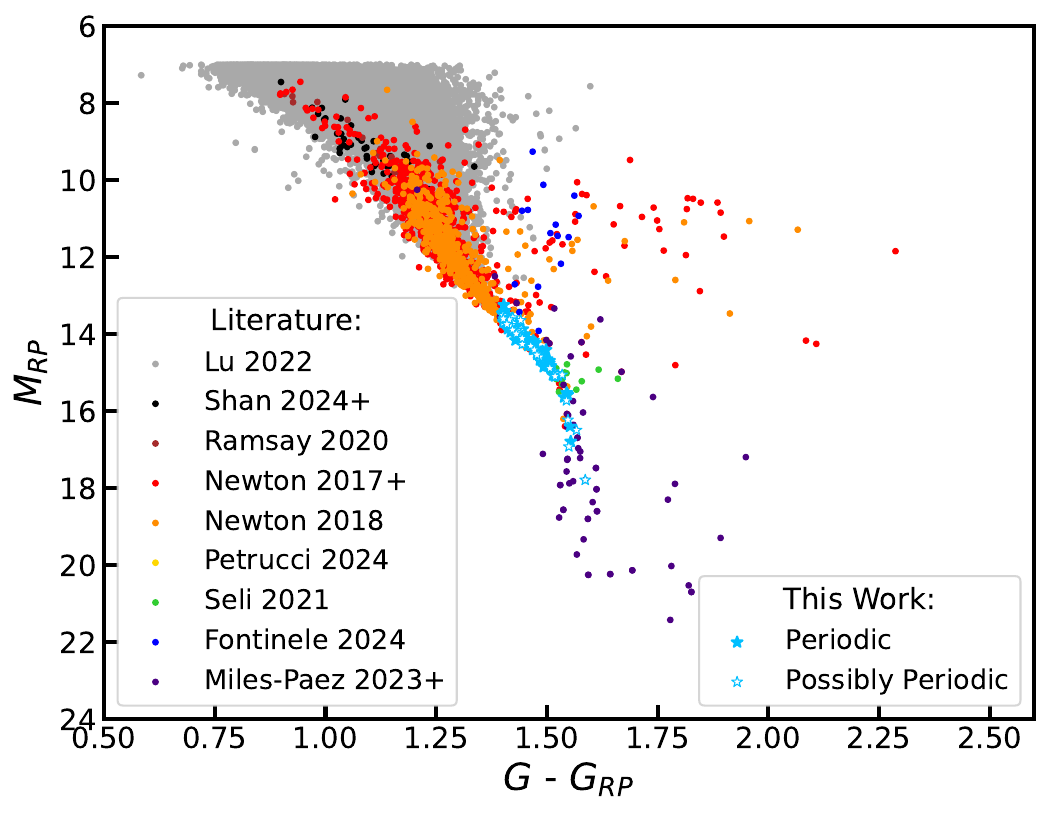}
        \includegraphics[width=0.48\linewidth]{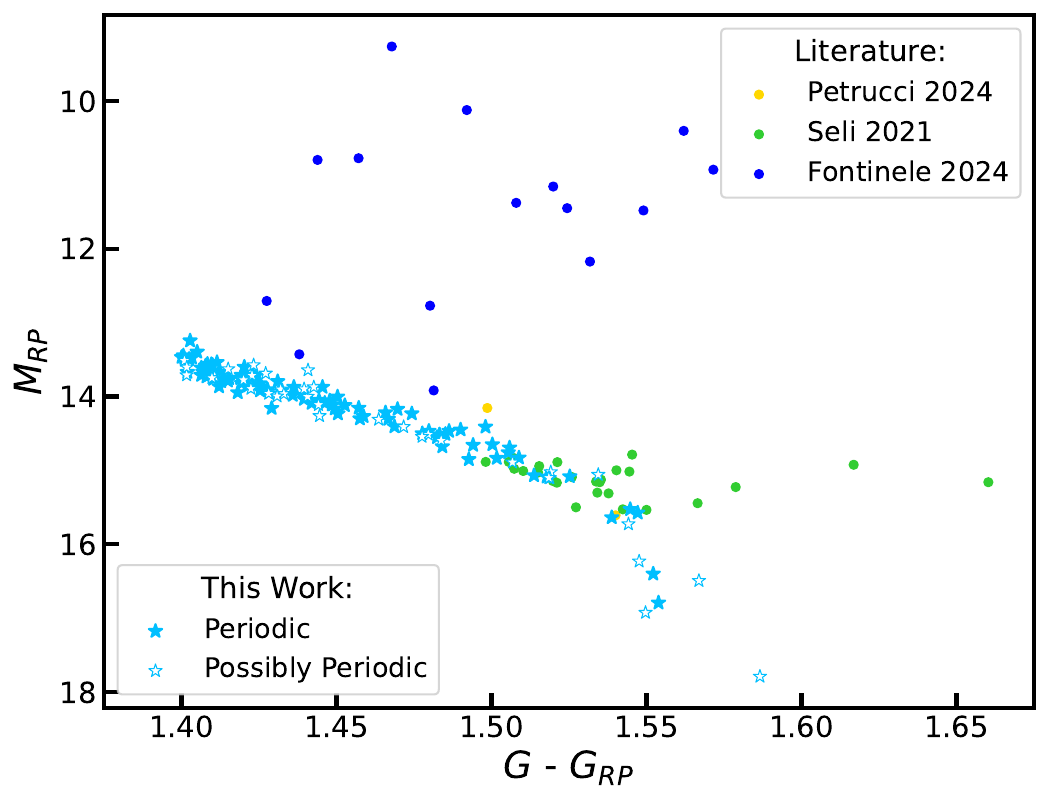}
	\caption{Color-magnitude diagrams showing all the rotation periods in the literature discussed in this work (left) and the rotation periods of the three studies specifically targeting late-M dwarfs in TESS (S21, P24, F24; right). Note that reddening and extinction are disregarded, as L22 is the only study with stars further than 150 pc. References in the legend with ``+'' indicate periods from multiple sources which were compiled by the named paper.}
	\label{fig:litcmd}
\end{figure*}

\begin{figure}
	\centering
	\includegraphics[width=0.85\linewidth]{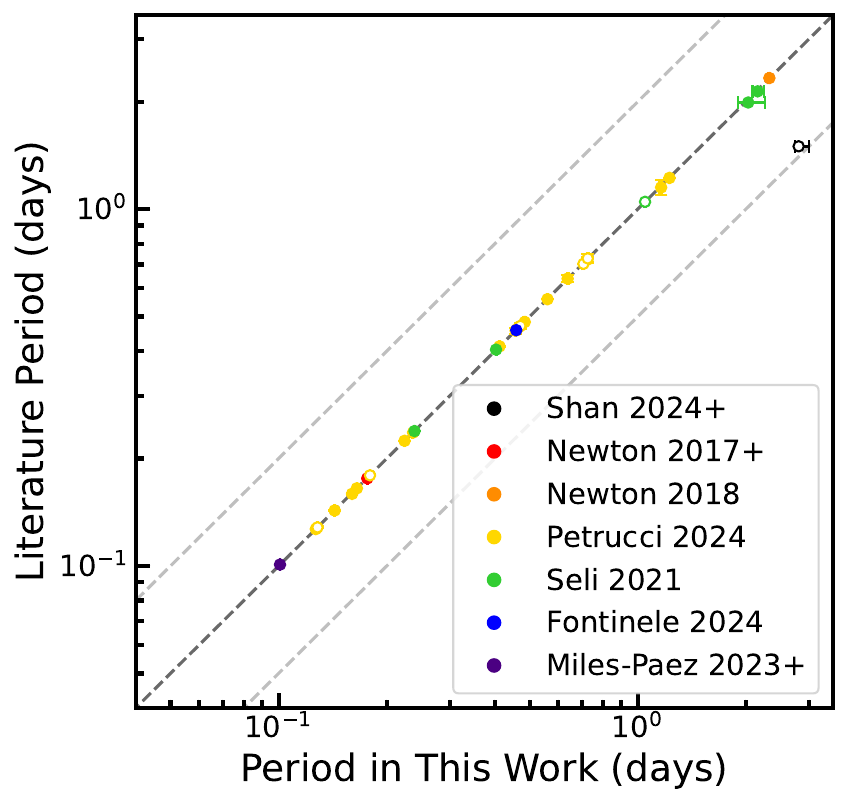}
	\caption{Comparison of periods determined in this study to previously published ones, where those exist in the literature (listed in Table~\ref{tab:litmatch}). Where multiple published periods exist for a star, we have adopted the most significant one. The gray dashed lines indicate the 1:1, 1:2, and 2:1 lines. Open circles indicate possible periods in this work, whereas closed circles are those which we consider real. Uncertainties in this work and in the literature (when available) are typically smaller than the data point. References in the legend with ``+'' indicate periods from multiple sources which were compiled by the named paper.}
	\label{fig:perlit}
\end{figure}

\subsection{Variability Amplitudes: 3\% Upper Limit}

For the most part, our results follow the amplitude detectability limits from \citet[][Figure \ref{fig:amptess}]{miles2023}. However, we do have 21 periodic stars below these limits: nine with ``real'' periodicities and 12 with ``possible'' periodicities. This is not unexpected, as our analysis considers multiple sectors and different reduction methods that may improve our ability to detect smaller amplitudes compared to \citet{miles2023}. 

We also find an upper limit to the observed variability amplitudes in TESS. The maximum detected periodic amplitudes for our $\lesssim$6 day periods are around 3\%, despite our being easily sensitive to greater variations. This is in agreement with findings from other studies of late-M dwarfs \citep{seli2021,petrucci2024}. 
%and follows a trend among cooler L and T dwarfs, where maximum variability amplitudes increase with decreasing temperature \citep{metchev2015}. 
The 3\% maximum variability amplitude among late-M dwarfs points to fairly uniform spot coverage and the lack of large discrete temperature abnormalities on the surface, as significant asymmetries would produce larger amplitudes. It also explains why we do not find any periodic variables at TESS magnitudes $>$17 since, in those cases, we would be sensitive only to greater amplitudes.

\subsection{Fraction of Variability per Number of Sectors Analyzed}
\label{sec:varfrac}
In Figure \ref{fig:hist_sectors}, we see that the fraction of periodic variables that we find increases with the number of available TESS sectors per star. As plotted with a red dotted line, the fraction of non-periodic stars per sector decreases with a slope of 0.03$\pm$0.01, with a statistically significant $p$-value of 0.028. This indicates that, as one might expect, we are more complete when there is more data available. Looking solely at the targets with 7 or more sectors of data, 16 of 40 (40.0$\pm$7.7\%) stars are variable. 
%This fraction increases to 47\% when looking at targets with 10 or more sectors, however, this subset contains only 17 stars (8 periodic).

The variability fraction can be compared to the sensitive brown dwarf variability study with the Spitzer Space Telescope of \citet{metchev2015}. They detect variability in 19 of 39 (49\%) single L3-T8 dwarfs with $\geq$0.2\% amplitudes in $\ sim$21-hour observing blocks. Given the comparable sensitivity and variability fraction, it may be that an empirical ceiling of $\sim$50\% observable variability exists for amplitudes above 0.1–0.2\% in late-M dwarfs. This would be set primarily by viewing geometry, as determined by the incompleteness analysis in \citet{metchev2015}.
Therefore, like with L and T dwarfs, spots may be ubiquitous in late-M dwarfs. 

\begin{longrotatetable}
\begin{deluxetable*}{cccccccccc}
        \tablecaption{Variability literature relevant to this study and discussed in Section \ref{sec:literature}. \label{tab:literature}}
\tabletypesize{\footnotesize}
\tablewidth{0pt}
\setlength{\tabcolsep}{2pt}
\tablehead{
\colhead{Survey} &  
\colhead{Additional reference} & 
\colhead{Primary} & 
\colhead{Primary Method(s)} & 
\colhead{SpT} & 
\colhead{Gaia $G-G_{RP}$} & 
\colhead{Mean $G-G_{RP}$} & 
\colhead{\# of Stars} & 
\colhead{\# of Rotation} & 
\colhead{Overlap with} \\
\colhead{} & 
\colhead{sources in compilations} & 
\colhead{Instrument(s)} & 
\colhead{for Determining Periodicity} & 
\colhead{Range} & 
\colhead{Range (mag)} & 
\colhead{(mag)} & 
\colhead{} & 
\colhead{Periods} & 
\colhead{This Work}
}
\startdata
 N17 & (1-22) & MEarth-North & MLM & $\sim$M3-M6 & 0.89-2.29 & 1.28 & 2202 & 530 & 2 \\
 N18 & & MEarth-South & MLM & $\sim$M3-M6 & 1.06-2.07 & 1.30 & 574 & 234 & 4 \\
 R20 & & TESS & LS & K9-M6 & 0.91-1.43 & 1.23 & 13,836 & 472 & 1 \\
 S21 &  & TESS & LS & $\sim$M7-M9 & 1.49-1.66 & 1.53 & 248 & 42 & 9 \\
 L22 & & ZTF & LS & Late K-M & 0.58-1.60 & 1.05 & $\sim7\times10^6$ & 40,553 & 0\\
 M23 & N17, (23-27, 28$^*$, 29$^*$) & TESS, 1-m class ground & LS + GP & M7-Y0 & & & 131 & 131 & 9 \\
 S24 & (1), (19), (30-54)  & TESS, ground-based & GLS & K7-M9 & 0.87-1.66 & 1.17 & 348 & 348 & 3 \\
 P24 & & TESS & LS + ACF & M4-L4 & 1.38-1.57 & 1.48 & 208 & 87 & 32 \\
 F24 & & TESS & FFT + LS + wavelet maps & M5-L3 & 1.32-1.75 & 1.50 & 250 & 71 & 1 \\
 \enddata
 \tablecomments{No $G-G_{RP}$ information is provided for M23 as many of the targets in the catalogue are too faint for Gaia photometry. $^*$brown dwarf rotation periods published after M23,  MLM = maximum likelihood method, LS = Lomb-Scargle periodogram, GP = Gaussian process, GLS = generalized Lomb-Scargle periodogram, ACF = autocorrelation function, FFT = fast Fourier transform. \\ \textbf{References:} N17 \citet{newton2017}, N18 \citet{newton2018}, R20 \citet{ramsay2020}, S21 \citet{seli2021}, M23 \citet{miles2023}, S24 \citet{shan2024}, P24 \citet{petrucci2024}, F24 \citet{fontinele2024}, (1) \citet{newton2016}, (2) \citet{hartman2011}, (3) \citet{krzeminski1969}, (4) \citet{pettersen1980a}, (5) \citet{pettersen1980b}, (6) \citet{busko1980}, (7) \citet{baliunas1983}, (8) \citet{benedict1998}, (9) \citet{alekseev1998b}, (10) \citet{alekseev1998y}, (11) \citet{robb1999}, (12) \citet{fekel2000}, (13) \citet{norton2007}, (14) \citet{kiraga2007}, (15) \citet{engle2009}, (16) \citet{shkolnik2010}, (17) \citet{hartman2010}, (18) \citet{messina2011}, (19) \citet{kiraga2012}, (20) \citet{mamajek2013}, (21) \citet{kiraga2013}, (22) \citet{mcquillan2013}, (23) \citet{miles2017a}, (24) \citet{miles2017b}, (25) \citet{tannock2021}, (26) \citet{miles2021}, (27) \citet{vos2022},  (28) \citet{rose2023}, (29) \citet{miles2025}, (30) \citet{andersson2022}, (31) \citet{amado2021}, (32) \citet{baroch2018}, (33) \citet{bluhm2020}, (34) \citet{caballero2022}, (35) \citet{alonso2019}, (36) \citet{dedrick2021}, (37) \citet{donati2023}, (38) \citet{dreizler2020}, (39) \citet{fouque2023}, (40) \citet{gorrini2023}, (41) \citet{irving2023}, (42) \citet{lafarga2021}, (43) \citet{luque2018}, (44) \citet{luque2022}, (45) \citet{medina2022}, (46) \citet{morin2008}, (47) \citet{morin2010}, (48) \citet{pass2023}, (49) \citet{raetz2020}, (50) \citet{suarez2015}, (51) \citet{suarez2017}, (52) \citet{suarez2018}, (53) \citet{stelzer2016}, (54) \citet{stock2020}}
\end{deluxetable*}

\end{longrotatetable}

\subsection{Comparison to Literature Periods}
\label{sec:literature}

We compare our results to a number of rotation period surveys in the literature that overlap with the color range of our target list. By comparing with these surveys, we can put our data into a broader context, from early M dwarfs to Y dwarfs. When multiple periods existed in the literature for a star, the period with the smallest uncertainty was used. Color-magnitude diagrams of the stars from the literature with rotation periods can be seen in Figure \ref{fig:litcmd}.

\begin{deluxetable*}{cccccc}
    \tablecaption{A comparison of rotation periods in this study and the literature (visualized in Figure \ref{fig:perlit}). \label{tab:litmatch}}
    \tablehead{\colhead{Gaia ID} & \colhead{RA} & \colhead{Dec} & \colhead{$P_{rot}$} & \colhead{Lit. $P_{rot}$} & \colhead{Lit.}\\
    \colhead{} & \colhead{(deg, J2016.5 epoch)} & \colhead{(deg, J2016.5 epoch)} & \colhead{(days)} & \colhead{(days)} & \colhead{source} \\}
\startdata
35227046884571776 & 43.2696424767906 & 16.8643738189774 &  & 97.56 & S24 \\ 
260343123440688128 & 72.2691005610588 & 51.6430882696422 &  & 0.7278$\pm$0.02108 & P24 \\ 
763973668623022464 & 170.679630526493 & 37.9301592669159 & 0.5605$^{+0.0025}_{-0.0029}$ & 0.5593$\pm$0.00041 & P24; N17 \\ 
847228998317017472 & 155.361347480149 & 50.9180885259734 & 1.0479$^{+0.0071}_{-0.0053}$ & 1.04848$\pm$0.00137 & S21 \\ 
1040681747033185152 & 130.224002332549 & 59.183785876575 & 0.2390$^{+0.0012}_{-0.0008}$ & 0.23882$\pm$0.00004 & S21 \\ 
1047188004010109440 & 155.694043791076 & 58.4260012180831 & 0.4570$^{+0.0021}_{-0.0019}$ & 0.4565$\pm$0.00848 & P24 \\ 
1113083282052337792 & 101.124561630198 & 72.1132592591049 & 0.4589$^{+0.0013}_{-0.0011}$ & 0.458 & F24 \\ 
1287312100751643776 & 217.178272272411 & 33.1744121056914 &  & 0.075$\pm$0.008 & MP23 \\ 
1454104436971779328 & 210.838758570404 & 30.1320167985817 & 0.6379$^{+0.0048}_{-0.0022}$ & 0.64009$\pm$0.00092 & S21; MP23; P24 \\ 
1552186307304378880 & 204.211044886255 & 47.8588993954501 & 0.10091$^{+0.00017}_{-0.00013}$ & 0.1009208$\pm$0.0000125 & MP23 \\ 
1604901876901703168 & 215.763328045861 & 51.7760046286171 & 0.4845$^{+0.0027}_{-0.0039}$ & 0.4828$\pm$0.00107 & P24 \\ 
1618010323247026560 & 217.650335793804 & 59.7242739296249 & 0.4122$^{+0.0017}_{-0.0017}$ & 0.4125$\pm$0.0007 & P24 \\ 
2349207644734247808 & 12.8426331466573 & -22.8595489615153 & 2.156$^{+0.089}_{-0.069}$ & 2.13867$\pm$0.00221 & S21 \\ 
2472387757755767168 & 20.204863249306 & -7.68489101058526 &  & 0.40031$\pm$0.00033 & S21 \\ 
2492866780297999232 & 33.554613532264 & -3.96280045849318 & 2.325$^{+0.013}_{-0.021}$ & 2.33 & N18 \\
260343123440688128 & 72.2691005610588 & 51.6430882696422 & 0.7256$^{+0.047}_{-0.033}$ & 0.72780$\pm$0.02108 & P24 \\
2631857350835259392 & 348.978521379539 & -6.46309671584034 & 0.12695$^{+0.00015}_{-0.00012}$ & 0.1269$\pm$0.00072 & P24 \\
2635476908753563008 & 346.626521627644 & -5.04352831942953 &  & 3.3 & S24 \\ 
2781513733917711616 & 11.3408953373176 & 16.5788804107873 &  & 0.1$\pm$0.004 & MP23 \\ 
2794735086363871360 & 9.07135464103575 & 18.3534101381391 & 0.12831$^{+0.00019}_{-0.00015}$ & 0.1283$\pm$0.0007 & P24; MP23 \\ 
3094447525008511488 & 121.487181241157 & 4.28432563249044 & 0.17654$^{+0.00038}_{-0.00043}$ & 0.176 & N17 \\ 
3257243312560240000 & 57.7501633018874 & -0.88123385827372 &  & 0.7289$\pm$0.0204 & P24 \\ 
3582675080520660992 & 185.965946799238 & -8.97997617559877 & 0.4712$^{+0.0017}_{-0.0021}$ & 0.4708$\pm$0.00898 & P24 \\ 
3701479918946381184 & 185.364824611618 & 2.95529877574394 & 0.17967$^{+0.00029}_{-0.00035}$ & 0.1795$\pm$0.00127 & P24 \\ 
3757613049856225792 & 162.055183999012 & -11.3428033197352 & 2.81$^{+0.20}_{-0.07}$ & 1.5 & S24 \\ 
3808159454810609280 & 162.176439325156 & 1.19847720790681 & 0.2368$^{+0.0005}_{-0.0014}$ & 0.2365$\pm$0.00107 & P24; MP23 \\ 
3873864513044635648 & 154.36091675488 & 7.32343468087821 & 0.4667$^{+0.0018}_{-0.0023}$ & 0.4672$\pm$0.00421 & P24 \\ 
3901165013100805248 & 187.436812761861 & 7.87744715487489 & 1.2256$^{+0.0045}_{-0.0064}$ & 1.2234$\pm$0.00195 & P24 \\ 
4752399493622045696 & 43.7719859347708 & -47.0167283562266 & & 0.308 & MP23 \\ 4860376345833699840 & 54.8985702067773 & -35.4275936398811 & 0.16019$^{+0.00027}_{-0.00040}$ & 0.1593$\pm$0.00098 & P24; MP23 \\ 
4871414343064541824 & 69.8913990777374 & -32.5976145049372 & 0.4610$^{+0.0013}_{-0.0009}$ & 0.4608$\pm$0.0038 & P24 \\ 
4923251780829182080 & 6.85074449089365 & -54.0292900479123 & 0.22422$^{+0.00070}_{-0.00046}$ & 0.2243$\pm$7e-05 & P24 \\ 
4929042942932181888 & 17.9460471480137 & -49.137801730274 &  & 2.4843$\pm$0.00802 & P24 \\ 
4967628688601251200 & 31.8108539638579 & -37.3634313226862 & 0.4030$^{+0.0019}_{-0.0010}$ & 0.40391$\pm$0.0002 & S21 \\ 
4971892010576979840 & 33.7874820098168 & -30.6686560652885 & 0.7058$^{+0.0039}_{-0.0082}$ & 0.7027$\pm$0.00064 & P24; S21 \\ 
4989399774745144448 & 15.7209623443364 & -37.6277093582328 & 2.03$^{+0.23}_{-0.13}$ & 1.99002$\pm$0.00607 & S21 \\ 
5072067381112863104 & 46.5074324064378 & -26.7967864956461 & & 11.067 & N18 \\
5461814875584141056 & 153.026285311875 & -30.8238199449872 & 0.7240$^{+0.0033}_{-0.0030}$ & 0.7221$\pm$0.01038 & P24 \\ 
5744451451968477952 & 140.570478647617 & -8.87001755434351 & 1.162$^{+0.006}_{-0.014}$ & 1.1504$\pm$0.05519 & P24 \\ 
6227871564690336512 & 226.065792785741 & -23.9327739753384 & 0.16518$^{+0.00026}_{-0.00025}$ & 0.165058$\pm$0.0000375 & MP23; P24 \\ 
6405457982659103872 & 333.46143166946 & -63.7020231646437 & 0.14327$^{+0.00031}_{-0.00014}$ & 0.14310$\pm$0.00003 & P24, R20 \\
6494861747014476288 & 358.844373543162 & -58.0011909780195 & & 166.165 & N18 \\
\enddata
\tablecomments{When multiple literature sources are present, the one listed first is the one used.}
\end{deluxetable*}

In this section, we detail the studies used, followed by a discussion of rotation period statistics, given the results in this work in the context of previous studies. Table \ref{tab:literature} lists all of the studies considered, and Table \ref{tab:litmatch} shows all of the periods that overlap with this work. Additionally, Figure \ref{fig:perlit} provides a visualization of how the periods in this work compare to those in the literature. We see complete agreement for all TESS-based periods and only one studied using a different instrument, which does not agree, but is a harmonic of the period we determine. What is not displayed are the targets for which either this work or the other study produced a period not found in the other. These stars are discussed in the following subsections.

\subsubsection{Newton et al.\ (2017, 2018) and References Therein}

\citet{newton2017} and \citet{newton2018}, N17 and N18 hereafter, studied the rotation period and activity of single mid-to-late M-dwarfs within 33 pc in the Northern and Southern hemispheres, respectively, as part of the MEarth Project \citep{irwin2008}.  N18 presents entirely new rotation periods, whereas N17 took 90\% of their data from \citet{newton2016}, supplemented with other works as seen in Table 2 of their paper. As these are ground-based surveys with multi-year long baselines, they could sample much longer periods than TESS, up to approximately 200 days.

As these studies focus on stars with stellar masses between approximately 0.10-0.33M$_{\odot}$, there is some overlap between these papers and our work. Of the 2202 stars in N17, 30 are in this work, but only two have rotation periods from N17. One agrees with our period, but the other does not. The discrepant star, Gaia DR3 763973668623022464, has also been studied in \citet[][see Section \ref{sec:P24}]{petrucci2024}, which agrees with our period of 0.5605$^{+0.0025}_{-0.0029}$ days as opposed to N17's 0.358-day period. As such, we trust the period we have determined.

N18 contained 574 stars, of which 12 matched our sample. Four of these targets had rotation periods from N18, though only two matched our results. Our analysis determined that the other two were undetectable. Gaia DR3 5072067381112863104 has a period of 11.067 days from N18, which is at the long end of what is detectable in our TESS analysis and lies near a strong peak in the window function of the data. Gaia DR3 6494861747014476288 has a reported period of 166.165 days in N18, far outside the capability of our TESS analysis to detect.

\subsubsection{Ramsay et.\ al. (2020)}

\citet[][R20 hereafter]{ramsay2020} %, R20 hereafter, 
is a study which uses PDCSAP data from TESS Sectors 1-13 to study low-mass stars ($\gtrsim$M0) selected from Gaia DR2, looking for ultrafast rotators ($P_{rot} < 0.3$ days) with an upper limit set at 1 day. However, R20 differs from our study in that it does not consider contamination from nearby stars as a potential source of periodicity and overall has looser constraints on period selection. We find that, as a result, some of the reported periods may not be accurate. For example, the fastest rotator reported in R20, Gaia DR3 5643174851828608128, with a period of 0.0603 days, is less than one arcminute ($<$3 TESS pixels) away from OGLE GD2329.12.00002, a known delta scuti star with the same period \citep{soszynski2021}.  As such, we only consider targets from R20 which pass our crowding check as detailed in Section \ref{sec:crowded}.

As R20 only studied targets brighter than $T$ = 14 mag, there is very little overlap with our sample. There is one target in common (Gaia DR3 6405457982659103872), for which R20 report the same period as found in this work.

\subsubsection{Seli et. al., (2021)}
\label{sec:S21}

\citet[][S21 hereafter]{seli2021} %, S21 hereafter, 
studied the activity (flares and rotational modulation) of TRAPPIST-1 analogue stars using the 30-minute cadence FFI data from Sectors 1-26 of TESS. S21 studied 248 stars within 0.5 magnitudes of TRAPPIST-1 on a Gaia DR2 $M_G$ vs. $G_{BP} - G_{RP}$ color-magnitude diagram. Of these, they found 42 rotation periods. As a TESS-based study, contamination could present an issue. S21 carefully considered contamination and used nearby stars as references in the processing of the light curves.

When the stars with rotation periods were matched with our work, nine were in our sample, for eight of which we confirmed the periods. The remaining 33 stars were either in crowded fields or were above our binary cutoff. There is one periodic variable from S21 that we can not confirm. Gaia DR3 2472387757755767168 has a reported period of 0.40031$\pm$0.00033 days, but this period does not have an amplitude SNR $>$ 5 in any of our methods.

\subsubsection{Lu et. al., (2022)}

\citet[][L22 hereafter]{lu2022} %, L22 hereafter, 
is the largest study of stellar rotation periods across the fully convective boundary. They used the Zwicky Transient Facility (ZTF) to determine the rotation period of 40,553 late-K to M dwarfs with periods $>$1 day. Their brightness limitation with ZTF was $G < 18$ mag; however, the faintest star with a detectable period in L22 has $G \sim 14$ mag. This causes a rapid drop-off in stars around M5. This highlights the importance of studies such as ours to expand our knowledge of rotation periods into the late-M dwarf regime. The objects with rotation periods in L22 do not overlap the 399 stars we consider but include four stars that did not pass our binary cutoff. 

\subsubsection{Miles-P\'{a}ez et. al. (2023) and References Therein}

\citet[][MP23 hereafter]{miles2023} is a study of rapidly rotating field ultracool dwarfs using a combination of TESS and 1-m class ground-based observations. MP23 published new periods for six stars and refined the periods of 11 known periodic stars using 2-minute PDCSAP TESS data or 30-minute TESS data reduced with \texttt{eleanor}. They also compiled a list of all 128 M7-Y0 stars with known rotation periods at the time (see Table \ref{tab:literature}).
%from Newton et al. (2017); Miles-Páez et al. (2017a,b); Tannock et al. (2021) and the new detections reported in MilesPáez (2021); Vos et al. (2022); Andersson et al. (2022) \textbf{properly cite}. 
We include this compilation in our plots simply as Miles-Paez 2023+. 

Of the 128 M7-Y0 dwarfs, nine were in this work. The rest were too faint, in crowded fields, or considered as potential binaries in this study. All TESS periods matched, though there were 4 from other non-TESS surveys that were either absent (3) or different (1) from our compilation. The discrepant target, Gaia DR3 3808159454810609280 (LSPM J1048+0111, L1), has a literature period of 4.71 hours (0.196 days), similar to the 0.2368$^{+0.0005}_{-0.0014}$ day period found in this work. This original period is from \citet{koen2003} and was determined from three nights of data with approximately 4.1 hours of observation each. Therefore, the new period from TESS is a revision of the previous one. For the three periods absent from this work (Gaia DR3 1287312100751643776, Gaia DR3 2781513733917711616 and Gaia DR3 4752399493622045696; see Table \ref{tab:litmatch}), there is no evidence of periodicity in TESS despite all having multiple sectors of data. In the case of Gaia DR3 4752399493622045696 (DENIS J025503.3-470049, L9), the lack of periodicity in TESS is likely due to the faintness of the object: it is the faintest in our sample ($T = 17.03$ mag). The other two targets are not as faint, but their reported periods come from $I$-band \citep[Gaia DR3 1287312100751643776, LHS 2924, M9;][]{martin1996} and Spitzer \citep[Gaia DR3 2781513733917711616, L2$\beta$;][]{vos2020} photometry, so there may be a wavelength dependence to their variability. As such, we cannot confirm the periods.

\subsubsection{Shan et. al., (2024) and References Therein}

\citet[][S24 hereafter]{shan2024} is a compilation of all rotation periods used for the CARMENES input sample, including new results from TESS (PDCSAP, SAP, and FFI) and ground-based observations (LCOGT, SuperWASP,  AstroLAB IRIS, Observatorio de Sierra Nevada [OSN], and Montsec Observatory). The entire list of references can be found in Table C.2 of S24. The CARMENES sample is designed to search for exoplanets (using the radial velocity method) around M-dwarfs and, therefore, contains M0-M9 stars. Only three of the 348 periodic variables in S24 are in our sample. The lack of overlap derives from the fact that the CARMENES sample is primarily dominated by early M-dwarfs. Of these, one (Gaia DR3 3757613049856225792, LP 731-58, M6.5, see Figure \ref{fig:perlit}) is a 1/2 harmonic of the period in our study (1.5 days from S24 vs. 2.81$^{+0.20}_{-0.07}$ days from this work), of which we trust the period we determined to be the true rotation period. Another is longer than the TESS baseline (Teegarden's Star; 97.56 days) and, therefore, could not be probed by this study. The final period reported is 3.3 days, which we do not see any evidence of in our data. However, this target, Gaia DR3 2635476908753563008, has only been observed in TESS Sector 70 and, therefore, would require more observations to confirm or reject the period.

\subsubsection{Petrucci et. al., (2024)}
\label{sec:P24}

\citet[][P24 hereafter]{petrucci2024} is a study that used 2-minute or 20-second TESS PDCSAP data to study the rotation periods and flare activity of ultracool dwarfs within 40 pc from SPECULOOS \citep{sebastian2021}. The spectral type range of this survey was M4-L4, and the $G - G_{RP}$ color range of periodic targets was 1.39-1.57 mag. 

There are 32 objects in P24 that overlap with the 399 targets studied in this work. The remaining 176 objects studied in P24 were omitted in this work, mostly (173 of the 176) due to being in crowded fields by our definition (see Section \ref{sec:crowded}). These stars are removed from our literature comparison in the same manner as potentially contaminated stars in R20. 
% These omitted stars are listed in Table \ref{tab:periods}.
Three stars were still in P24 but not in our work: one (Gaia DR3 5471345889049823744) because it does not have a Gaia parallax and is therefore not in the GCNS from which our targets were selected, and the other two (Gaia DR3 3562157781229213312 and Gaia DR3 1502523188143833088) because they have RUWE values $>$1.4, indicating likely unresolved binarity.

Of the overlapping 32 objects, 24 agree with this work (21 with matching periods, and three determined to be non-periodic in both surveys), six have periods in this work but not in P24, and two have a period in P24 but not in this work. Both of the periods missing in this work (see Table \ref{tab:litmatch}) are eliminated in our TL verification step as being likely contaminated. As such, we do not believe that these periods are accurate.

\subsubsection{Fontinele et. al., (2024)}

\citet{fontinele2024}, F24 hereafter, also study the rotation periods of ultracool dwarfs using TESS PDCSAP data. F24 specifically uses the 2-minute cadence data. The input catalogue for F24 is the objects of Table 4 of \citet{sarro2023}. These objects are all members of BANYAN $\Sigma$ \citep{gagne2018b} groups or hierarchical mode association clustering \citep{li2007} clusters within 150 pc of the Sun. As such, this sample is skewed towards younger objects, which are most likely excluded from our sample due to their position on a color-magnitude diagram falling in the binary main sequence or brighter. F24 found periods for 71 (17 ambiguous and 54 unambiguous) of the 250 stars that they studied. Only one target (Gaia DR3 1113083282052337792) overlapped with this work and was consistent with our results. As with R20 and P24, this study does not consider contamination from nearby sources, and as such, those which fail our crowding check are removed from the analysis.

\subsection{Period-Spectral Type Relation: a Lower Envelope to Rotation Periods}

\begin{figure}
	\centering
	\includegraphics[width=1\linewidth]{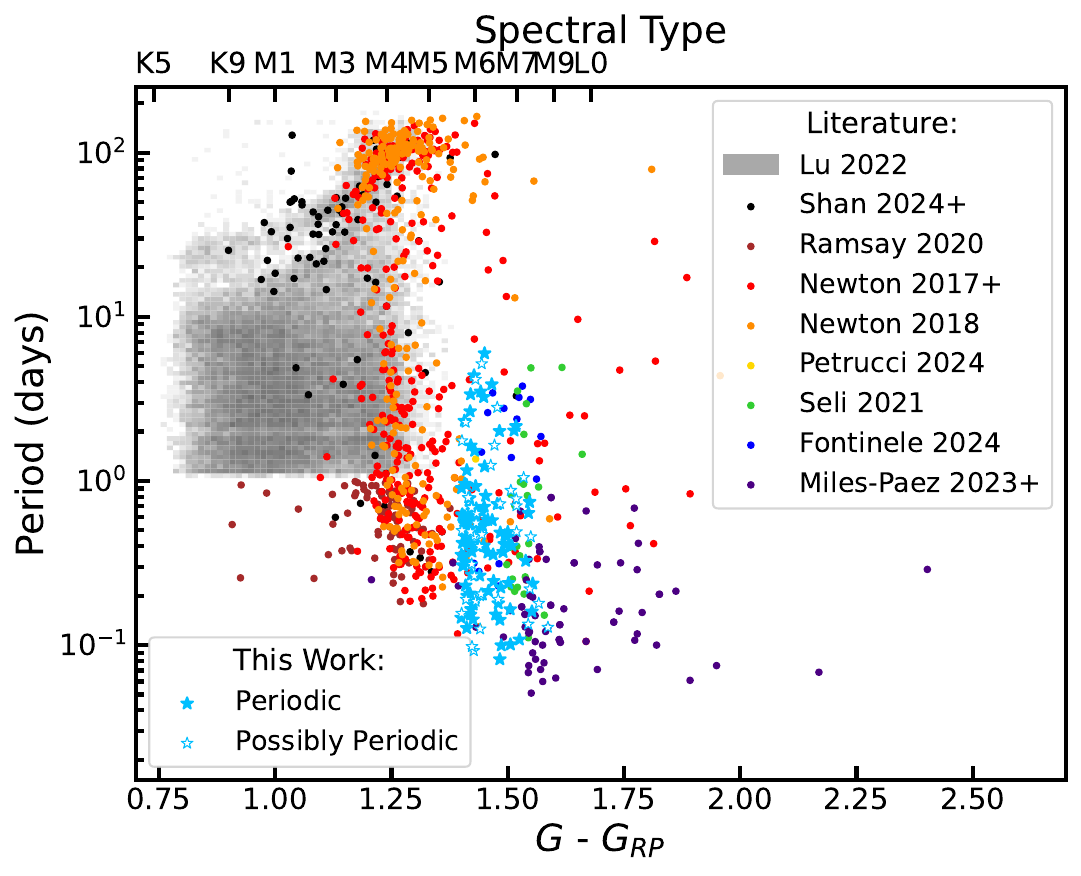}
    \includegraphics[width=1\linewidth]{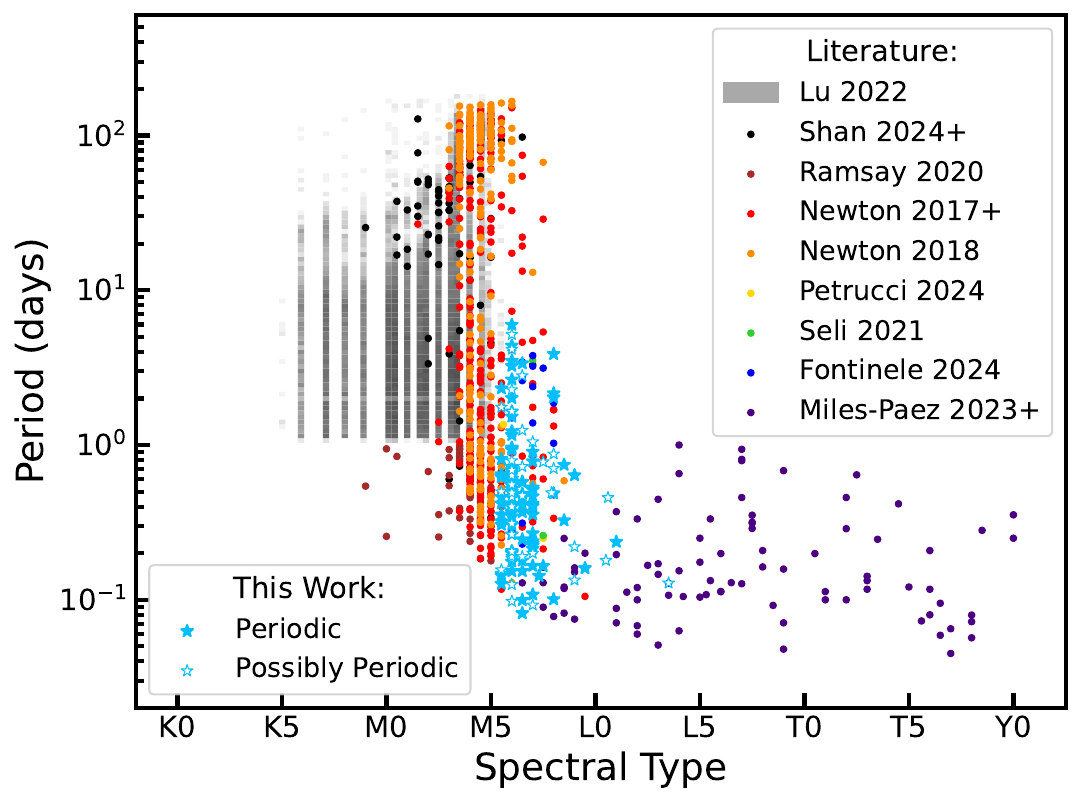}
	\caption{\textit{Top:} A comparison of Gaia $G - G_{RP}$ color (and correlated spectral type from \citet{pecaut2013}) to rotation period of M dwarfs and brown dwarfs. \textit{Bottom:} Period vs.\ spectral type for periodic targets in this work and in the literature. Where spectral types were unavailable, they were calculated from $G - G_{RP}$ color down to $G - G_{RP}$ = 1.6 mag and $J - H$ for fainter objects \citep{pecaut2013}. While the lower boundary on period appears to be astrophysical, the upper boundary for $>$M6 dwarfs stems from the difficulty of observing longer periods of faint objects. L22 is represented by a 2D histogram due to the large number of stars. References in the legend with ``+" indicate periods from multiple sources which were compiled by the named paper.}
	\label{fig:pergrplit}
\end{figure}

We combine our results with those from the previous works (from mid-M dwarfs to brown dwarfs, Section~\ref{sec:literature}) to create a wider context for the variability in our late-M dwarf sample. The distribution of these periods against Gaia $G-G_{RP}$ color is shown in the top panel of Figure \ref{fig:pergrplit}. In order to include brown dwarfs that are too optically faint in  Gaia, we also plot the period against spectral type (the bottom panel of Figure \ref{fig:pergrplit}). When spectral types were not available, we computed approximate spectral types using the conversion from \citet{pecaut2013} as seen in Section \ref{sec:spt}. Gaia $G - G_{RP}$ color was used for $G - G_{RP} \leq$ 1.6 mag, beyond which we used 2MASS $J-H$ color. 

Figure \ref{fig:pergrplit} reveals a lower envelope of periods that decreases from around 2.4 hours (0.1 days) for M4--M5 dwarfs, to $\approx$2 hours for M6--M9 late-M dwarfs, and $\approx$1 hour for cooler objects. This trend appears to extend to spectral types earlier than M4. However, there are very few data points at $<$ 1-day periods for $<$M4 since the largest study in this regime (L22) did not include periods shorter than 1 day. The result is consistent with previous studies \citep{somers2017, tannock2021, miles2023, pass2023} and indicates a limit to the maximum speed at which stars below the fully convective boundary can spin. 

\citet{tannock2021} note that this maximum speed is only about 35\% of what is expected from purely rotational break-up arguments. They posit that other considerations may play a more important role in determining structural integrity, such as the stability limit imposed by oblateness or the magnetic dynamo preventing further spin-up. Above the substellar boundary, \citet{miles2023} also consider more vigorous convection due to thermonuclear energy generation as a possible explanation for the speed limit as this could counteract spin-up.

There are two stars that do fall below this lower envelope in the $<$M4 regime from R20. They are Gaia DR3 2320602926320492672, an M0 dwarf with a 0.2568 day period, which is a known eclipsing binary \citep{samus2017}, and Gaia DR3 5138244683886239104, an M4 dwarf with a 0.2387 day period whose light curve also suggests that it is an eclipsing binary. Due to how short these periods are, it is likely that the secondary components in these binaries are tidally locked.
%, and thus the orbits trace the rotation period, which is why they are included in this analysis. 
As the stars are tidally locked, the orbital period is the same as the rotation period of the primary components. These rotation periods are shorter than those of single M-dwarfs, as the tidal forces between the binary stars that lead to tidal locking prevent spin-down by magnetic braking and, therefore, the stars retain their rapid rotation
%remain rapidly rotating 
\citep{fleming2019}.

While it may appear that there is an additional trend with the upper envelope of periods, this stems primarily from the difficulty of observing longer periods of fainter targets. As shown in the comparison to the literature, most studies on late-M dwarfs have been done with TESS, which tends to be biased against periods longer than a few days, as discussed in Sections \ref{sec:per_det} and \ref{sec:rot_stats}. Additionally, brown dwarf studies tend to focus on rapid rotators due to the difficulty and time required to observe long periods of very faint objects.

\subsection{A 76\% Increase in Rotation Periods Under 1 Day}

\begin{figure}
	\centering
	\includegraphics[width=0.8\linewidth]{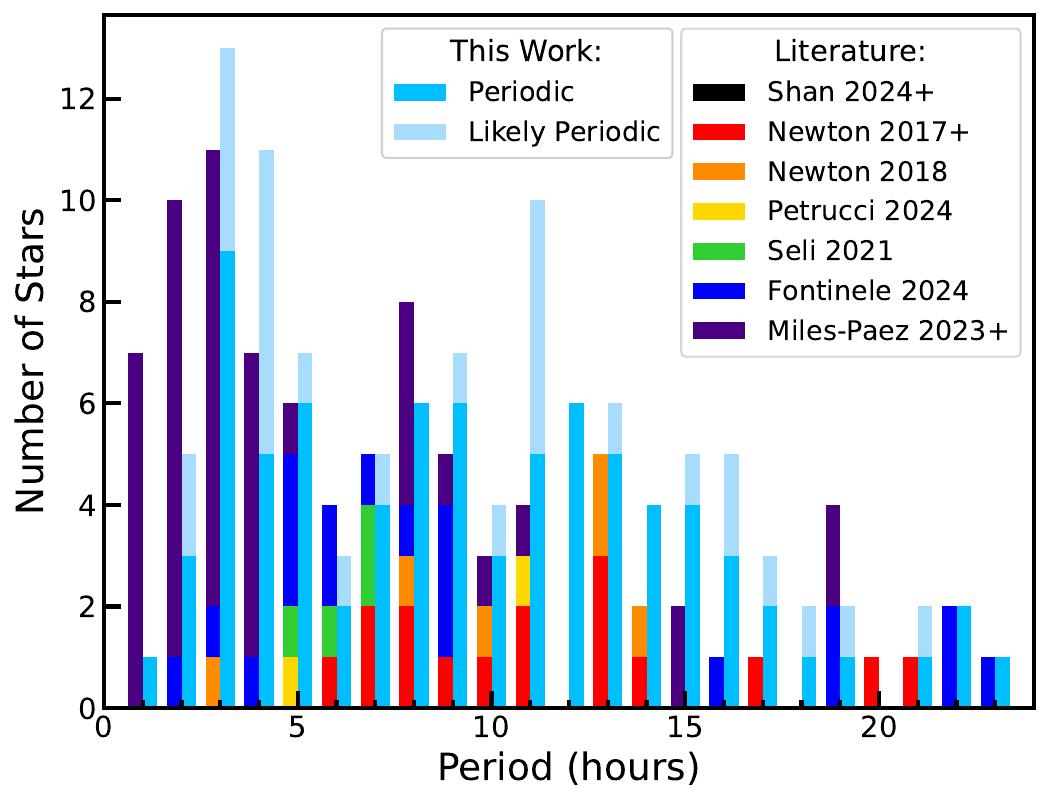}
	\caption{A histogram of late-M dwarf periods from this work and the literature, for objects with 1.4 mag $ < G-G_{RP} <$ 1.6 mag and periods less than one day. After accounting for previously known periods, our study expands the number of known $<$1-day periods by 76\%.}
	\label{fig:periodhist}
\end{figure}

Given our use of a 2-day moving average for detrending the TESS data in some aspects of our analysis, our results are most complete for periods shorter than a day. Figure \ref{fig:periodhist} shows the period distribution of such rapidly rotating 1.4 mag $ < G - G_{RP} <$ 1.6 mag ($\approx$M6--M9) dwarfs: the color range which contains all periodic stars in our sample. We see no obvious trend outside of the bias of TESS for shorter periods. Of the 199 late-M dwarfs with $<$1 day periods in this color range, 110 come from this work, and 86 have not been previously published. This represents a 76\% increase in known rapidly rotating late-M dwarfs.

\begin{figure}
	\centering
	\includegraphics[width=1\linewidth]{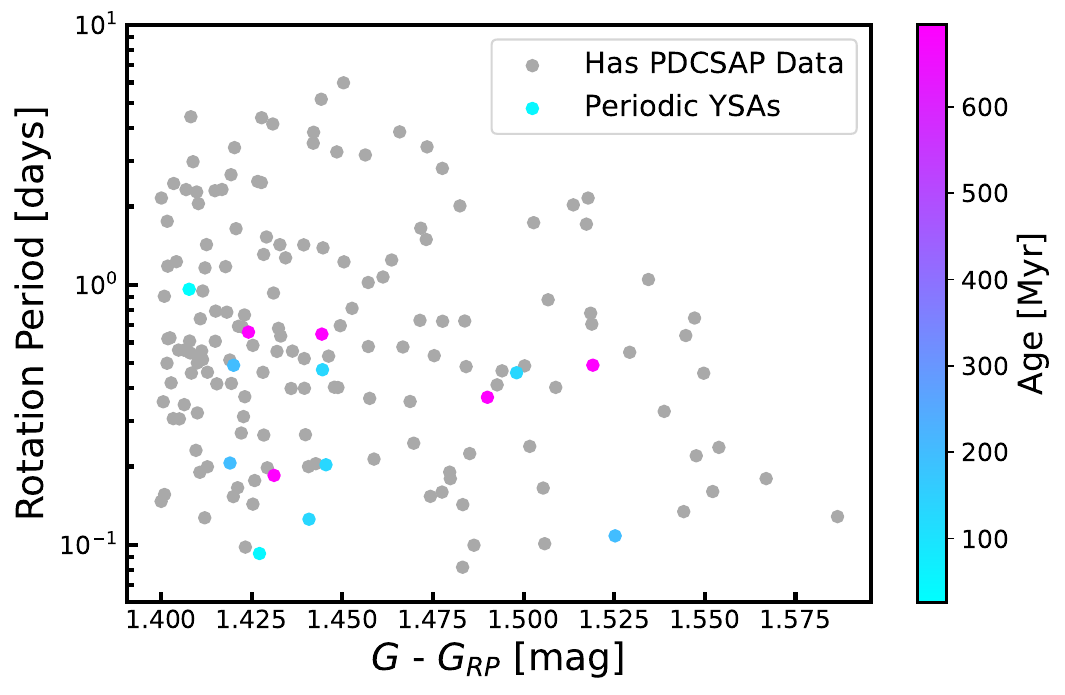}
	\caption{%Non-variable (black) or variable (color) late-M dwarf members of young stellar associations (YSAs) compared against our sample of 399 candidate single uncontaminated TESS late-M dwarfs with PDCSAP data
 %    %A comparison to targets in young stellar associations (YSAs) vs.\ field dwarfs 
 %    on a Gaia $G - G_{RP}$ color-magnitude diagram. The color of the point indicates the rotation period of the periodic young stellar association targets, while the size indicates the age of the young stellar association.
 Periodic late-M dwarf members of young stellar associations (YSAs, colored) compared against our sample of 399 candidate single uncontaminated TESS late-M dwarfs with PDCSAP data (gray) on a plot of rotation period vs. Gaia $G-G_{RP}$ color. The color of the point indicates the age of the young stellar association.}
	\label{fig:ysas}
\end{figure}

\begin{figure}
	\centering
	\includegraphics[width=0.8\linewidth]{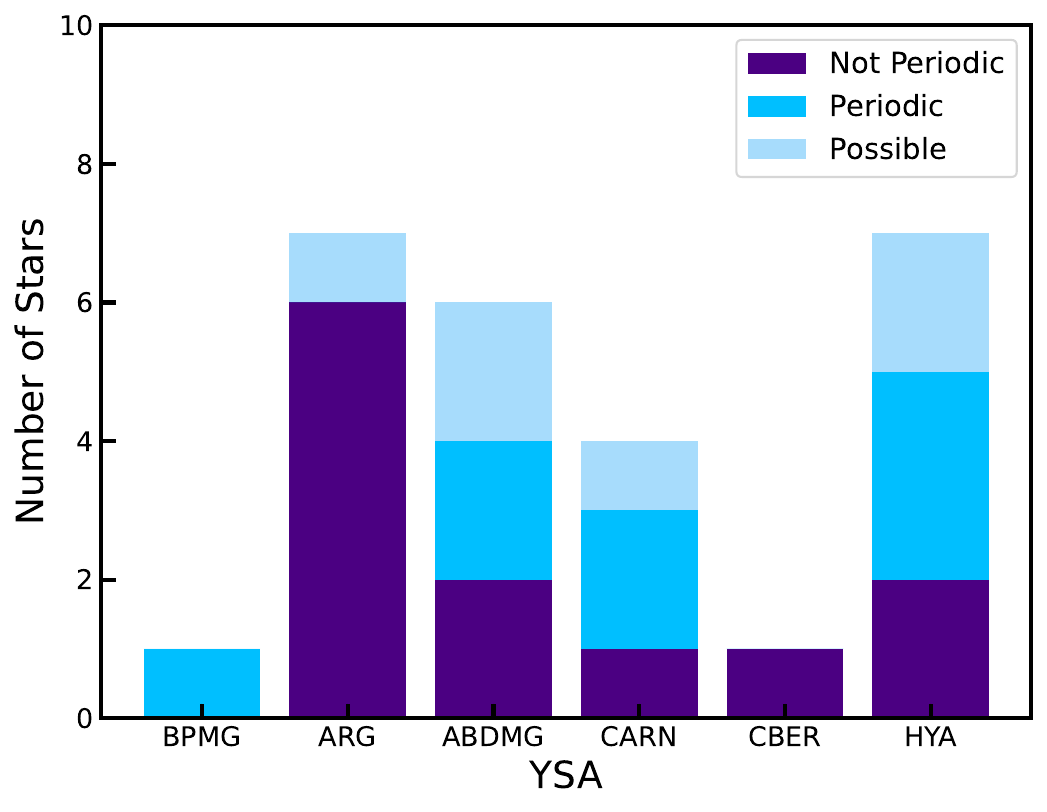}
	\caption{Distribution of periodic, possibly periodic, and not periodic stars in each young stellar association, from youngest to oldest.}
	\label{fig:ysabar}
\end{figure}

\subsection{Young Stellar Association Membership: Younger Late-M Dwarfs are More Commonly Variable}
\label{sec:membership}

\begin{deluxetable*}{cccccccccc}
    \tablecaption{Young stellar association memberships based on space motions and BANYAN $\Sigma$ \citep{gagne2018b}\label{tab:ysas}}
    \tablehead{\colhead{Gaia ID} & \colhead{RA} & \colhead{Dec} & \colhead{RV} & \colhead{Association} & \colhead{Association} & \colhead{Age} & \colhead{Ref} & \colhead{$P_{\rm rot}$} & \colhead{Period} \\ 
\colhead{} & \colhead{(deg, J2016.5)} & \colhead{(deg, J2016.5)} & \colhead{source} & \colhead{} & \colhead{probability (\%)} & \colhead{(Myr)} & \colhead{} & \colhead{(days)} & \colhead{status} 
}
\startdata
39612036694609152 & 58.834798365 & 14.658132847 &   & HYA & 99.76 & 695$^{+85}_{-67}$ & 1 &    &  not real \\ 
49324572659092992 & 64.236079392 & 20.876563189 &   & HYA & 99.74 & 695$^{+85}_{-67}$ & 1 &    & not real  \\ 
57083830512424448 & 55.229665494 & 19.496347844 &   & HYA & 99.63 & 695$^{+85}_{-67}$ & 1 & 0.6461$^{+0.0035}_{-0.0029}$ & real \\ 
65638443294980224 & 58.506768145 & 23.275780323 &   & HYA & 98.43 & 695$^{+85}_{-67}$ & 1 &  0.4907$^{+0.0039}_{-0.0025}$  &  possible \\ 
145168558871003776 & 69.271730757 & 22.857121735 &   & HYA & 99.51 & 695$^{+85}_{-67}$ & 1 &  0.1848$^{+0.0019}_{-0.0008}$ & possible \\ 
147258180719392512 & 71.688092698 & 24.610897642 &   & HYA & 97.18 & 695$^{+85}_{-67}$ & 1 & 0.6578$^{+0.0031}_{-0.0072}$ & real \\ 
1092729981791423232 & 133.35297065 & 65.552829843 &   & ABDMG & 69.34 & 133$^{+15}_{-20}$ & 2 &  & not real \\ 
1113083282052337792 & 101.12456163 & 72.113259260 &   & ABDMG & 99.54 & 133$^{+15}_{-20}$ & 2 & 0.4589$^{+0.0013}_{-0.0011}$ & real \\ 
1247310768216133888 & 209.76356327 & 20.807923900 &   & ARG & 81.43 & 45-50 & 3 &  & not real \\  
1439919263800588928 & 268.66867292 & 64.136002617 &   & ABDMG & 83.2 & 133$^{+15}_{-20}$ & 2 &    & not real \\ 
1590615235127351424 & 222.65154356 & 47.39971920 &   & CARN & 67.24 & 200 & 4 & 0.20619$^{+0.00053}_{-0.00033}$ & real \\ 
1603770170199226368 & 218.505205 & 50.663554739 & 7 & ARG & 66.63 & 45-50 & 3 &    & not real \\ 
2404524452684958208 & 343.63247830 & -15.995739533 &   & ABDMG & 83.32 & 133$^{+15}_{-20}$ & 2 & 0.20307$^{+0.00046}_{-0.00025}$ & real \\ 
2519880818919830784 & 33.989415887 & 5.6348557997 &   & ABDMG & 97.62 & 133$^{+15}_{-20}$ & 2 & 0.12535$^{+0.00034}_{-0.00051}$ & possible \\ 
2747910730136768768 & 7.979450940 & 6.8297386428 &   & CARN & 81.79 & 200 & 4 & 0.10819$^{+0.00031}_{-0.00037}$ & real \\ 
2781513733917711616 & 11.340895337 & 16.578880411 & 8 & ARG & 99.91 & 45-50 & 3 &    & not real  \\ 
2827319663210149120 & 354.18417485 & 21.893980056 &   & ARG & 54.45 & 45-50 & 3 &    & not real  \\ 
3410856112139584000 & 68.806963374 & 20.133481828 &   & HYA & 99.74 & 695$^{+85}_{-67}$ & 1 & 0.3692$^{+0.0011}_{-0.0015}$ & real \\ 
3582675080520660992 & 185.96594680 & -8.9799761756 &   & ABDMG & 99.65 & 133$^{+15}_{-20}$ & 2 & 0.4712$^{+0.0017}_{-0.0021}$ & possible \\ 
3655776282891553664 & 220.86562211 & 3.2818038328 &   & CARN & 73.72 & 200 & 4 &    & not real \\ 
3860610076465867520 & 156.35919905 & 5.210601996 & 7 & ARG & 92.52 & 45-50 & 3 & 0.09251$^{+0.00014}_{-0.00009}$ & possible \\ 
3949840839539031936 & 183.78959302 & 18.136523481 &   & CARN & 50.69 & 200 & 4 & 0.4902$^{+0.0043}_{-0.0046}$ & possible \\ 
3961053551744388352 & 191.13153175 & 25.789103162 &   & CBER & 99.94 & 562 & 5 &    & not real \\ 
4772152597972257664 & 82.096982072 & -52.306914031 &   & BPMG & 89.83 & 26 & 6 & 0.9612$^{+0.0068}_{-0.0032}$ & real \\ 
4873243032763432064 & 72.754459818 & -34.036862152 &   & ARG & 99.77 & 45-50 & 3 &    & not real  \\ 
6822023214270050432 & 334.154837088 & -19.478026885 &   & ARG & 51.05 & 45-50 & 3 &    & not real \\ 
\enddata
\tablecomments{HYA = Hyades open cluster, ABDMG = AB Doradus moving group, ARG = Argus association, CARN = Carina-Near moving group, CBER = Coma Berenices, BPMG = $\beta$ Pictoris moving group.\\\textbf{References:} (1) \citet{galindo2022}, (2) \citet{gagne2018a}, (3) \citet{zuckerman2018}, (4) \citet{zuckerman2006}, (5) \citet{silaj2014}, (6) \citet{malo2014}, (7) \citet{jonsson2020}, (8) \citet{faherty2016}}
\end{deluxetable*}

We checked our targets using the BANYAN $\Sigma$ tool \citep{gagne2018b} to determine their membership in young stellar associations. As they are all within 100 pc and we removed overluminous targets, we found, as expected, that the majority of stars have solutions consistent with field stars and thus should have ages on the order of 1 Gyr or higher. However, 26 objects of our 399 have a probability greater than 50\% of being in a young stellar association. An important caveat is that only 55 of the 399 targets had published radial velocities; therefore, the membership probability is tentative for many of the stars. As seen in %the middle panel of 
Figure \ref{fig:ysas}, the ages of the young stellar associations range from 26$\pm$3 Myrs \citep[$\beta$ Pictoris moving group,][]{malo2014} to 695$^{+85}_{-67}$ Myrs \citep[Hyades open cluster,][]{galindo2022}. A full list of our sample of objects in young stellar associations can be seen in Table \ref{tab:ysas}.

Of these objects, 14 have rotation periods (eight real, six possible; Figure~\ref{fig:ysabar}). Considering binomial uncertainties, this is 54$\pm$10\% of the young stellar association members, which can be compared to the total fraction of 31.6$\pm$2.4\% among just the field objects.  
% \sout{A breakdown of the membership of stars and their rotation status can be seen in Figure \ref{fig:ysabar}.} 
This indicates that periodic variability may be more commonly detected with TESS in young stars. This is likely because young stars are more magnetically active and therefore display more and higher amplitude variability, increasing our ability to detect them.

There are not enough periodic young stellar association targets %in YSAs studied 
in this work to robustly compare rotation periods to age. 
% \sout{However, this indicates that periodic variability may be more commonly detected with TESS in young stars. This is likely because young stars are more magnetically active and therefore display more and higher amplitude variability, increasing our ability to detect them.}
We only note that the period distribution of the 14 likely periodic late-M dwarf targets in young stellar associations is similar to that of the older late-M dwarfs in our sample, with a notable lack of $>$1 day periods among the young objects in Figure 27. However, the small sample size of young objects in our study precludes a robust conclusion.

\section{Summary and Conclusions}
\label{sec:summary_conclusions}
We have defined a robust sample of candidate single late-M dwarfs in TESS based on their Gaia and 2MASS photometry. In ensuring completeness to all stars with $\geq$M6 ($G-G_{RP}\geq1.40$ mag) photometric spectral types, our sample also includes several dozen M5--M5.5 dwarfs. At the cool end, our sample includes eight L dwarfs as late as L9.
We determined rotation periods for 133 (95 real and 38 possible) candidate single late-M dwarfs with TESS PDCSAP photometry out of 399 total with amplitudes ranging from 0.08\%--2.71\%. This increases the number of known rotation periods under 1 day for late-M dwarfs dwarfs by 76\%. In doing so, we compared the SPOC pipeline with PDCSAP data to \texttt{eleanor}, and found that while PDCSAP is better for identifying periods, \texttt{eleanor} is a good choice when PDCSAP data is unavailable, at least for periods $<$1 day.

Furthermore, we tested \texttt{TESS-Localize} on our mostly uncontaminated sample and found that not only can it identify contamination from stars $>$5 pixels away, but it is also a useful tool to determine whether a periodic signal is a systematic trend.

We found good agreement with other TESS studies when we compared our results to the literature. However, heavier scrutiny in every step of the period determination process allowed us to identify some previously reported periods as spurious due to contamination or systematics in the TESS data. 

We do not find any $>$3\% amplitudes, pointing to fairly uniform spot coverage on late-M dwarf photospheres. Furthermore, we find that the fraction of variable stars increases with the number of available TESS sectors, reaching up to $\sim 50\%$. This may represent an empirical limit set by viewing geometry and suggests that spot coverage is likely ubiquitous in late-M dwarfs, similar to what is observed in L and T dwarfs.

We also found a lower envelope for rotation periods that decrease with increasing color, in agreement with \citet{miles2023}. Finally, we note that a higher fraction of late-M dwarfs in young stellar associations had periods detectable by TESS compared to field stars.

% 

% Also note that the akcnowlodgment environment does not support long amounts of text. If you have a lot of people and institutions to acknowledge, do not use this command. Instead, create a new \section{Acknowledgments}.
\begin{acknowledgments}
% \section*{Acknowledgements}
We would like to thank the anonymous referee for their insightful comments that improved the scientific quality of this paper. This work was funded by the Canadian Space Agency under the FAST-AO 2019 (grant \#19FAWESB40) and FAST-AO 2021 (grant \#21FAUWOB12) programs. Additional funding was provided through a Canada Graduate Scholarship--Masters from the National Science and Engineering Research Council of Canada and an Ontario Graduate Scholarship. TESS data was obtained from the Mikulski Archive for Space Telescopes maintained by the Space Telescope Science Institutite (STScI). STScI is operated by the Association of Universities for Research in Astronomy, Inc., under NASA contract NAS5-26555. This research has made use of the SIMBAD database, operated at CDS, Strasbourg, France. PAMP acknowledges the use of the grant RYC2021-031173-I funded by MCIN/AEI/10.13039/501100011033 and by the ``European Union NextGenerationEU/PRTR''. We also would like to thank Dr.\ Megan Tannock for providing comments on the final draft of the paper.
\end{acknowledgments}


\vspace{5mm}
\section*{Data Availablity}
All TESS SPOC PDCSAP light curves used in this paper can be found at MAST  \citep{10.17909/t9-nmc8-f686,10.17909/t9-wpz1-8s54}.

\facilities{TESS \citep{ricker2015}, Gaia \citep{vallenari2023}}


% \texttt{numpy}, \texttt{scipy}, \texttt{matplotlib}, \texttt{pandas}, \texttt{astropy}, and \texttt{lightkurve}

\software{numpy \citep{harris2020},
          scipy \citep{virtanen2020},
          matplotlib \citep{hunter2007},
          pandas \citep{reback2020},
          astropy \citep{2013A&A...558A..33A,2018AJ....156..123A},
          lightkurve \citep{cardoso2018},
          eleanor \citep{feinstein2019},
          TESS-Localize \citep{higgins2023},
          PyAstronomy \citep{pya},
          wotan \citep{hippke2019},
          dynesty \citep{speagle2020},
          celerite2 \citep{celerite1, celerite2}
          }



% 


\bibliography{references}{}
\bibliographystyle{aasjournal}



\end{document}